\documentclass[12pt]{article}
\pdfoutput=1
\usepackage{bbm}
\usepackage{color}
\usepackage{authblk}
\usepackage{latexsym}
\usepackage{epsfig,amssymb,euscript, mathrsfs,cite}
\usepackage{amsmath}
\usepackage{mathrsfs} 
\usepackage{enumerate}
\definecolor{MyDarkBlue}{rgb}{0.15,0.15,0.45}
\usepackage[linktocpage=true]{hyperref}
\usepackage{subcaption}

\usepackage{mathtools}

\hypersetup{
colorlinks=true,
citecolor=MyDarkBlue,
linkcolor=MyDarkBlue,
urlcolor=MyDarkBlue,
pdfauthor={},
pdftitle={},
pdfsubject={hep-th}
}

\newsavebox{\ns}
\newsavebox{\dbrane}
\newsavebox{\dbshort}

\def\be{\begin{equation}}
\def\ee{\end{equation}}
\def\bea{\begin{eqnarray}}
\def\eea{\end{eqnarray}}

\newcommand{\nn}{\notag \\}

\def\cO{{\mathcal O}}
\def\eq#1 { \begin{equation} #1 \end{equation} }

\newcommand\R{\mathbb{R}}
\newcommand\Z{\mathbb{Z}}

\newcommand{\tr}{\mathrm{tr}}

\newlength{\sswidth}

\numberwithin{equation}{section}       % equation numbers in each section

\begin{document}

\begin{titlepage}

\vfill

\begin{flushright}
Imperial/TP/2020/JG/03\\
ICCUB-20-XXX
\end{flushright}

\vfill

\begin{center}
   \baselineskip=16pt
   {\Large\bf Spatially modulated and supersymmetric\\ mass deformations of $\mathcal{N}=4$ SYM}
  \vskip 1.5cm
  %\vskip 1.5cm
Igal Arav$^1$, K. C. Matthew Cheung$^1$, Jerome P. Gauntlett$^1$\\
Matthew M. Roberts$^1$ and Christopher Rosen$^2$\\
     \vskip .6cm
             \begin{small}\vskip .6cm
      \textit{$^1$Blackett Laboratory, 
  Imperial College\\ London, SW7 2AZ, U.K.}
        \end{small}\\
     \vskip .6cm
             \begin{small}\vskip .6cm
      \textit{$^2$Departament de F\'isica Qu\'antica i Astrof\'isica and Institut de Ci\`encies del Cosmos (ICC), \\
      Universitat de Barcelona, Mart\'i Franqu\`es 1, ES-08028, \\Barcelona, Spain.}
        \end{small}\\
                       \end{center}
\vfill

\begin{center}
\textbf{Abstract}
\end{center}
\begin{quote}
We study mass deformations of $\mathcal{N}=4$, $d=4$ SYM theory that are spatially modulated in one
spatial dimension and preserve some residual supersymmetry. We focus on generalisations of 
$\mathcal{N}=1^*$ theories and show that it is also possible, for suitably chosen
supersymmetric masses, to preserve $d=3$ conformal symmetry associated with a co-dimension one interface.  
Holographic solutions can be constructed using $D=5$ theories of gravity that arise
from consistent truncations of $SO(6)$ gauged supergravity and hence type IIB supergravity. 
For the mass deformations that preserve $d=3$ superconformal symmetry we construct
a rich set of Janus solutions of $\mathcal{N}=4$ SYM theory which have the same coupling constant on either side of the interface.
Limiting classes of these solutions give rise to RG interface solutions with $\mathcal{N}=4$ SYM on one side of the interface
and the Leigh-Strassler (LS) SCFT on the other, and also to a Janus solution for the LS theory. 
Another limiting solution is a new supersymmetric $AdS_4\times S^1\times S^5$ solution of type IIB supergravity.
\end{quote}

\vfill

\end{titlepage}

\tableofcontents

\newpage

\section{Introduction}\label{sec:intro}

Mass deformations of $\mathcal{N}=4$ $d=4$ SYM theory that preserve some supersymmetry have been extensively studied 
and are associated with rich dynamical features under RG flow 
(see e.g. \cite{Parkes:1982tg,Namazie:1982br,Leigh:1995ep,Freedman:1999gp,Girardello:1999bd,Dorey:1999sj,Polchinski:2000uf,Aharony:2000nt,Bobev:2018eer,Petrini:2018pjk,Bobev:2019wnf}). In this paper
we will explore mass deformations of $\mathcal{N}=4$ SYM theory that are spatially modulated in one of the three spatial dimensions
and yet still preserve some supersymmetry. A particularly interesting sub-class of such deformations also preserve
conformal symmetry with respect to the remaining three spacetime dimensions and  describe
co-dimension one, superconformal interfaces. 

The investigations of this paper are somewhat analogous to those that have been carried out in the context of ABJM theory. 
It is known that the homogeneous (i.e. spatially independent) mass deformations of ABJM theory
\cite{Hosomichi:2008jb,Gomis:2008vc} can be generalised to mass deformations
that depend on one of the two spatial coordinates and preserve 1/2 of the supersymmetry \cite{Kim:2018qle}. 
Further generalisations, preserving less supersymmetry, were subsequently analysed in \cite{Kim:2019kns}. Holographic descriptions of such deformations, preserving 1/4 of the supersymmetry of $D=11$ supergravity, were first constructed in \cite{Gauntlett:2018vhk} using a Q-lattice construction \cite{Donos:2013eha}.
The results of \cite{Gauntlett:2018vhk} included novel solutions that are holographically dual to boomerang RG flows which
flow from ABJM theory in the UV back to ABJM theory in the IR. The Q-lattice construction of \cite{Gauntlett:2018vhk} was substantially generalised in \cite{Arav:2018njv}, where it was shown that there is a novel class of $D=11$ supergravity solutions, again preserving 1/4 of the supersymmetry, 
which can be obtained by simply solving the Helmholtz equation on the complex plane. In addition to presenting a new set of solutions describing boomerang RG flows, the
construction of \cite{Arav:2018njv} also included the Janus solutions of \cite{DHoker:2009lky}. Finite temperature generalisations, using the Q-lattice construction, have been discussed in \cite{Ahn:2019pqy,Hyun:2019juj}.

Before continuing, we pause to note that there are various usages of  ``Janus" in the literature. 
In this paper it will refer to a co-dimension one, planar, conformal interface that has the same CFT on either side of the interface  (or the same up to a discrete parity symmetry). This includes the rich set of examples associated with
$\mathcal{N}=4$ SYM theory that are obtained by varying the coupling constant and theta angle as in, for example,
\cite{Bak:2003jk,Clark:2004sb,Clark:2005te,DHoker:2006vfr,DHoker:2006qeo,DHoker:2007zhm,DHoker:2007hhe,Gaiotto:2008sd,Suh:2011xc,Bobev:2020fon}. 
For these Janus configurations the CFT is being deformed by exactly marginal operators away from the interface, and in some cases there
are also additional sources for relevant operators located on the interface itself. For the Janus solutions of $D=11$ supergravity 
in \cite{DHoker:2009lky} the ABJM theory is deformed by relevant operators located on the interface, while for those in
\cite{Arav:2018njv,Bobev:2013yra} the ABJM theory is also deformed by relevant scalar operators that, generically, 
have spatial dependence away from the interface (see also \cite{Herzog:2019bom}).
 
In this paper we will show that there are interesting new supersymmetric Janus configurations of $\mathcal{N}=4$ SYM theory that arise from spatially modulated fermion and boson mass deformations but  {\it with the same coupling constant and theta angle on either side of the interface}.
In addition to these Janus solutions, we will also construct novel
holographic solutions dual to conformally invariant, co-dimension one interfaces, separating two different CFTs.
In these configurations the two CFTs are related by Poincar\'e invariant RG flow, so we refer to these as ``RG interfaces" (see \cite{Brunner:2007ur,Gaiotto:2012np}) and they are further discussed in \cite{1807495}. 

To determine which spatially modulated mass deformations of $\mathcal{N}=4$ SYM theory preserve supersymmetry, 
we deploy the technology developed in \cite{Festuccia:2011ws} (for theories with less supersymmetry, one can use the simpler approach of \cite{Anderson:2019nlc}). As in \cite{Maxfield:2016lok} we couple the theory to off-shell conformal supergravity and then take the Planck mass to infinity so that the supergravity fields become non-dynamical. In this limit one is left with an $\mathcal{N}=4$ supersymmetric field theory
coupled to the supergravity fields, which are now viewed as background couplings. 
The background couplings which preserve supersymmetry can then be determined by analysing the
supersymmetry transformations of the field theory coupled to the supergravity theory. 

We will focus our investigations on generalising 
the class of homogeneous mass deformations known as the
$\mathcal{N}=1^*$ theories. 
Recall that the field content of 
$\mathcal{N}=4$ SYM, in terms of an $\mathcal{N}=1$ language, consists of a vector multiplet coupled to three chiral superfields $\Phi_a$. 
Deforming the theory by adding a superpotential of the form $\delta\mathcal{W}\sim \sum_{a=1}^3m_a \tr \Phi_a \Phi_a$, where
$m_a$ are constant, complex mass parameters, defines the $\mathcal{N}=1^*$ class of theories. Three cases of particular interest are
(i) the ``one mass model" with (say) $m_1=m_2=0$, 
(ii) the  ``equal mass model", with $m_1=m_2=m_3$,  and
(iii) the $\mathcal{N}=2^*$ theory with (say) $m_1=m_2$ and $m_3=0$. 

We will show that all of these $\mathcal{N}=1^*$ theories can be generalised so that the mass parameters depend
on one of the three spatial coordinates while generically preserving $\mathcal{N}=1$ Poincar\'e supersymmetry with respect to the
remaining $d=3$ spacetime dimensions. For the case of the $\mathcal{N}=2^*$ theory there is an 
enhancement to $\mathcal{N}=2$ Poincar\'e supersymmetry in $d=3$.
Furthermore, it is possible
to suitably choose the mass parameters so that the $\mathcal{N}=1$ Poincar\'e supersymmetry is enhanced to
an $\mathcal{N}=1$ or $\mathcal{N}=2$ superconformal supersymmetry in $d=3$, respectively. 
This latter class of deformations thus defines a class of Janus 
configurations of $\mathcal{N}=4$ SYM theory, which have the novel feature that the coupling constant and the theta angle take 
the same value on either side of the interface, in contrast to previously constructed Janus configurations of $\mathcal{N}=4$ SYM.
At this point it is worth emphasising that our field theory results concerning supersymmetric Janus configurations of $\mathcal{N}=4$ SYM with constant coupling across the interface 
are complementary to the classification carried out in \cite{DHoker:2006qeo}, for which it was assumed that the coupling constant varies across the interface 
and that any additional deformations are proportional to the spatial derivative of the coupling constant.

The deformations we consider can also be studied holographically by constructing solutions of type IIB supergravity.
A convenient way to construct such solutions is to first construct them within the context of $D=5$ maximal gauged supergravity \cite{Gunaydin:1984qu,Pernici:1985ju,Gunaydin:1985cu} and then uplift them to $D=10$ using \cite{Lee:2014mla,Baguet:2015sma}. In fact, for the deformations we consider, we can utilise the consistent truncations of $D=5$ maximal gauged supergravity
discussed in \cite{Bobev:2013cja,Bobev:2016nua}, that couple the metric to a number of scalar fields. 
In particular, there is a consistent truncation model that is suitable for studying the mass deformations
for each of the three $\mathcal{N}=1^*$ theories mentioned above.

We will first derive the BPS equations that are relevant for spatially modulated mass deformations
of $\mathcal{N}=4$ SYM theory that preserve $ISO(1,2)$ symmetry. In this case the BPS equations are partial differential equations in two variables.
For this class of solutions we will also carry out a detailed
analysis of holographic renormalisation which is rather involved. There is a set of finite counterterms that one needs to consider in order
to have a supersymmetric renormalisation scheme. By demanding that the energy density of these BPS configurations is a total spatial derivative, 
thus leading to vanishing total energy, imposes some constraints on the counterterms (this is a complementary approach to
the ``Bogomol'nyi trick" of \cite{Freedman:2013ryh,Bobev:2013cja,Bobev:2016nua}). 
However, determining the full set of conditions required for a supersymmetric scheme is left to future work.

We will then focus on the BPS equations for the special subclass of solutions associated with Janus configurations. 
By writing the $D=5$ metric ansatz using a foliation by $AdS_4$ slices, the BPS equations become a set of ODEs that we then numerically solve
for each of the three consistent truncations. For each of the three models we find Janus solutions that approach the $\mathcal{N}=4$ SYM 
$AdS_5$ vacuum
on either side of the interface. We also find solutions that approach the $AdS_5$ vacuum on one side and are singular on the other,
as well as solutions that are singular on both sides, whose physical interpretation is unclear.

Additionally, for the one mass model we find new types of solutions that are further explored 
in \cite{1807495}. Recall that
homogeneous mass deformations in the one mass model induce an RG flow to the Leigh-Strassler (LS) fixed point \cite{Leigh:1995ep}. 
From the gravity side, within the truncation we consider, in addition to the
$\mathcal{N}=4$ SYM $AdS_5$ vacuum solution, there
are two additional  $AdS_5$ solutions, related by a $\mathbb{Z}_2$ symmetry, which we will denote LS$^\pm$, and each dual to the LS fixed point. Here we will construct novel solutions
that are dual to superconformal RG interfaces, approaching the $\mathcal{N}=4$ SYM  $AdS_5$ solution on one side and one of the two LS $AdS_5$ solutions 
on the other. We will also construct solutions that approach LS$^+$ $AdS_5$ on one side of the interface and LS$^-$ $AdS_5$
on the other, giving rise to Janus solutions of the Leigh-Strassler SCFT.

Some of the new explicit solutions that we construct here, including both the Janus and the RG interface solutions, are analogous to the supergravity solutions found in \cite{Bobev:2013yra} associated with ABJM theory. Here we will also obtain more detailed information on the sources and expectation values of various operators by using our holographic renormalisation results.

We also find a particularly interesting new feature for the equal mass model. This model is the most complicated to analyse since it has
four real scalar fields instead of three. Furthermore, one of the scalar fields is the dilaton dual to the coupling constant of
$\mathcal{N}=4$ SYM. While there are certainly rich Janus solutions for which
the coupling constant is different on either side of the interface, we focus our attention on solutions where it has the same value. We find a novel
class of Janus solutions that, surprisingly, approach a solution that is periodic in a bulk coordinate. By compactifying this coordinate
one then obtains a supersymmetric $AdS_4\times S^1$ solution. After uplifting to type IIB this is a new supersymmetric
$AdS_4\times S^1\times S^5$ solution that will be further explored in \cite{toappear3}.

The plan of the rest of the paper is as follows. In section \ref{sec:two} we determine the conditions for spatially dependent mass deformations
of $\mathcal{N}=4$ SYM theory to preserve supersymmetry, focussing on the three classes of $\mathcal{N}=1^*$ theories.
In section \ref{sugratrunc} we introduce the supergravity truncation of $D=5$ maximal gauged supergravity
\cite{Bobev:2013cja,Bobev:2016nua} that couples the metric to ten scalars, as well as three further truncations that are relevant for studying
the three classes of $\mathcal{N}=1^*$ theories. In sections \ref{secfour} and \ref{bpsjanus} we will present the BPS equations relevant for spatially dependent mass deformations that preserve $d=3$ superPoincar\'e and superconformal invariance, respectively.
In section \ref{susyjansol} we present and discus various new supergravity solutions, including the new Janus solutions as well
as the solutions dual to superconformal RG interfaces involving the LS fixed point for the one mass model and the novel
$AdS_4\times S^1$ solution for the equal mass model.
In section \ref{susyjansol} we conclude the paper with some discussion. We have also included three
appendices. Appendix \ref{deriveio21bps} contains some details on the derivation of the BPS equations.
Appendix \ref{appb} discusses in some detail the holographic renormalisation scheme that we use and
appendix \ref{appc} develops this further for the class of Janus solutions.

\section{Supersymmetric mass deformations of $\mathcal{N}=4$ SYM}\label{sec:two}

The coupling of $\mathcal{N}=4$ SYM to off-shell conformal supergravity \cite{Bergshoeff:1980is} was carried out in \cite{deRoo:1985np,deRoo:1984zyh,Bergshoeff:1985ms}. 
In \cite{Maxfield:2016lok} it was highlighted that this setup can be utilised to study supersymmetric deformations of
$\mathcal{N}=4$ SYM. Furthermore, some supersymmetric deformations of $\mathcal{N}=4$ SYM, including some known Janus 
configurations involving non-trivial profiles for the coupling constant and theta angle, were recovered in this language in \cite{Maxfield:2016lok}. 
Here we will use this formalism\footnote{An alternative approach, presumably equivalent to \cite{Maxfield:2016lok}, was given in
\cite{Choi:2017kxf,Choi:2018fqw}.} to study a new class of spatially dependent mass deformations that generalize the
homogeneous $\mathcal{N}=1^*$ mass deformations.  

We emphasise that in this section we 
use a ``{\it mostly plus}" $(-,+,+,+)$ convention for the metric. This should be contrasted with our later usage of a ``{\it mostly minus}" convention when we discuss supergravity solutions.

The possible bosonic deformations of $\mathcal{N}=4$ SYM are parametrised by the bosonic auxiliary fields of the off shell conformal supergravity
theory, which transform in particular representations of $SU(4)_R$, the global $R$-symmetry of the undeformed theory. 
The deformations transforming in the {\bf 1} of $SU(4)_R$ are associated with placing $\mathcal{N}=4$ SYM on a curved manifold as well as spatially modulating the gauge coupling and theta angle, encapsulated in the complex parameter $\tau\equiv \frac{\theta}{2\pi}+i\frac{4\pi}{g^2}$. In addition there are deformations $E_{ij}$ in the {\bf 10} of the $SU(4)_R$, 
$D^{ij}{}_{kl}$ in the ${\bf 20'}$, both Lorentz scalars, as well as a one-form $V_\mu^i\,_j$ and a
two-form $T_{\mu\nu}^{ij}$ transforming in the ${\bf 15}$ and {\bf 6}, respectively. 
In this paper, our primary focus will concern spatially dependent mass deformations of the bosonic and fermionic fields
involving $E_{ij}$ and $D^{ij}{}_{kl}$ and so in the following we set 
\begin{align}\label{nofifsix}
V_\mu^i\,_j=0,\quad T_{\mu\nu}^{ij}=0,\quad \tau=\text{constant}\,.
\end{align}
We note that the components $E_{ij}$ and $D^{ij}{}_{kl}$ are both complex and satisfy
\begin{align}\label{consDE}
E_{ij}&=E_{ji}\,,\nn
D^{ij}{}_{kl}&=-D^{ji}{}_{kl}=-D^{ij}{}_{lk}\,,\nn
(D^{kl}{}_{ij})^*&=D^{ij}{}_{kl}=\frac{1}{4}\epsilon^{ijmn}\epsilon_{klpq}D^{pq}{}_{mn}\,,\nn
D^{ij}{}_{kj}&=0\,,
\end{align}
with $i,j, \dots=1,\dots,4$. 

To see how these background fields couple to $\mathcal{N}=4$ SYM we recall that the field content of the latter
is given by gauge fields $A_\mu$,
fermions $\psi_i$, transforming in the ${\bf 4}$ of the $SU(4)_R$ and bosons $\phi^{ij}$, satisfying
$(\phi^{ij})^*=\phi_{ij}=\frac{1}{2}\epsilon_{ijkl}\phi^{kl}$, transforming in the {\bf 6}. 
The deformed action, in flat spacetime, is given by\footnote{Note that we largely follow the conventions and notation of \cite{deRoo:1985np,deRoo:1984zyh}. Thus, $\psi^i$ is a chiral spinor satisfying $\gamma_5\psi^i=+\psi^i$ transforming in the ${\bf\bar{4}}$ of $SU(4)$. 
The conjugate spinor, $\psi_i$, defined by $\psi_i\equiv B(\psi^i)^*$ (in contrast to the notation used in \cite{Maxfield:2016lok}) where $B^{-1}\gamma_aB=\gamma_a^*$, has the opposite chirality, $\gamma_5\psi_i=-\psi_i$, and transforms in
the ${\bf 4}$ of $SU(4)$. Note that we have changed the sign of $\left(M_\psi \right)_{ij}$ in \eqref{expdefsps} compared with \cite{Maxfield:2016lok}, in agreement with eq. (10) of \cite{deRoo:1985np}. In addition, the dependence on $\tau$ in 
\eqref{defsymact} is obtained from  \cite{deRoo:1985np,deRoo:1984zyh} by writing $\phi^a$, which can be used to parametrise $SU(1,1)/U(1)$, as $\phi^1=\frac{1}{2\sqrt{\text{Im}\tau/4\pi}}(1-i\frac{\tau}{4\pi})$ and
$\phi^2=\frac{1}{2\sqrt{\text{Im}\tau/4\pi}}(1+i\frac{\tau}{4\pi})$, again in contrast to \cite{Maxfield:2016lok}.}
\begin{align}\label{defsymact}
S=\int \!\! d^4y\,\tr\Big( &-\frac{1}{4g^2} F_{\mu \nu}F^{\mu \nu}-\frac{\theta}{32\pi^2} F_{\mu \nu} \ast F^{\mu \nu} -\tfrac{1}{2} \mathcal{D}_\mu \phi^{ij} \mathcal{D}^\mu \phi_{ij}- \bar \psi_i \gamma^\mu \mathcal D_\mu \psi^i  \nn
&-g{} \phi_{ij} [\bar \psi^i,{\psi^j}] -g{} \phi^{ij} [\bar {\psi_i},{\psi_j}] + \tfrac{1}{2}g^2[{\phi^{ij}},{\phi_{jk}}][{\phi^{kl}},{\phi_{li}}] \nn
& +\tfrac{1}{2} \phi_{ij} \left( M_\phi \right)^{ij}{}_{kl} \phi^{kl}  +\tfrac{1}{2} \bar \psi^i \left( M_\psi \right)_{ij} \psi^j  +\tfrac{1}{2}
 \bar \psi_i \left( \bar M_{\psi} \right)^{ij}  \psi_j \cr
& -\tfrac{2}{3}g\left(\bar M_{\psi} \right)^{kl} \phi^{ij} [{\phi_{ik}},{\phi_{jl}}]  
-\tfrac{2}{3}g \left(M_\psi \right)_{kl} \phi_{ij} [{\phi^{ik}},{\phi^{jl}}] \Big)\,.
\end{align}
Here the first two lines are essentially the undeformed action with
$F_{\mu \nu} = \partial_\mu A_\nu - \partial_\nu A_\mu + [{A_\mu},{A_\nu}]$, 
$\mathcal{D}_\mu \phi^{ij} = \partial_\mu \phi^{ij} + [{A_\mu},{\phi^{ij}}]$ and
$\mathcal{D}_\mu \psi^i  = \partial_\mu \psi^i  + [{A_\mu},{\psi^i}]$.
In the third and fourth lines we have used the following mass matrices for the bosons and fermions:
\begin{align}\label{expdefsps}
\left( M_\phi \right)^{ij}{}_{kl} &=  \frac{1}{2}D^{ij}{}_{kl}
-\frac{1}{12} \delta^{[i}{}_k \delta^{j]}{}_l \left( \bar E^{mn} E_{mn}\right), \cr
\left(M_\psi \right)_{ij} &= -\frac{1}{2} E_{ij} , \qquad
\left(\bar M_{\psi} \right)^{ij} = -\frac{1}{2}  \bar E^{ij} \,,
\end{align}
and ${\bar E}^{ij}\equiv (E_{ij})^*$.

At this stage $E_{ij}$ and $D^{ij}{}_{kl}$ can have arbitrary dependence on the the spacetime coordinates. The supersymmetry transformations of the matter fields for this deformed theory are given by\footnote{Note that $\epsilon^i$, $\eta^i$ both transform
in the ${\bf \bar{4}}$ of $SU(4)$ and satisfy the chirality conditions
$\gamma_5\epsilon^i=+\epsilon^i$, $\gamma_5\eta^i=-\eta^i$. The conjugate spinors $\epsilon_i$, $\eta_i$ transform in
the ${\bf 4}$ of $SU(4)$ with $\gamma_5\epsilon_i=-\epsilon_i$, $\gamma_5\eta_i=+\eta_i$.
We note that \eqref{trsfssnf} can be obtained from eq. (5) of \cite{deRoo:1985np}.}
\begin{align}\label{trsfssnf}
\delta A_\mu &= g{} \left( \bar\epsilon^i \gamma_\mu \psi_i + \bar\epsilon_i \gamma_\mu \psi^i \right) , \cr
\delta \psi^i & = -{1\over 2g{} }F_{\mu \nu} \gamma^{\mu \nu} \epsilon^i - 2 \mathcal{D}_\mu \phi^{ij} \gamma^\mu \epsilon_j + \bar E^{ij}\phi_{jk} \epsilon^k  - 2g{}  [{\phi^{ij}},{\phi_{jk}}] \epsilon^k - 2 \phi^{ij} \eta_j, \cr
\delta \phi^{ij} &= 2\bar \epsilon^{[i} \psi^{j]} - \epsilon^{ijkl} \bar\epsilon_k \psi_l\,.
\end{align}
Here $\epsilon^i$, $ \eta^i$ parametrise possible Poincar\'e and superconformal supersymmetries, respectively. Such supersymmetries will
only be present provided that there are solutions to the following equations 
\begin{align}\label{masteqs}
0  =& E_{ij}\epsilon^j\,,\nn
0  =& -\frac{1}{2}\varepsilon^{ijlm}\partial_\mu E_{kl}\gamma^\mu{\epsilon}_m + D^{ij}\,_{kl}\epsilon^l +\frac{1}{2}E_{kl}\bar{E}^{l[i}\epsilon^{j]}- \frac{1}{6} E_{ml}\bar{E}^{ml}\delta^{[i}_k \epsilon^{j]} -\frac{1}{6}E_{ml}\bar E^{m[i}\delta^{j]}_k\epsilon^l
\nn&-{ 1\over 2} \epsilon^{ijlm} E_{kl}\eta_m\,,\nn
0  =&2 \partial_\mu\epsilon^i- \gamma_\mu \eta^i\,,
\end{align}
which can be obtained from eq. (4.9),(4.10) of \cite{Bergshoeff:1980is}.

Notice that a complete basis of solutions to the last line of \eqref{masteqs} is given by
\begin{align}\label{spandsc}
\epsilon^i &= \mathrm{constant},\quad \eta^i = 0\,, \nn
\epsilon^i&=\frac{1}{2} y^\mu \gamma_\mu \eta^i,\quad \eta^i=\mathrm{constant}.
\end{align}
In the following, when we refer to solutions to the background equations \eqref{masteqs} with a given $\epsilon^i$ we
will always mean solutions as in the first line of \eqref{spandsc}, which are the Poincar\'e supersymmetries.
When referring to a solution with a given $\eta^i$, we will mean a solution as in the second line with an associated $\epsilon^i(y)$, 
which are the super conformal supersymmetries.

Here we will not attempt to find the most general solution to \eqref{masteqs}.
Instead we will focus on generalising some known homogeneous (i.e. spatially independent) mass deformations that,
moreover, can be studied holographically within the context of known truncations of $N=8$, $SO(6)$ gauged supergravity in $D=5$.
Specifically, we will consider the homogeneous $\mathcal{N}=1^*$ deformations and then 
allow for an additional dependence on just one of the three spatial coordinates.

To cast the $\mathcal{N}=1^*$ deformations in the present formalism we first recall that 
in terms of $\mathcal{N}=1$ language
the field content of 
$\mathcal{N}=4$ SYM  consists of a 
vector multiplet, which includes the gauge-field and the gaugino,
coupled to three chiral superfields $\Phi_a$ which transform
in the {\bf 3} of $SU(3)$ in the decomposition $SU(3)\times U(1)\subset SU(4)_R$.  The homogeneous $\mathcal{N}=1^*$ deformations are obtained by adding
to the superpotential 
the term $\delta\mathcal{W}\sim \sum_{a=1}^3 m_a \tr \Phi_a \Phi_a$, with
$m_a$ complex. 
This deformation gives rise to masses for the bosons and fermions in the chiral multiplets, 
but there is no mass deformation for the gaugino, consistent with preserving $\mathcal{N}=1$ supersymmetry. 
In the present formalism
these $\mathcal{N}=1^*$ deformations are associated with fermion mass deformations of the form
\begin{align}\label{diagE}
E_{ij} = \operatorname{diag}(m_1,m_2,m_3,0),
\end{align}
as well as associated bosonic mass deformations, parametrised by both $E_{ij}$ and
certain components of $D^{ij}{}_{kl}$ that are needed to preserve supersymmetry, as we shortly recall below.

Here we will generalise the $\mathcal{N}=1^*$ deformations by allowing $m_a$ to depend on one of the spatial coordinates: $m_a=m_a(y)$.
Let us first analyse the generic case with distinct $m_a\ne 0$, before discussing some special subcases. 
From the first line of \eqref{masteqs} we see that we can, generically, preserve $\mathcal{N}=1$ Poincar\'e supersymmetry
of the form
\begin{align}\label{sgenep}
\epsilon = (0,0,0,\epsilon^4)\,.
\end{align}
In the homogeneous case, with $m_a$ constant, it is not difficult to see that the middle equation of \eqref{masteqs} can be satisfied by
choosing
\begin{align}\label{deemassone}
D^{14}{}_{14}=D^{23}{}_{23}=\frac{1}{12}\left(|m_2|^2+|m_3|^2-2|m_1|^2\right)\,,\nn
D^{24}{}_{24}=D^{13}{}_{13}=\frac{1}{12}\left(|m_3|^2+|m_1|^2-2|m_2|^2\right)\,,\nn
D^{34}{}_{34}=D^{12}{}_{12}=\frac{1}{12}\left(|m_1|^2+|m_2|^2-2|m_3|^2\right)\,.
\end{align}
If we now allow $m_a=m_a(y)$, taking $i=1$, $j=2$ and $k=3$ in \eqref{masteqs}, for example, we also need to satisfy
\begin{equation}
 D^{12}\,_{34}\epsilon^4 -\frac{1}{2}\partial_y m_3 \gamma^y {\epsilon}_4 = 0\,,
\end{equation} 
and we recall here that $\epsilon_4$ is the spinor conjugate to $\epsilon^4$: $\epsilon_4=B(\epsilon^4)^*$.
This can be solved after imposing the following projection on the Poincar\'e supersymmetry parameters
\begin{equation}\label{genproj}
\gamma^y {\epsilon}_4 = e^{i\sigma} \epsilon^4\,,
\end{equation}
where $\sigma$ is a constant.
Indeed, we find that these and in fact all components of \eqref{masteqs} are satisfied by choosing
\begin{align}
D^{12}\,_{34}=(D^{34}\,_{12})^*&=\frac{1}{2}e^{i\sigma}\partial_y m_3, \nn
D^{23}\,_{14}=(D^{14}\,_{23})^*&=\frac{1}{2}e^{i\sigma}\partial_y m_1, \nn
D^{31}\,_{24}=(D^{24}\,_{31})^*&=\frac{1}{2}e^{i\sigma}\partial_y m_2\,,
\end{align}
as well as \eqref{deemassone}, with all
other components zero.

The projection condition \eqref{genproj} breaks half of the supersymmetry of the $\mathcal{N}=1^*$ theories leaving two
supercharges. As the deformation only depends on one of the three spatial dimensions, we have preserved Poincar\'e invariance
in the remaining $d=3$ spacetime dimensions. Thus, generically, the above deformation preserves 
$\mathcal{N}=1$ Poincar\'e supersymmetry in $d=3$. For special choices of $m_a(y)$ we can preserve 
$\mathcal{N}=1$ conformal supersymmetry in $d=3$. To see this, we take
\begin{align}
\eta^i=(0,0,0,\eta^4)\,.
\end{align}
Then again considering, for example, $i=1$, $j=2$ and $k=3$ in \eqref{masteqs}, as well as recalling the second line of
\eqref{spandsc}, we are led to the condition
\begin{align}
[2m_3+m_3'\gamma^y (y^\mu\gamma_\mu)]\eta_4=m_3'e^{i\sigma}(y^\mu\gamma_\mu)\eta^4
\end{align}
This can be solved, as well as all other conditions, by imposing the projection condition 
\begin{align}
\gamma^y \eta_4 = -e^{i\sigma} \eta^4\,.
\end{align}
and choosing
\begin{align}
m_a = \frac{\lambda_a}{y}\,,
\end{align}
for arbitrary complex constants $\lambda_a$.

Obviously these mass sources are singular at $y=0$, which is the location of the interface.
It is interesting to point out that we can choose different mass sources on either side of the interface and still
preserve the same supersymmetry. In particular, for
the conformally invariant case we can take
\begin{align}\label{mixedsce}
m_a&= \begin{cases}
      \frac{\lambda_a}{y},\qquad y>0,  \\
      \frac{\tilde\lambda_a}{y},\qquad y<0, 
    \end{cases}
\end{align}
with $\lambda_a$ and $\tilde \lambda_a$ independent constants. We will see such sources arise in the supergravity solutions 
that we construct later in the paper\footnote{\label{insref}
By suitable analytic continuation of the deformations that preserves conformal invariance, one can make
contact with the  mass deformations of the $\mathcal{N}=1^*$, $\mathcal{N}=2^*$ theories when placed on the round four-sphere as studied
in \cite{Bobev:2013cja,Bobev:2016nua}. Also note that the deformations we consider which are not conformally invariant do not involve any additional components of $E_{ij}$ and $D^{ij}{}_{kl}$, which is a useful observation in constructing supergravity solutions, as we discuss in the next section.}.

Let us now further consider three special cases that we will focus on in the sequel.

\subsection{$\mathcal{N}=1^*$ one mass model}\label{onemasmod}
For this case we assume that only one of the $m_a$ is non-zero, say $m_3$. We therefore consider
a fermion mass matrix $E_{ij}$ of the form
\begin{equation}\label{onemassee}
E = \operatorname{diag}(0,0,m,0).
\end{equation}
In the homogeneous case, with $m$ independent of $y$, we can preserve 
$\mathcal{N}=1$ supersymmetry in $d=4$ of the form \eqref{sgenep}
by also turning on
\begin{align}
D^{\alpha4}\,_{\alpha4} &= D^{\alpha3}\,_{\alpha3} = \frac{1}{12}|m|^2,\qquad\text{no sum on $\alpha\in \{1,2\}$}\,,\nn
D^{12}\,_{12} &=D^{34}\,_{34}= -\frac{1}{6}|m|^2\,.
\end{align}
These homogeneous deformations preserve a global $SU(2)\times U(1)_R$ symmetry. 
To see this we first decompose $SU(3)\times U(1)_1\subset SU(4)_R$ with the $SU(3)$ acting on each of the indices $i,j\in\{1,2,3\}$
in $E_{ij}$. We then further decompose $SU(2)\times U(1)_2\subset SU(3)$ to find that the global symmetry preserving
\eqref{onemassee} consists of this $SU(2)$ factor as well as a diagonal subgroup $U(1)_R\subset U(1)_1\times U(1)_2$.
Notice that the spinor \eqref{sgenep} parametrising the $\mathcal{N}=1$ Poincar\'e supersymmetry is charged under this $U(1)_R$
so it is in fact an $R$-symmetry of the $\mathcal{N}=1^*$  theory.

When $m=m(y)$, we can preserve $\mathcal{N}=1$ Poincar\'e supersymmetry in $d=3$ satisfying
\eqref{genproj} with
\begin{equation}
D^{12}\,_{34}=(D^{34}\,_{12})^*=\frac{1}{2}e^{i\sigma}\partial_y m\,.
\end{equation}
Notice that when $m=m(y)$ the $U(1)_R$ $R$-symmetry of the $\mathcal{N}=1^*$  theory is broken
and we are left only with an $SU(2)$ global symmetry. 
If we choose $m = \frac{\lambda}{y}$ then, in addition, we preserve $\mathcal{N}=1$ superconformal supersymmetry.

\subsection{$\mathcal{N}=1^*$ equal-mass model}\label{eqmassmod}
For this case, we assume $m_1=m_2=m_3$ so that the fermion mass matrix $E_{ij}$ takes the form
\begin{equation}
E_{ij}=\operatorname{diag}(m,m,m,0)\,.
\end{equation}
In the homogeneous case, with $m$ independent of $y$, taking $D^{ij}\,_{kl} = 0$
we preserve $\mathcal{N}=1$ supersymmetry in $d=4$ of the form \eqref{sgenep}.
By again considering the decomposition 
$SU(3)\times U(1)\subset SU(4)_R$ with the $SU(3)$ acting on each of the indices $i,j\in\{1,2,3\}$
in $E_{ij}$, we see that these homogeneous mass deformations maintain an $SO(3)\subset SU(3)$ global symmetry of the parent
$\mathcal{N}=4$ SYM theory.

When $m=m(y)$, we can preserve $\mathcal{N}=1$ Poincar\'e supersymmetry in $d=3$ satisfying
\eqref{genproj} with
\begin{align}
D^{\alpha4}\,_{\beta 4} &=D^{\alpha\beta}\,_{\gamma\delta}=0,\nn
D^{\alpha\beta}\,_{\gamma4} &= (D^{\gamma4}\,_{\alpha\beta})^*=\frac{1}{4}\epsilon^{\alpha\beta\delta}\epsilon_{\gamma\epsilon\phi}D^{\epsilon\phi}{}_{\delta4}=\frac{1}{2}\varepsilon^{\alpha\beta\gamma}e^{i\sigma}\partial_y m,
\end{align}
where, in this subsection, we have used the indices $\alpha,\beta,\gamma,\dots \in \{ 1,2,3\}$.
We observe that the spatially dependent deformations maintain the
$SO(3)$ global symmetry of the homogeneous case. If we choose $m = \frac{\lambda}{y}$ then, in addition, we preserve
 $\mathcal{N}=1$ superconformal supersymmetry.

\subsection{$\mathcal{N}=2^*$ model}\label{subsec:N=2Case}
For this case we assume that one of the masses is zero, say $m_3=0$, and the remaining two are equal
$m_1=m_2$. Thus, we consider a fermion mass matrix $E_{ij}$ of the form
\begin{equation}
E_{ij}=\operatorname{diag}(m,m,0,0).
\end{equation}
We again first consider the homogeneous case with $m$ independent of $y$. 
By choosing
\begin{align}
D^{12}\,_{12} &= D^{34}\,_{34} = \frac{1}{6} |m|^2\,,\nn
D^{\alpha p}\,_{\alpha p} &= -\frac{1}{12}|m|^2,\qquad \text{no sum on $\alpha\in \{1,2\}$ or $p\in \{3,4\}$}\,,
\end{align} 
we find that there is now an enhancement to $\mathcal{N}=2$ supersymmetry of the form
\begin{equation}\label{n2psu}
\epsilon = (0,0,\epsilon^3,\epsilon^4)\,.
\end{equation} 
These deformations preserve an $SU(2)_R\times U(1)\subset SU(4)_R$ global symmetry with $SU(2)_R$ the 
$R$-symmetry. 
To see this we can decompose $SU(2)_1\times SU(2)_2\times U(1)\subset SU(4)_R$ with $SU(2)_1$ and $SU(2)_2$ acting on
the indices $i,j\in\{1,2\}$ and $i,j\in\{3,4\}$, respectively. Then $SU(2)_R$ is $SU(2)_2$, and clearly rotates the 
$\mathcal{N}=2$ supersymmetry parameters in \eqref{n2psu}. The $U(1)\subset SU(2)_1$ symmetry acts as an $SO(2)$ rotation of the 
$1,2$ directions and leaves \eqref{n2psu} inert.

There can also be an enhancement of supersymmetry when $m=m(y)$ compared with the previous cases. 
From \eqref{masteqs} with $(i,j)=(1,p)$, with $p\in \{3,4\}$,
and $k=2$, we find the condition
\begin{equation}\label{n2condproj}
\frac{1}{2} \varepsilon^{pq}\partial_y m \gamma^y {\epsilon}_q + D^{1p}\,_{2q} \epsilon^q = 0\,,
\end{equation}
and $q\in \{3,4\}$.
To solve this we can consider a general projection condition of the form
\begin{equation}
\gamma^y {\epsilon}_p = M_{pq} \epsilon^q,
\end{equation}
where $M_{pq}$ is some constant $2\times2$ matrix. The consistency with the complex conjugate of this condition requires that $M$ satisfies
$\bar{M}^{pq} M_{qr} = \delta^p_r$.
If we define
\begin{equation}
\tilde{M}^p\,_r = \varepsilon^{pq} M_{qr}\,,
\end{equation}
then \eqref{n2condproj} implies
\begin{equation}
D^{1p}\,_{2q}=-\frac{1}{2}\partial_y m \tilde{M}^p\,_q\, .
\end{equation}
The tracelessness condition for $D$, in \eqref{consDE}, now requires that $\tilde{M}$ is traceless (and therefore $M$ is symmetric). $\tilde{M}$ is therefore a traceless matrix in $U(2)$. The remaining components of $D$ can then be inferred from \eqref{consDE}.

Note that the choice of the matrix $M$ breaks the $SU(2)_R$ $R$-symmetry of the homogeneous deformations down to a $U(1)_R$. This is expected since the spatially deformed solution preserves $\mathcal{N}=2$ Poincar\'e supersymmetry in $d=3$ and so we expect 
an $SO(2)=U(1)$ $R$-symmetry. The overall global symmetry is $U(1)_R\times U(1)$.
If we choose $m=\frac{\lambda}{y}$ then we preserve $\mathcal{N}=2$ superconformal symmetry in $d=3$ with
\begin{equation}
\eta_i = (0,0,\eta_3,\eta_4)\,,
\end{equation} 
and
\begin{equation}
\gamma^y {\eta}_p = -M_{pq} \eta^q\,.
\end{equation}

\section{Supergravity truncations}\label{sugratrunc}

To study the spatially dependent mass deformations holographically we would like to construct suitable
solutions of type IIB supergravity \cite{Schwarz:1983qr,Howe:1983sra}. A convenient way to do this is to construct solutions of maximally supersymmetric
$SO(6)$ gauged supergravity in $D=5$ and then uplift the solutions to $D=10$ using the results of \cite{Lee:2014mla,Baguet:2015sma}. The $D=5$ theory has 42 scalar fields, parametrising the coset
$E_{6(6)}/USp(8)$, which transform as ${\bf 1}$+${\bf 1}$, ${\bf 10}$+$ {\bf \overline{10}}$ and ${\bf 20'}$ with respect to $SO(6)$.
It is thus still rather unwieldy and so one seeks suitable consistent truncations of the $D=5$ theory. 

In fact, for general
constant, complex mass parameters $m_a$, associated with the $\mathcal{N}=1^*$ theories, there is an additional truncation that can be utilised,
as discussed in \cite{Bobev:2016nua}, which can also be used\footnote{A Euclidean version of the
same truncated theory was used in \cite{Bobev:2016nua} to study mass deformations of $\mathcal{N}=1^*$, $\mathcal{N}=2^*$ theories
on a four-sphere. Thus, by analytic continuation one can expect that the same truncation can be used to study the conformally invariant mass deformations that we study here. From footnote \ref{insref} we can also expect that
this truncation can be used for more general mass deformations.} when $m_a=m_a(y)$.
Specifically, one keeps the fields of $SO(6)$ gauged supergravity that are invariant under a $(\mathbb{Z}_2)^3$ symmetry of 
the $SO(6)\times SL(2,\mathbb{R})$ symmetry of the theory. This leads to an $\mathcal{N}=2$ $D=5$ gauged supergravity theory 
coupled to two vector multiplets and four hypermultiplets. This theory contains eighteen scalar fields which parametrise the coset
\begin{align}\label{etnscs}
\mathcal{M}_{18} = SO(1,1)\times SO(1,1) \times \frac{SO(4,4)}{SO(4)\times SO(4)}\,.
\end{align}
Schematically, these 18 scalar fields are dual to the following operators in $\mathcal{N}=4$ SYM theory:
\begin{align}\label{opfieldmapz}
\Delta=4:\qquad  \varphi,\quad\tilde\varphi&\quad \leftrightarrow \quad \tr F_{\mu\nu} F^{\mu\nu},\quad \tr F_{\mu\nu} *F^{\mu\nu} \,, \nn
\Delta=3:\qquad \qquad  \phi_i &\quad \leftrightarrow \quad \tr(\chi_i\chi_i+\text{cubic in $Z_i$})\,, \qquad i=1, 2, 3 \,, \nn
   \phi_4&\quad \leftrightarrow \quad \tr(\lambda \lambda+\text{cubic in $Z_i$})\,, \nn
\Delta=2:\qquad \qquad  \alpha_i &\quad \leftrightarrow \quad \tr(Z_i^2)\,, \qquad\qquad\qquad\qquad i=1, 2, 3\,,\nn
\beta_1 &\quad \leftrightarrow \quad \tr(|Z_1|^2+|Z_2|^2 -2 |Z_3|^2 )\,,\nn
   \beta_2 &\quad \leftrightarrow \quad \tr(|Z_1|^2-|Z_2|^2)\,.
   \end{align}
Here $\varphi$, $\tilde\varphi$ are real and are the ${\bf 1}$+${\bf 1}$ irreps of $SO(6)$ mentioned above. 
The fields $\phi_i$, $\phi_4$ are complex and arise from the ${\bf 10}$+${\bf \overline{10}}$ irreps. The three complex scalar fields
$\alpha_i$ and the two real scalars $\beta_1,\beta_2$, which parametrise the $SO(1,1)\times SO(1,1)$ factors in the scalar manifold $\mathcal{M}_{18}$,
arise from the ${\bf 20'}$ irrep\footnote{For reference, we note that 
$\beta_1,\beta_2$ are the two real scalars that appear in the $\mathcal{N}=2$ gauged supergravity model coupled to 
two vector multiplets\cite{Cvetic:1999xp}, often called the STU model.
If we supplement the STU model with complex $\phi_i,\phi_4$ we can obtain the charged cloud
truncation considered in \cite{Bobev:2010de}; the scalars in this truncation parametrise the coset
$SO(1,1)\times SO(1,1) \times \left[\frac{SU(1,1)}{U(1)} \right]^4$, but it is a different set of scalars of $SO(6)$ gauge supergravity
than those kept in \eqref{tensctrunc}. It is also different to the truncation of \cite{Chong:2004ce,Liu:2007rv}, which has scalars parametrising the same coset, but does not contain any scalars in the ${\bf 10}$ of $SO(6)$, dual to fermion mass deformations.}.
For the $\mathcal{N}=4$ SYM operators appearing on the right hand side of \eqref{opfieldmapz}, written in an $\mathcal{N}=1$ language, 
we note that $Z_i$ and $\chi_i$ are the bosonic and fermionic components of the chiral superfields $\Phi_i$ while
$\lambda$ is the gaugino of the vector multiplet. We also recall that the supergravity modes do not 
capture the Konishi operator $\tr(|Z_1|^2+|Z_2|^2 + |Z_3|^2 )$. 
Having source terms for the three complex scalars $\phi_i$ with $i=1,2,3$ is dual to deforming $\mathcal{N}=4$ SYM by the
the three fermion masses $m_a$ given in \eqref{diagE} (up to normalisation). Thus, allowing for spatially dependent sources for these
$\phi_i$ as well as suitable source terms for $\alpha_i$, $\beta_1$ and $\beta_2$, dual to bosonic masses 
we can holographically study spatially dependent mass deformations with
arbitrary complex $m_a(y)$ that we discussed in the last section. As far as we are aware, however, this
$\mathcal{N}=2$ $D=5$ gauged supergravity theory has not been explicitly constructed.

If we restrict to deformations for which the mass parameters $m_a(y)$ are all real, we can make further progress.
Indeed, as discussed in \cite{Bobev:2016nua}, we can then utilise a further truncation that just keeps the metric and ten scalar fields which parametrise the coset
\begin{align}\label{tensctrunc}
\mathcal{M}_{10} =  SO(1,1)\times SO(1,1) \times \left[\frac{SU(1,1)}{U(1)} \right]^4 \,.
\end{align}
This is achieved by truncating the $\mathcal{N}=2$ $D=5$ gauged supergravity theory using an additional $\mathbb{Z}_2$ symmetry, which 
lies in a certain $\left[ O(6) \times SL^\pm(2, \R) \right] / \Z_2 $ subgroup,  which is the actual symmetry group of ${\cal N} = 8$ gauged supergravity \cite{Pilch:2000fu}.
Although not a supergravity theory, this truncation can be used to obtain supersymmetric solutions of $SO(6)$ gauged supergravity and hence type IIB supergravity. The 10 real scalars consist of $\varphi$, $\phi_i$, $\phi_4$, $\alpha_i$ and $\beta_1,\beta_2$ all
of which are now real and dual to the obvious Hermitian generalisations of the operators  given in
\eqref{opfieldmapz}. We will refer to $\varphi$ as the ``dilaton".

As already noted, the two scalars $\beta_1,\beta_2$ parametrize the $SO(1,1)\times SO(1,1)$ 
factor in $\mathcal{M}_{10}$. The remaining eight scalars of this truncation, parametrising $\left[\frac{SU(1,1)}{U(1)} \right]^4$, can be packaged into
four complex scalar fields $z^A$ via
\begin{align}
     z^1 &= \tanh \Big[ \frac{1}{2} \big( \alpha_1 + \alpha_2 + \alpha_3 + \varphi 
       -i \phi_1 -i \phi_2 -i \phi_3 +i \phi_4 \big) \Big] \,,\nn
            z^2 &= \tanh \Big[ \frac{1}{2} \big( \alpha_1 - \alpha_2 + \alpha_3 - \varphi 
       -i \phi_1 +i \phi_2 -i \phi_3 -i\phi_4 \big) \Big] \,,\nn
     z^3 &= \tanh \Big[ \frac{1}{2} \big( \alpha_1 + \alpha_2 - \alpha_3 - \varphi 
       -i\phi_1 -i\phi_2 +i \phi_3 -i \phi_4 \big) \Big] \,,\nn
     z^4 &= \tanh \Big[ \frac{1}{2} \big( \alpha_1 - \alpha_2 - \alpha_3 + \varphi 
       -i \phi_1 +i \phi_2 +i\phi_3 +i\phi_4 \big) \Big] \,.
     \end{align}
The gravity-scalar part of the Lagrangian can be written as
\begin{align}\label{bulklag}
\mathcal{L} = -\frac{1}{4}R +3(\partial\beta_1)^2+(\partial\beta_2)^2+ \frac{1}{2}\mathcal{K}_{A\bar{B}}\partial_{\mu}z^{A}\partial^{\mu}\bar{z}^{\bar{B}} - \mathcal{P}  \,,
 \end{align}
where  ${\cal P}$ is the scalar potential  and $\mathcal{K}$ is the K\"ahler potential given by
\begin{align}\label{kpot}
\mathcal{K}=-\sum_{A=1}^4\log(1-z^A\bar z^A)\,.
\end{align}
The scalar potential can be conveniently derived from a superpotential-like quantity
\begin{align}\label{superpotlike}
\mathcal{W} \equiv ~&\frac{1}{L}e^{2\beta_1+2\beta_2}\left(1+z^1z^2+z^1z^3+z^1z^4+z^2z^3+z^2z^4+z^3z^4+z^1z^2z^3z^4\right)\nn
+ &\frac{1}{L}e^{2\beta_1-2\beta_2}\left(1-z^1z^2+z^1z^3-z^1z^4-z^2z^3+z^2z^4-z^3z^4+z^1z^2z^3z^4\right)\nn
 +&\frac{1}{L}e^{-4\beta_1}\left(1+z^1z^2-z^1z^3-z^1z^4-z^2z^3-z^2z^4+z^3z^4+z^1z^2z^3z^4\right)\,,
\end{align}
via
\begin{align}
   \mathcal{P} = \frac{1}{8}e^{\mathcal{K}}\left[\frac{1}{6}\partial_{\beta_1}\mathcal{W}\partial_{\beta_1}\overline{\mathcal{W}}
     +\frac{1}{2}\partial_{\beta_2}\mathcal{W}\partial_{\beta_2}\overline{\mathcal{W}}+\mathcal{K}^{\bar{B} A}    
      \nabla_{A}\mathcal{W}\nabla_{\bar{B}}\overline{\mathcal{W}} -\frac{8}{3}\mathcal{W}\overline{\mathcal{W}}\right]\,,
 \end{align}
where $\mathcal{K}^{\bar{B}A}$ is the inverse of ${\cal K}_{A \bar B}$ and 
$ \nabla_{A}\mathcal{W}\equiv \partial_A\mathcal{W}+\partial_A \mathcal{K}\mathcal{W}$.

The ten scalar model is invariant under $\mathbb{Z}_2\times S_4$ discrete symmetries acting on the bosonic fields, which also leave $\mathcal{W}$ invariant (we will not make precise
the discrete action on the fermions).
First, it is invariant under
the $\mathbb{Z}_2$ symmetry
\begin{align}\label{genz2}
z^A\to -z^A\,,\quad\Leftrightarrow\quad \{\phi_i,\phi_4,\alpha_i,\varphi\}\to- \{\phi_i,\phi_4,\alpha_i,\varphi\}\,.
\end{align}
Second, it is also invariant under an $S_3$ permutation symmetry which acts on $(z^2,z^3,z^4)$ as well as $\beta_1,\beta_2$
and is generated by two elements. The first generator acts via
\begin{align}\label{gens31}
z^2 &\rightarrow -z^4, \quad z^4 \rightarrow - z^2\,, \quad\Leftrightarrow\quad \phi_1 \leftrightarrow \phi_2\,,\quad \alpha_1 \leftrightarrow \alpha_2\,,\nn
\beta_2 &\rightarrow - \beta_2\,,
\end{align}
and the second generator acts via
\begin{align}\label{gens32}
z^3 &\rightarrow -z^4,\quad  z^4 \rightarrow - z^3\,,
\quad\Leftrightarrow\quad \phi_1 \leftrightarrow \phi_3\,,\quad \alpha_1 \leftrightarrow \alpha_3\,,\nn
\beta_1 &\rightarrow -\frac{1}{2}(\beta_1 + \beta_2)\,,\quad \beta_2 \rightarrow \frac{1}{2}(\beta_2 - 3 \beta_1)\,.
\end{align}
The action on $\phi_i$, $\alpha_i$ makes it clear that these generates $S_3$.
In addition, there is also an invariance under the interchange of pairs of the $z^A$:
\begin{align}\label{gens33}
z^1 &\leftrightarrow z^4,\quad  z^2 \leftrightarrow z^3\,,
\, &\Leftrightarrow\quad  (\phi_2,\phi_3)\to -(\phi_2,\phi_3)\,,\quad(\alpha_2,\alpha_3)\to -(\alpha_2,\alpha_3)\,,\nn
z^1 &\leftrightarrow z^2,\quad  z^3 \leftrightarrow z^4\,,
\, &\xLeftrightarrow{\mathbb{Z}_2}\quad  (\phi_1,\phi_3)\to -(\phi_1,\phi_3)\,,\quad(\alpha_1,\alpha_3)\to -(\alpha_1,\alpha_3)\,,\nn
z^1 &\leftrightarrow z^3,\quad  z^2 \leftrightarrow z^4\,,
\, &\xLeftrightarrow{\mathbb{Z}_2}\quad  (\phi_1,\phi_2)\to -(\phi_1,\phi_2)\,,\quad(\alpha_1,\alpha_2)\to -(\alpha_1,\alpha_2)\,,
\end{align}
where the equivalence in the last two uses \eqref{genz2}. Together with \eqref{genz2}-\eqref{gens32} these generate 
$\mathbb{Z}_2\times S_4$ as observed in \cite{Bobev:2020ttg}. The $S_4$ acts by permuting $(z^1,-z^2,-z^3,z^4)$ as well as transforming $\beta_1,\beta_2$ and we also point out that $\phi_4$ is left inert by the $S_4$ action.
We also point out that the $D=5$ theory is invariant under shifts of the dilaton
\begin{align}\label{dilshift}
\varphi\to\varphi+c\,,
\end{align}
where $c$ is a real constant\footnote{The transformation acts via fractional linear transformations
on the $z^A$ and changes $\mathcal{K}\to\mathcal{K}+f+\bar f$ and $\mathcal{W}\to e^{-f}\mathcal{W}$ with 
$f=f(z^A)$. One can check that this leaves 
$\mathcal{P}$ and hence the Lagrangian \eqref{bulklag} invariant.
One can also check that the BPS conditions \eqref{bpsten} are covariant provided that $\varepsilon_1\to e^{f/4-\bar f/4}\varepsilon_1$ and\
$\varepsilon_2\to e^{-f/4+\bar f/4}\varepsilon_2$.}.

Using the conventions of \cite{Bobev:2016nua} (also see appendix \ref{deriveio21bps}) a 
solution to the equations of motion of this model is supersymmetric provided that one can find a pair of symplectic Majorana
spinors $(\varepsilon_1, \varepsilon_2)$, with $\varepsilon_2 = -i\gamma^4\varepsilon_1^*$, that obey
\begin{align}\label{bpsten}
    \nabla_\mu \varepsilon_1 
  +\mathcal{A}_\mu \varepsilon_1
          - \frac{1}{6} e^{\mathcal{K}/2} \overline{\mathcal{W}} \gamma_\mu \varepsilon_2
    &= 0\,,\nn
\gamma^{\mu}\partial_{\mu}z^{A} \varepsilon_1 +\frac{1}{2} e^{\mathcal{K}/2}\mathcal{K}^{\bar{B} A}\left(\nabla_{\bar{B}}\overline{\mathcal{W}}\right) \varepsilon_2 &= 0\;, \nn
 3 \gamma^{\mu}\partial_{\mu}\beta_1 \varepsilon_1+\frac{1}{4}e^{\mathcal{K}/2}\left(\partial_{\beta_1}\overline{\mathcal{W}}\right) \varepsilon_2 &= 0\;, \nn
  \gamma^{\mu}\partial_{\mu}\beta_2 \varepsilon_1+\frac{1}{4}e^{\mathcal{K}/2}\left(\partial_{\beta_2}\overline{\mathcal{W}}\right) \varepsilon_2 &= 0\;,
 \end{align}
where we have defined
\begin{align}\label{caydef}
\mathcal{A}_\mu\equiv  -\frac{1}{4} \big[ \partial_A \mathcal{K} \partial_\mu z^A 
          -   \partial_{\bar B} \mathcal{K} \partial_\mu \bar{z}^{\bar B}  \big]\,.
\end{align}
We observe that the equations for preservation of supersymmetry given in \eqref{bpsten}
preserve the discrete symmetries \eqref{genz2}-\eqref{dilshift}.

The $AdS_5$ vacuum solution dual to $\mathcal{N}=4$ SYM theory,
is obtained when all of the scalars vanish. 
{For this solution, the discrete symmetries \eqref{genz2}-\eqref{gens33} }
are associated with discrete $R$-symmetries of 
$\mathcal{N}=4$ SYM theory.

Following \cite{Bobev:2016nua}, we now identify additional truncations of the ten scalar model
that can be used for real, spatially dependent mass deformations associated with each of the 
three cases considered in the last section.

\subsection{$\mathcal{N}=1^*$ one mass model}\label{sec:onemass}
This model is obtained by taking the limit where two of the masses vanish, which we take to be $m_1=m_2=0$ as in section \ref{onemasmod} and now with $m_3$ real.
Starting with the ten scalar model \eqref{tensctrunc}, we must have source terms for $\phi_3$ and $\alpha_3$. It turns out to be consistent to 
set $\phi_1=\phi_2=0$, $\alpha_1=\alpha_2=0$ as well as 
$\varphi=\phi_4=0$ and $\beta_2=0$:
\begin{align}\label{sscalp}
z^1 = z^2 =- z^3 = -z^4\,~~\text{and}~~ \beta_2=0\,,
\end{align} with
\begin{align}
     z^1 &= \tanh \big[ \frac{1}{2} \big( \alpha_3
       -i \phi_3  \big) \big] \,.
     \end{align}
This results in a three-scalar model with fields $z^1$, and $\beta_1$, which
we will use to construct supersymmetric Janus solutions later. 
{The discrete symmetries \eqref{genz2}-\eqref{gens33}} reduce to just the $\mathbb{Z}_2$ symmetry generated by $z^1\to -z^1$.

An important feature of this model is that in addition to the $AdS_5$ vacuum solution with vanishing scalars, and dual
to $\mathcal{N}=4$ SYM theory, there are also two other $AdS_5$ solutions, labelled LS$^\pm$. These two solutions
are related by the $\mathbb{Z}_2$ symmetry \eqref{genz2} and given by
\begin{align}\label{lsfpts}
z^1 = \pm i (2-\sqrt{3})\,,\qquad \beta_1 = -\frac{1}{6} \log(2)\,,\qquad
\tilde L= \frac{3}{2^{5/3}} L,
\end{align}
where $\tilde L$ the radius of the $AdS_5$ space for both LS$^\pm$ solutions.
When uplifted to type IIB these fixed point solutions preserve $SU(2)\times U(1)_R$ global symmetry and are each holographically dual
to the $\mathcal{N}=1$ SCFT found by Leigh and Strassler in \cite{Leigh:1995ep}.
By examining the linearised fluctuations of the scalar fields about the LS$^\pm$ vacua, we find
that $\alpha_3$ is dual to an irrelevant operator $\mathcal{O}^{\Delta=2+\sqrt{7}}_{\alpha_3}$ with conformal dimension $\Delta=2+\sqrt{7}$. The linearised modes involving 
$\phi_3$ and $\beta_2$ mix, and after diagonalisation we find modes that are dual to one relevant and one irrelevant operator in the LS SCFT, which we label
$\mathcal{O}^{\Delta=1+\sqrt{7}}_{\phi_3,\beta_2}$ and $\mathcal{O}^{\Delta=3+\sqrt{7}}_{\phi_3,\beta_2}$ with dimensions
$\Delta=1+\sqrt{7}\sim3.6$ and $\Delta=3+\sqrt{7}$, respectively.

Note that if we set $\alpha_3=0$, in this $D=5$ model we obtain a model with two real scalars
which is the same as that used to construct the homogeneous RG flows associated with
the $\mathcal{N}=1^*$ one mass model. These RG flows, which preserve $SU(2)\times U(1)_R$ global
symmetry, flow to the Leigh-Strassler fixed point \cite{Leigh:1995ep} in the IR and were constructed in
\cite{Freedman:1999gp} and uplifted to type IIB in \cite{Pilch:2000fu}. This gravitational model, with
$\alpha_3=0$, preserves the $SU(2)\times U(1)_R$ global symmetry\footnote{\label{foot8}For orientation,
note that if one keeps an $SU(2)\times U(1)\subset SU(3)\subset SO(6)$ invariant sector of $SO(6)$ gauged supergravity, one obtains
an $\mathcal{N}=2$ supergravity coupled to one vector multiplet and one hypermultiplet \cite{Pilch:2000fu}.
The five scalars parametrise the coset $SO(1,1)\times SU(2,1)/[SU(2)\times U(1)]$.
With $\beta_2$ the $SO(1,1)$ factor, the remaining coset is obtained by supplementing $\phi_3$ with a complex partner,
associated with a complex fermion mass, and two more scalars $\varphi$, $\tilde\varphi$ dual to operators as in
\eqref{opfieldmapz}. } 
and since the $U(1)_R$ is broken when the mass deformations
are spatially modulated, as discussed in section \ref{onemasmod}, it cannot be used in this context.

\subsection{$\mathcal{N}=1^*$ equal-mass model}\label{eqmassctrunc}
For this model we have $m_1=m_2=m_3\equiv m$, as in section \ref{eqmassmod}, and here we are considering $m$ to be real.
We must have ${\phi}_1={\phi}_2={\phi}_3$ as well as ${\alpha}_1 = {\alpha}_2= {\alpha}_3$ and both non-zero, associated with
the sources for the fermion and boson mass deformations. It turns out to be inconsistent to further set the gaugino condensate ${\phi}_4$ or  the scalar $\varphi$ to zero. However, it is consistent to set $\beta_1=\beta_2 = 0$. Thus, setting
\begin{align}
\label{eqmasstrunc}
   z^4=-z^3=-z^2\,,~~~ \text{and}\qquad
   \beta_1=\beta_2 = 0\,,
\end{align}
in the ten scalar model \eqref{tensctrunc}
leads to a model with four scalars,  parametrised
by $(z^1,z^2)$ 
with
\begin{align}\label{4scmod}
     z^1 &= \tanh \big[ \frac{1}{2} \big( 3\alpha_1 + \varphi 
       -i 3\phi_1+ i \phi_4 \big) \big] \,,\nn
            z^2 &= \tanh \big[ \frac{1}{2} \big( \alpha_1 - \varphi 
       -i \phi_1 -i\phi_4 \big) \big] \,.
     \end{align}
The discrete symmetries \eqref{genz2}-\eqref{gens33} reduce to the symmetry generated by $(z^1,z^2)\to -(z^1, z^2)$.
This truncation is invariant under shifts of the dilaton \eqref{dilshift}.
The K\"ahler potential \eqref{kpot} is now
\begin{align}\label{kpotequal}
\mathcal{K}=-\log(1-z^1\bar z^1)-3\log(1-z^2\bar z^2)\,,
\end{align}
and an explicit expression for the potential $\mathcal{P}$ can be found in (3.8) of \cite{Bobev:2019wnf}.
We use this four scalar model to construct supersymmetric Janus solutions later.

We note that this four scalar model can be further truncated to give a theory with two real
scalars, by setting $\alpha_1=\varphi=0$. This theory keeps $\phi_1$, associated with real $SO(3)\subset SU(3)_R$ invariant fermion masses,
and the gaugino condensate field $\phi_4$. This model is in fact the same model as that used by GPPZ
\cite{Girardello:1999bd} to construct RG flows associated with homogeneous $SO(3)$ invariant mass deformations
(and uplifted to type IIB in \cite{Petrini:2018pjk,Bobev:2018eer} extending \cite{Pilch:2000fu}); 
see (3.12) of \cite{Bobev:2019wnf} for the explicit field redefinition. It cannot be used for spatially dependent
masses, however.

For the equal mass model, with spatially dependent complex masses, there is in fact an alternative consistent truncation that can be used.
By keeping an $SO(3)\subset SU(3)\subset SO(6)$ invariant sector of $SO(6)$ gauge supergravity, one can obtain
an $\mathcal{N}=2$ supergravity coupled to two hypermultiplets \cite{Pilch:2000fu,Kol:2016ucd}. The 8 scalars of this theory
parametrise the quaternionic-K\"ahler manifold
\begin{align}
\mathcal{M}_{SO(3)} =  \frac{G_{2(2)}}{SU(2)\times SU(2)} \,.
\end{align}
The scalar fields of this model consist of adding a complex partner to the four real scalars $\alpha_1,\varphi,\phi_1,\phi_4$ of the four-scalar model \eqref{4scmod}. Although we will not utilise this truncation in this paper, it is a natural arena for additional investigations of spatially
dependent masses for the equal mass model.

\subsection{$\mathcal{N}=2^*$ model}\label{n2modeltrunc}
This model is obtained by setting two of the masses to be equal and one to be zero and specifically we consider
$m_1=m_2\ne 0$ and $m_3=0$, as in section \ref{subsec:N=2Case}, now with $m_1$ real. To study this case we can consistently 
set ${\phi}_1={\phi}_2$, ${\alpha}_1 = {\alpha}_2$  and $\beta_1\ne 0$, while imposing ${\alpha}_3 = {\phi}_3={\phi}_4={\varphi}=\beta_2=0$
in the ten scalar model \eqref{tensctrunc}. Equivalently, we set
\begin{align}
z^1 = z^3\,,\qquad   z^2=z^4=\beta_2=0\,.
\end{align}
with
\begin{align}
     z^1 &= \tanh \big[ \alpha_1 
       -i \phi_1 \big] \,,
             \end{align}
leading to a three-scalar model, parametrised by $z^1$ and $\beta_1$, discussed in \cite{Bobev:2013cja},
which
we will use to construct supersymmetric Janus solutions later.
This model is invariant under the discrete symmetry generated by $z^1\to -z^1$.

Note that if we set $\alpha_1=0$, we obtain a gravitational model with two real scalars
which is the same\footnote{One should make the identification between scalar fields here and in \cite{Pilch:2000ue}
via $\cos2\phi_1= 1/\cosh 2\chi$.} as that used to construct the RG flows associated with
the homogeneous $\mathcal{N}=2^*$ deformations in \cite{Pilch:2000ue}. These RG flows preserve $SU(2)_R\times U(1)$ global
symmetry. Thus, this gravitational model
cannot be used\footnote{Similarly, the $\mathcal{N}=4$ gauge supergravity
theory discussed in \cite{Freedman:1999gp} with 11 scalars parametrising the coset $SO(1,1)\times SO(5,2)/[SO(5)\times SO(2)]$
cannot be used to study spatially dependent $\mathcal{N}=2^*$ mass deformations. Note that this truncation is obtained by considering
an $SU(2)$ invariant sector of $SO(6)$ gauged supergravity. If one further restricts to an $SU(2)\times U(1)$ invariant sector then
one should get the $\mathcal{N}=2$ gauged supergravity mentioned in footnote \ref{foot8}.} to study spatially dependent
mass deformations for the $\mathcal{N}=2^*$ model since, as discussed in section \ref{subsec:N=2Case},
the spatial dependence breaks $SU(2)_R\times U(1)$ down to $U(1)_R\times U(1)$.

\section{Supersymmetric mass deformations with $ISO(1,2)$ symmetry}\label{secfour}

In this section we will discuss the BPS equations that are associated with  
supersymmetric mass deformations with $ISO(1,2)$ symmetry. Furthermore, in appendix \ref{appb} we
will analyse holographic renormalisation for this class of solutions, which will be useful in future studies of these solutions
as well as when we discuss physical properties of the supersymmetric Janus solutions, as a special sub-class.
We leave the details in the appendix \ref{appb}, but we highlight here that there are a number of interesting
issues, including a large number of possible finite counterterms, subtleties in obtaining a supersymmetric renormalisation
scheme and interesting source terms which appear in the conformal anomaly.

In the context of the ten scalar truncation discussed in section \ref{sugratrunc}, we 
consider the ansatz
\begin{align}\label{iso21metans}
ds^2 = e^{2A}(dt^2 - dy_1^2 - dy_2^2) - e^{2V} dx^2 - N^2 dr^2\,,
\end{align}
with $A,V,N$ and the scalars $z^A$, $\beta_1$, $\beta_2$, functions of $(x,r)$ only. This ansatz preserves
$ISO(1,2)$ symmetry associated with the coordinates $t,y_1,y_2$. The coordinates $r,x$, together, parametrise both
the remaining field theory direction, upon which the mass deformations depend, as well as the holographic radial coordinate.
There is some residual gauge freedom in this ansatz, associated with reparametrising $(r,x)$ and in practise we will
find it convenient to fix this in different ways in the sequel.

In appendix \ref{deriveio21bps} we derive a set of BPS equations associated with the preservation, in general, of 1/4 of the supersymmetry. We use the orthonormal frame $(e^{0},e^{1},e^{2},e^{3},e^{4})= (e^A dt, e^A dy_1, e^A dy_2, e^V dx, Ndr)$ and note that the supersymmetry transformations are parametrised by pair of symplectic Majorana spinors
$\epsilon_1$ and $\epsilon_2$. We find that
the Killing spinors are independent of $t,y_1,y_2$ and satisfy the following projection condition 
\begin{align}\label{basicprojtext}
\gamma^{012}\epsilon_1&=-i\kappa \epsilon_1\,,
\end{align}
with $\kappa=\pm 1$,
which implies $\gamma^{012}\epsilon_2=i\kappa \epsilon_2$ as a result of the 
Majorana condition $\epsilon_2=-i\gamma^4\epsilon_1^*$, as well as 
\begin{align}\label{basicprojtexttwo}
\gamma^4\epsilon_1=e^{i\xi}\epsilon_2\,,
\end{align}
where $\xi$ is a function of $(x,r)$. It is also worth noting that we then have 
$\epsilon_1^*=ie^{-i\xi}\epsilon_1$.
The associated system of BPS equations are then given by 
\begin{align}\label{bpsderxi1appatext}
e^{-V}\partial_x A+i\kappa N^{-1}\partial_r A-\frac{i\kappa}{3}e^{\mathcal{K}/2}e^{-i\xi}\bar{\mathcal{W}}&=0\,,\nn
-e^{-V}\partial_x\xi-\kappa N^{-1}\partial_r V
+2ie^{-V}\mathcal{A}_{x}
+\frac{\kappa}{3}e^{\mathcal{K}/2}\text{Re}(e^{-i\xi}\bar{\mathcal{W}})&=0\,,\nn
-N^{-1}\partial_r\xi+\kappa N^{-1}e^{-V}\partial_x N
+2iN^{-1}\mathcal{A}_{r}
+\frac{1}{3}e^{\mathcal{K}/2}\text{Im}(e^{-i\xi}\bar{\mathcal{W}})&=0\,,
\end{align}
where we recall the definition of $\mathcal{A}_\mu$ given in \eqref{caydef},
as well as 
\begin{align}\label{fermvsremaintext}
i\kappa e^{i\xi}\left(e^{-V}\partial_x +i\kappa N^{-1}\partial_r \right)z^A&=\frac{1}{2}e^{\mathcal{K}/2}\mathcal{K}^{\bar B A}\nabla_{\bar B}\bar{\mathcal{W}}\,,\nn
i\kappa e^{i\xi}\left(e^{-V}\partial_x +i\kappa N^{-1}\partial_r \right)\beta_1&=\frac{1}{12}e^{\mathcal{K}/2}
\partial_{\beta_1}\bar{\mathcal{W}}\,,\nn
i\kappa e^{i\xi}\left(e^{-V}\partial_x +i\kappa N^{-1}\partial_r \right)\beta_2&=\frac{1}{4}e^{\mathcal{K}/2}
\partial_{\beta_2}\bar{\mathcal{W}}\,.
\end{align}
The dependence of the Killing spinor on $(x,r)$ can be determined and we
find that they are given by
$\epsilon_1=e^{A/2}e^{i\xi/2}\eta_0$ where $\eta_0$ is a constant spinor satisfying the projection given in \eqref{basicprojtext}.
We note that these BPS equations are not all independent, and there is also an issue of consistency, given the reality 
of various functions entering these equations, a point we return to below.
Observe that these BPS equations are invariant under 
\begin{align}
r\to-r,\qquad x\to -x,\qquad \xi\to\xi+\pi\,.
\end{align}

It is interesting to point out that if we choose the gauge $N=e^V$, then the equations can be written in a simplified form, analogous to
what was seen in \cite{Arav:2018njv}. 
We introduce the complex coordinate
$w=r-i\kappa x$ and the $(1,0)$ form $B$ defined by
\begin{align}
B\equiv \frac{1}{6}e^{i\xi+V+\mathcal{K}/2}{\mathcal{W}}d w\,.
\end{align}
The equations
\eqref{bpsderxi1appatext} can then be cast in the form
\begin{align}\label{parelbeetext}
\partial A&=B\,,\nn
\bar \partial B&= - \mathcal{F}B\wedge \bar B\,,
\end{align}
where $\mathcal{F}$ is a real quantity just depending on $\mathcal{W}$, $\mathcal{K}$ given by
\begin{equation}\label{ceffdef}
\mathcal{F} \equiv 1-\frac{3}{2}\frac{1}{|\mathcal{W}|^2}\nabla_A\mathcal{W}\mathcal{K}^{A\bar B}\nabla_{\bar B}\bar{\mathcal{W}}
-\frac{1}{4}|\partial_{\beta_1}\log \mathcal{W}|^2
-\frac{3}{4}|\partial_{\beta_2}\log \mathcal{W}|^2\,,
\end{equation}
and $\partial$, $\bar \partial$ are the holomorphic and anti-holomorphic exterior derivatives.
Similarly, \eqref{fermvsremaintext} become
\begin{align}
\bar\partial z^A&=-\frac{3}{2}(\bar{\mathcal{W}})^{-1}\mathcal{K}^{\bar B A}\nabla_{\bar B}\bar{\mathcal{W}}\bar B\,,\nn
\bar\partial \beta_1&=-\frac{1}{4}(\bar{\mathcal{W}})^{-1}\partial_{\beta_1}\bar{\mathcal{W}}\bar B\,,\nn
\bar\partial \beta_2&=-\frac{3}{4}(\bar{\mathcal{W}})^{-1}\partial_{\beta_2}\bar{\mathcal{W}}\bar B\,.
\end{align}

 Interestingly, as we show in appendix \ref{deriveio21bps}, we can use this formulation of the BPS equations to show that the consistency of the BPS equations
requires a non-trivial condition on $\mathcal{W}$, which we give in \eqref{eq:BPSConsistencyConditions}. Furthermore, we can show that
the specific $\mathcal{W}$ that appears in the ten scalar truncation, given in \eqref{superpotlike}, does in fact satisfy this condition. 
We expect the underlying reason for this is that we are working with a theory that comes from a truncation of a supersymmetric theory.

\section{BPS equations for Janus solutions}\label{bpsjanus}
We now consider a particular sub-class of the BPS configurations that we considered in the last subsection.
The ansatz for the $D=5$ metric is given by
\begin{align}\label{metjanus}
ds_5^2=e^{2A_J} ds^2(AdS_4)-N^2dr^2\,,
\end{align}
where $A_J=A_J(r)$, $N=N(r)$ and we take the scalar fields $\beta_1,\beta_2,z^A$ to be functions of $r$ only.
Here $ds^2(AdS_4)$ is the metric on $AdS_4$ of radius $\ell$, given, for example, in Poincar\'e coordinates by
\begin{align}\label{ads4poinc}
ds^2(AdS_4)=\ell^2\left[-\frac{dx^2}{x^2}+\frac{1}{x^2}\left(dt^2-dy_1^2-dy_2^2\right)\right]\,.
\end{align}
The factor of $\ell$ can be absorbed after redefining $A_J$, but it is convenient\footnote{Specifically,
in the resulting BPS equations that we write down below, if we take $\ell\to \infty$ we obtain the BPS equations for 
ordinary Lorentz invariant RG flows with metric $ds_5^2=e^{2A(r)}ds^2(\mathbb{R}^{1,3})-dr^2$. One can also make contact with
the BPS flow equations in Euclidean signature of \cite{Bobev:2016nua}, 
for an ansatz $ds_5^2=e^{2A(r)}ds^2(S^4)-dr^2$, after taking $\ell^2\to -\ell^2$.} to keep it explicit.
Notice that we recover the metric on $AdS_5$ with radius $L$ if $N=1$ and
\begin{align}
e^{A_J}=\frac{L}{\ell}\cosh\frac{r}{L}\,.
\end{align}

We can obtain the BPS equations for the Janus solutions as a special sub-class of those considered in the last section.
Specifically, if we take
\begin{align}\label{janmetchoice}
e^{V}=e^{A}=\ell e^{A_J}x^{-1}\,,
\end{align}
then the metric ansatz \eqref{iso21metans} precisely gives \eqref{metjanus}. From the first and third BPS equations
in \eqref{bpsderxi1appatext} we then get
 \begin{align}\label{Apbps}
N^{-1}\partial_r A_J + \frac{i\kappa}{\ell} e^{-A_J} 
- \frac{e^{-i\xi}}{3}e^{\mathcal{K}/2}\overline{\mathcal{W}} &=0\,,\nn
i\partial_r \xi
+2\mathcal{A}_r
          -\frac{i}{3}\text{Im}\left(N e^{-i\xi}e^{\mathcal{K}/2}\overline{\mathcal{W}}\right)  &=0\,,
\end{align}
respectively, with the second equation in \eqref{bpsderxi1appatext} implied by the first of these. Furthermore, from
\eqref{fermvsremaintext} we get the remaining BPS equations 
\begin{align}
\label{eq:ten_scalar_BPS}
N^{-1}\partial_r z^A + \frac{e^{-i\xi}}{2}e^{\mathcal{K}/2}\mathcal{K}^{A\bar B} \nabla_{\bar B} \overline{\mathcal{W}} = 0 ,\nn
N^{-1}\partial_r \beta_1 + \frac{e^{-i\xi}}{12}e^{\mathcal{K}/2}\partial_{\beta_1}\overline{\mathcal{W}}=0,  \nn
N^{-1}\partial_r \beta_2 + \frac{e^{-i\xi}}{4}e^{\mathcal{K}/2}\partial_{\beta_2}\overline{\mathcal{W}}=0\,. 
 \end{align}

 We can also obtain the Poincar\'e type Killing spinors for the Janus solutions directly from those given in the last section and
 we find 
 \begin{align}\label{bpstextprojection}
 \varepsilon_1=e^{i\xi/2+A_J/2}\ell^{1/2}\frac{1}{\sqrt{x}}\eta_0,\qquad \gamma^{012}\eta_0&=-i\kappa \eta_0\,,
 \end{align}
 where $\eta_0$ is a constant spinor, and
$\epsilon_2=e^{-i\xi}\gamma^4\epsilon_1$. In addition there are also superconformal type Killing spinors
 of the form 
\begin{align}
\varepsilon_1= \frac{1}{\sqrt{\ell}}\left[\sqrt{x}+\frac{1}{\sqrt x}\left(t\gamma_0+y_1\gamma_1+y_2\gamma_2\right)\gamma^3\right]e^{i\xi/2+A_J/2}\eta_0\,,
\end{align}
where $\eta_0$ is a constant spinor satisfying 
\begin{align}
\gamma^{012}\eta_0&=-i\kappa \eta_0\,,
\end{align}
and again
$\epsilon_2=e^{-i\xi}\gamma^4\epsilon_1$.
Observe that the BPS equation \eqref{Apbps}, \eqref{eq:ten_scalar_BPS} are invariant under the transformation 
\begin{align}
r\to -r,\qquad \xi\to \xi+\pi,\qquad \kappa\to -\kappa\,,
\end{align}
but note that the latter changes the projection on the Killing spinor.
They are also invariant under 
\begin{align}\label{z2twosymgm1}
r\to-r,\qquad z^A\to \bar z^A,\qquad\xi\to -\xi+\pi\,.
\end{align}

Further insight into these BPS equations can be obtained by choosing the gauge $N=e^{A_J}$ and then
recasting them in a manner similar to what we did in the last section.
Specifically, if we define
\begin{align}
B_r\equiv \frac{1}{6}e^{i\xi+A_J+\mathcal{K}/2}{\mathcal{W}}\,,
\end{align}
then we obtain\footnote{The equations below can be immediately obtained from 
those in \eqref{ay13}-\eqref{eq:BPSConsistencyConditions} if we choose $e^V=N=e^{A_J}$ and $e^A=\ell e^{A_J-x/\ell}$, which 
also gives rise to the metric ansatz \eqref{metjanus} with $N=e^{A_J}$, but, in contrast to 
\eqref{janmetchoice} and \eqref{ads4poinc}, the metric for the $AdS_4$ sections are in horospherical coordinates rather than Poincar\'e type coordinates.}
 the following BPS equations:
\begin{align}
\label{eq:RewrittenJanusBPSAeq}
\partial_r A_J - \frac{i\kappa}{l}&=2B_r, \\
\partial_r B_r &= 2 \mathcal{F} B_r \bar{B}_r\,,\label{eq:RewrittenJanusBPSrBeq}
\end{align}
where $\mathcal{F}$ is the real quantity just depending on $\mathcal{W}$, $\mathcal{K}$ given in \eqref{ceffdef},
as well as
\begin{align}
\label{eq:RewrittenJanusBPSzeq}
\partial_r z^A &= -3 \mathcal{K}^{A\bar{B}} \frac{\nabla_{\bar{B}}\overline{\mathcal{W}}}{\overline{\mathcal{W}}} \bar{B}_r \,, \nn
\partial_r \beta_1 &= - \frac{1}{2}  \partial_{\beta_1}\log\overline{\mathcal{W}} \bar{B}_r \, ,\nn
\partial_r \beta_2 &= - \frac{3}{2} \partial_{\beta_2}\log\overline{\mathcal{W}} \bar{B}_r \, .
\end{align}

Now the right hand side of \eqref{eq:RewrittenJanusBPSrBeq} is real, which implies that $\operatorname{Im}(B)$ is constant, in agreement with \eqref{eq:RewrittenJanusBPSAeq}. To further examine the consistency of the equations, given the reality of $\beta_1$ and $\beta_2$, we first note that for any function
$\overline{\mathcal{G}}(\bar{z}^A,\beta_1,\beta_2)$ which depends only on the scalar fields and is anti-holomorphic in the four complex scalars $z^A$, using \eqref{eq:RewrittenJanusBPSAeq}-\eqref{eq:RewrittenJanusBPSzeq} we can deduce
\begin{equation}\label{lemmatext}
\partial_r (\overline{\mathcal{G}} \bar{B}_r) = 2 (\hat{\mathcal{O}}\overline{\mathcal{G}}) B_r \bar{B}_r \,,
\end{equation}
where $\hat{\mathcal{O}}$ is a differential operator on the scalar manifold defined as
\begin{equation}
\hat{\mathcal{O}}\overline{\mathcal{G}} \equiv
\mathcal{F}\overline{\mathcal{G}}
-\frac{3}{2} \mathcal{K}^{\bar{A}B} \frac{\nabla_B \mathcal{W}}{\mathcal{W}} \partial_{\bar{A}} \overline{\mathcal{G}} 
-\frac{1}{4} \partial_{\beta_1}\log\mathcal{W}\partial_{\beta_1} \overline{\mathcal{G}} - \frac{3}{4} \partial_{\beta_2}\log\mathcal{W}\partial_{\beta_2} \overline{\mathcal{G}} .
\end{equation}
Then, taking the $r$ derivative of the last two equations in \eqref{eq:RewrittenJanusBPSzeq}, we
obtain the following necessary conditions for these set of equations to be consistent with $\beta_i$ being real:
\begin{equation}
\label{eq:BPSConsistencyConditionstext}
\operatorname{Im}\left( \hat{\mathcal{O}} \partial_{\beta_i} \log \overline{\mathcal{W}} \right) = 0\,, \qquad (i=1,2).
\end{equation}
Notice that these conditions do not involve $B$, just the scalar fields, and hence they are necessary conditions on $\mathcal{K}$ and $\mathcal{W}$.
One can explicitly check that these conditions are satisfied for \eqref{kpot} and \eqref{superpotlike} in the ten scalar model.

It is also not difficult to see that if \eqref{eq:BPSConsistencyConditionstext} is satisfied, then it is sufficient for a solution to exist,
given a set of starting values for $z^A,\beta_i,B$ satisfying the condition
\begin{equation}
\label{eq:BPSStartingConditiontext}
\operatorname{Im}\left(\partial_{\beta_i} \log \overline{\mathcal{W}} \bar{B}_r \right) = 0 \qquad (i=1,2) .
\end{equation}
Indeed, by taking the taking the $r$ derivative of the expression on the left hand side, using \eqref{lemmatext} and
given that \eqref{eq:BPSConsistencyConditionstext} holds, we see that
\eqref{eq:BPSStartingConditiontext} is guaranteed to be satisfied along the flow. 
Furthermore, for any starting values of $z^A, \beta_i$, one can always choose a starting value of $B_r$ that satisfies \eqref{eq:BPSStartingConditiontext}, and then solve the equations. 

From the above arguments, given \eqref{eq:BPSConsistencyConditionstext} is satisfied one can also conclude the following:
\begin{itemize}
\item If the starting values of $z^A, \beta_i$ are such that $\partial_{\beta_i} \log\mathcal{W} = 0$ or $\operatorname{Im}(\partial_{\beta_i} \log\mathcal{W}) \neq 0 $ (for $i=1,2$), then given a chosen value of $\frac{\kappa}{l}$ one can always find a starting value for $\operatorname{Re}(B_r)$ such that \eqref{eq:BPSStartingConditiontext} is satisfied and solve the equations. 
It is then guaranteed from \eqref{eq:BPSStartingConditiontext}
that along each point in the flow, either $ \partial_{\beta_i} \log\mathcal{W} = 0$ or $\operatorname{Im}(\partial_{\beta_i} \log\mathcal{W}) \neq 0$.
\item Conversely,  a choice of starting values with $\partial_{\beta_i} \log\mathcal{W} \neq 0$ and real, for either $\beta_1$ or $\beta_2$, is consistent with equations \eqref{eq:RewrittenJanusBPSzeq} and \eqref{eq:RewrittenJanusBPSrBeq} but is incompatible with equation \eqref{eq:RewrittenJanusBPSAeq} since it requires $\operatorname{Im}(B_r) = 0$.
\item From the last two equations in \eqref{eq:RewrittenJanusBPSzeq} it is clear that the turning point for $\beta_i$ corresponds to a point in which $\partial_{\beta_i} \log\mathcal{W} = 0$. 
\item For a turning point of $A_J$, from \eqref{eq:RewrittenJanusBPSAeq} we have $\operatorname{Re}(B_r)=0$ and therefore \eqref{eq:BPSStartingConditiontext} implies that at this point we must have $\operatorname{Re}(\partial_{\beta_i} \log\mathcal{W})=0$. 
Thus, at this turning point we are just free to specify initial conditions for the $z^A$ which implies that the family
of solutions is of dimension twice the number of $z^A$ that are active.
\end{itemize}

Note that we have proved these results for the flows using the gauge $N=e^{A_J}$. however, they are 
gauge invariant results for the flows and hence, are also valid for the gauge $N=1$ that we use to numerically construct the solutions
in the next section.

\section{Supersymmetric Janus Solutions}\label{susyjansol}
In this section we present various solutions to the BPS equations \eqref{Apbps}, \eqref{eq:ten_scalar_BPS}
that we derived in the previous section, including families of Janus solutions. 
More precisely, we will do this for each of three different consistent truncations of the ten scalar model
that we discussed in section \ref{sugratrunc}. Before doing that we make some general comments
concerning how we obtain the field theory sources and expectation values, both with $AdS_4$ boundary metric and with flat boundary metric,
using the holographic renormalisation scheme that we outline in detail in appendices \ref{appb} and \ref{appc}.

\subsection{Preliminaries}\label{prelim}

Let us focus on the Janus solutions which describe a planar co-dimension one interface in $\mathcal{N}=4$ SYM that is supported
by spatially dependent mass sources. These solutions have a metric of the form given in
\eqref{metjanus}:
\begin{align}
ds^2=e^{2A_J} ds^2(AdS_4)-dr^2\,,
\end{align}
where we have now set $N=1$ for convenience,
with 
\begin{align}\label{ads4poinct}
ds^2(AdS_4)=\frac{\ell^2}{x^2}\left(-{dx^2}+dt^2-dy_1^2-dy_2^2\right)\,.
\end{align}
Notice that in the gauge $N=1$, the BPS equations are invariant under shifts of the radial coordinate
\begin{align}\label{rshift}
r\to r+constant\,.
\end{align}

It is illuminating to first recall that the $\mathcal{N}=4$ SYM $AdS_5$ vacuum solution with this $AdS_4$ slicing is given by
\begin{align}
e^{A_J}=\frac{L}{\ell}\cosh\frac{r}{L}\,,
\end{align}
with vanishing scalar fields. If we now employ the coordinate 
transformation
\begin{align}\label{eq:coordinates_transformation}
x=\sqrt{y_3^2+L^2e^{-2\rho/L}},\quad e^{r/L}={e^{\rho/L}}\frac{y_3+\sqrt{y_3^2+L^2e^{-2\rho/L}}}{L}\,,
\end{align}
we obtain the $AdS_5$ metric, with flat-slicing, given by
\begin{align}
ds^2_5=e^{2\rho/L}(dt^2-dy_1^2-dy_2^2-dy_3^2)-d\rho^2\,.
\end{align}
In the $(\rho,y_3)$ coordinates the conformal boundary is reached at $\rho\to\infty$ and has a flat boundary metric
with coordinates $(t,y_i)$.
On the other hand in the $(r,x)$ coordinates the conformal boundary has three components: two half spaces
$r\to\pm\infty$ at $x\ne 0$, associated with $y_3>0$ and $y_3<0$, respectively,
joined together at the planar interface at $x=0$ and finite $r$, associated with $y_3=0$.
As $r\to\pm\infty$ we naturally obtain the $AdS_4$ metric on the two half spaces.
A few more details are provided in appendix \ref{appc} and we have also illustrated the set-up there in figure \ref{fig:janus}.

\subsubsection{Janus solutions: field theory on $AdS_4$}
The Janus solutions of $\mathcal{N}=4$ SYM that we construct approach the $\mathcal{N}=4$ SYM $AdS_5$ vacuum as $r\to\pm\infty$ but
with additional mass sources. Analogous to the discussion for the $AdS_5$ vacuum solution itself, 
the conformal boundary of these Janus solutions consists of three components: two half spaces,
with $AdS_4$ metrics, joined together at a planar interface along the boundary of the $AdS_4$. Note that the boundary at $x=0$ is not a standard asymptotically  locally $AdS_5$ region, as the  scalars are not approaching an extremum of the potential, but only at $r = \pm \infty.$ 

Let us first
consider the $r\to\infty$ end of the interface, returning to the $r\to-\infty$ end in section \ref{otherendtext}.
As $r\to\infty$ we
demand that we have the schematic expansion 
\begin{align}\label{asexp10scjan}
A_J&=\frac{r}{L}+A_0+\cdots+{A}_{(v)}e^{-4r/L}+\cdots\,,\nn
\phi_i&=\phi_{i,(s)}e^{-r/L}+\dots+{\phi}_{i,(v)}e^{-3r/L}+\cdots\,,\qquad i=1,\dots,4\,,\nn
\alpha_i&=\alpha_{i,(s)}\frac{r}{L}e^{-2r/L}+{\alpha}_{i,(v)}e^{-2r/L}+\cdots\,,,\qquad i=1,\dots,3\,,\nn
\beta_i&=\beta_{i,(s)}\frac{r}{L}e^{-2r/L}+{\beta}_{i,(v)}e^{-2r/L}+\cdots\,,\qquad i=1,\dots,2\,,\nn
\varphi&=\varphi_{(s)}+\cdots+{\varphi}_{(v)}e^{-4r/L}+\cdots\,.
\end{align}
Recall that in the $N=1$ gauge the BPS equations have the residual shift symmetry in $r$ \eqref{rshift}. By shifting the radial coordinate via
$r\to r-A_0 L$ we can always remove the constant $A_0$ term and we shall do so in the following. In particular all the expressions for the expectation values and sources given below are obtained with
\begin{align}\label{aztext}
A_0=0\,.
\end{align}
The various other coefficients in this expansion, which are all constant, are constrained by the BPS equations, as we detail below.
The constants $\phi_{i,(s)}$, $\alpha_{i,(s)}$, $\beta_{i,(s)}$, $\varphi_{(s)}$ are associated with constant sources for the mass deformations of $\mathcal{N}=4$ SYM when placed on $AdS_4$. Recalling from \eqref{opfieldmapz}
that these are sources for operators of conformal dimension $\Delta=3,2,2,4$, respectively, it
is useful to note that the field theory sources on $AdS_4$ that are invariant under a Weyl rescaling of the $AdS_4$ radius $\ell$ 
are given by $\ell\phi_{i,(s)}$, $\ell^2\alpha_{i,(s)}$, $\ell^2\beta_{i,(s)}$,
$\varphi_{(s)}$.
In this paper we will not discuss deformations that involve the coupling constant of $\mathcal{N}=4$ SYM, and so we will always set
\begin{align}
\varphi_{(s)}=0\,.
\end{align}
The BPS equations then imply that these sources satisfy 
\begin{align}\label{bpsrelscetext}
\alpha_{i,(s)}&=-\kappa \frac{L}{\ell} \phi_{i,(s)}\,,\qquad i=1,\dots,3\,,\nn
\beta_{1,(s)}&=\frac{1}{3}\left(\phi_{1,(s)}^2+\phi_{2,(s)}^2-2\phi_{3,(s)}^2\right)\,,\nn
\beta_{2,(s)}&=\phi_{1,(s)}^2-\phi_{2,(s)}^2\,,\nn
\phi_{4,(s)}&=0\,.
\end{align}
Notice that these relations respect the field theory scaling dimensions of the sources on $AdS_4$ that we just mentioned above.

Similarly, the constants ${\phi}_{i,(v)}$, ${\alpha}_{i,(v)}$, ${\beta}_{i,(v)}$,
${\varphi}_{(v)}$ in \eqref{asexp10scjan}, with suitable contributions from the sources, give rise to the expectation values of the scalar
operators. We will give explicit expressions for these in each of the specific truncations below. Here, to illustrate, we just highlight a simple example:
using the renormalisation scheme discussed in appendix \ref{appb}, we find that for $\mathcal{N}=4$ SYM on $AdS_4$ we have
 \begin{align}
\langle\mathcal{O}_{\alpha_i}\rangle
&=\frac{1}{4\pi GL}\Big({\alpha}_{i,(v)}-2\delta_\alpha \alpha_{i,(s)}\Big)\,.
\end{align}
Here $\delta_\alpha$ is an undetermined constant that parametrises a finite counterterm, which we haven't fixed.
As we will see below it is intimately connected with a novel feature of the expectation values of the operators in flat spacetime.
We also note that due to the structure of the conformal anomaly $\ell^2\langle\mathcal{O}_{\alpha_i}\rangle$ is not invariant under rescalings of $\ell$, 
as one might have expected, a point we return to below.

\subsubsection{Janus solutions: field theory on flat spacetime}
We are primarily interested in obtaining the sources and expectation values
for operators of $\mathcal{N}=4$ SYM in flat spacetime, as in section \ref{sec:two}. 
Now the metric on $AdS_4$ in \eqref{ads4poinct}
is conformal to flat spacetime. Thus, we can obtain the relevant quantities in flat spacetime from those on $AdS_4$
by simply performing a Weyl transformation with Weyl factor  $x^2/\ell^2$. However, while the sources transform
covariantly under Weyl transformations, the expectation values do not due to the presence
of source terms appearing in the conformal anomaly $\mathcal{A}$ (similar to \cite{Petkou:1999fv,Bianchi:2001kw}), schematically given by
 \begin{align}\label{traceteeconfanomtext}
&8\pi GL\mathcal{A}
=-\frac{L^4}{8}\Big(R_{ab}R_{}^{ab}-\frac{1}{3}R^2\Big)-{L^2}\sum_{i=1}^4\Big[(\nabla\phi_{i,(s)})^2   +\frac{1}{6}R  \phi_{i,(s)}^2   \Big]\nn
&-\sum_{i=1}^3\alpha_{i,(s)}^2-6\beta_{1,(s)}^2-2\beta_{2,(s)}^2+\frac{8}{3 }\sum_{i=1}^4\phi_{i,(s)}^4-\frac{8}{3}\sum_{1\leq i<j\leq 4}^4\phi_{i,(s)}^2\phi_{j,(s)}^2+\cdots
\end{align}
where the dots refer to extra terms involving finite counterterms (see \eqref{tracetee},\eqref{traceteeconfanom}). 

In fact we can obtain the relevant results within holography by carrying out a bulk coordinate transformation
so that
as we approach the $r\to\infty$ component of the conformal boundary (say) it has a flat metric. 
Indeed for this component of the conformal boundary we can use the coordinate transformation of the form
\begin{align}\label{asbfgghtext}
e^{r/L} &= \frac{y_3}{\ell} e^{\rho/L}  + \frac{L^2}{4 \ell y_3}e^{-\rho/L} + \cO(e^{-3\rho/L}/y_3^3)\,,\nn
x &= y_3 + \frac{L^2}{2 y_3}e^{-2\rho/L} + \cO(e^{-4\rho/L}/y_3^3)\,,
\end{align}
with $y_3>0$. Substituting this into \eqref{asexp10scjan} then leads to an expansion of the bulk fields as $\rho\to\infty$, which one can find in appendix \ref{appc}. Having done this, we can then employ\footnote{\label{ft13}To be clear, to do this one should use the results of appendix \ref{appb} by
replacing the coordinates $(r,x)$ there with $(\rho,y_3)$.}
 the holographic renormalisation scheme for $ISO(1,2)$ invariant configurations discussed in appendix \ref{appb}, in order to read off the sources and the expectation values with the field theory now on flat spacetime.

 The non-trivial sources for the dual scalar operators in $\mathcal{N}=4$ SYM
 theory now have the expected dependence on the spatial coordinate $y_3$ (still with $y_3>0$) that we saw in section \ref{sec:two}:
\begin{align}\label{fspace}
\frac{\ell\phi_{i,(s)}}{y_3},\quad &\frac{\ell^2 \alpha_{i,(s)}}{y_3^2}, \quad i=1,\dots,3\nn
&\frac{\ell^2\beta_{i,(s)}}{y_3^2}\,,\quad i=1,,2\,,
\end{align}
with $\phi_{4,(s)}=\varphi_{(s)}=0$.
Recalling that the numerators in these expressions are the scale invariant field theory sources on $AdS_4$, we see that in flat spacetime these
field theory sources have scaling dimensions 1, 2, 2 associated with operators that have dimensions $\Delta=3,2,2$, respectively.
Furthermore, when combined with the BPS relations \eqref{bpsrelscetext}, these expressions are in alignment with those that we derived in section \ref{sec:two} for
each of the three different $\mathcal{N}=1^*$ truncations. 

Due to the structure of the conformal anomaly, the expressions for the expectation
values are more involved. To illustrate, here we just note that we have
 \begin{align}\label{delalphalog}
\langle\mathcal{O}_{\alpha_i}\rangle
&=\frac{1}{4\pi GL}\frac{\ell^2}{y_3^2}\Big({\alpha}_{i,(v)}+\alpha_{i,(s)}
\log(\frac{y_3}{\ell e^{2\delta_\alpha}})\Big)\,,
\end{align}
and give explicit expressions for the other expectation values for the three subtruncations below.
We highlight the appearance of the novel $\log(y_3)$ term that appears in this expectation value.
 Notice that performing a scaling of the $y_3$ coordinate is associated with 
a shift in $\delta_\alpha$, which parametrises a finite counterterm. We can certainly choose a renormalisation scheme in which we set $\delta_\alpha=0$.
However, there are additional similar finite counterterms that appear in expectation values of other operators, as
we will see in each of the specific truncations below, and we have not been able
to find a simple argument that would fix all of them in a way that is consistent with supersymmetry. 
Given the log terms appearing in the expectation values we expect that there will be at least one set of supersymmetric finite 
counterterms that one is free to add. We leave further investigation on this issue to future work.

From the above results we can conclude that under a Weyl transformation of the $AdS_4$ boundary metric of the form 
$h_{ab}\to \Lambda^2 h_{ab}$, with $\Lambda=x/l$,
the sources transform covariantly with $\phi_{i(s)}\to \Lambda^{-1}\phi_{i(s)}$, $\alpha_{i(s)}\to \Lambda^{-2}\alpha_{i(s)}$ and 
$\beta_{i(s)}\to \Lambda^{-2}\beta_{i(s)}$. However, the expectation values do not transform covariantly due to the anomaly and, for example,
we have
$\langle\mathcal{O}_{\alpha_i}\rangle\to \Lambda^{-2}\langle\mathcal{O}_{\alpha_i}\rangle+\frac{\alpha_{i(s)}}{4\pi G L}\Lambda^{-2}\log\Lambda$.
The transformation properties for all the expectation values can be obtained from \eqref{dim2weylt}-\eqref{dim4weylt}. 
It is worth emphasising that these results imply that some care is required in comparing expectation values of operators 
on $AdS_4$ for solutions with different values of the $AdS_4$ radius $\ell$ due to this non-covariant rescaling.
In practice, in all of our numerics we have set $\ell=1$ (as well as $L=1$).

\subsubsection{The $r\to-\infty$ end of the conformal boundary}\label{otherendtext}
The analysis above concerned the component of the conformal boundary for the Janus solutions with $AdS_4$ slicing at $r\to\infty$. There is a similar analysis for the component at $r\to -\infty$, which by assumption, is again approaching the $\mathcal{N}=4$ SYM $AdS_5$ vacuum. 
Firstly, we can consider the expansion
given in \eqref{asexp10scjan} after replacing $r\to -r$:
\begin{align}\label{asexp10scjanotherend}
A_J&=-\frac{r}{L}+\tilde A_0+\cdots+\tilde{A}_{(v)}e^{4r/L}+\cdots\,,\nn
\phi_i&=\tilde\phi_{i,(s)}e^{r/L}+\dots+\tilde{\phi}_{i,(v)}e^{3r/L}+\cdots\,,\qquad i=1,\dots,4\,,\nn
\alpha_i&=-\tilde\alpha_{i,(s)}\frac{r}{L}e^{2r/L}+\tilde{\alpha}_{i,(v)}e^{2r/L}+\cdots\,,,\qquad i=1,\dots,3\,,\nn
\beta_i&=-\tilde\beta_{i,(s)}\frac{r}{L}e^{2r/L}+\tilde{\beta}_{i,(v)}e^{2r/L}+\cdots\,,\qquad i=1,\dots,2\,,\nn
\varphi&=\tilde\varphi_{(s)}+\cdots+\tilde{\varphi}_{(v)}e^{4r/L}+\cdots\,,
\end{align}
and we can and will set 
\begin{align}\label{tilaz}
\tilde A_0=0\,,
\end{align} 
by shifting the radial coordinate\footnote{When one numerically constructs a solution, one generically finds that $A_0$ and $\tilde A_0$ in 
\eqref{asexp10scjan} and \eqref{asexp10scjanotherend} are non-zero and not equal. In order to utilise our holographic renormalisation results
with $A_0=\tilde A_0=0$, one 
needs to shift the radial coordinate by different constants at
$r=\pm\infty$.}. 
As we show in appendix \ref{appc2},
the BPS equations then imply that the coefficients are related as in the $r\to\infty$ case, but after taking
$\kappa\to-\kappa$. We emphasise that we are not changing the projections on the preserved Killing spinors in doing this.
So, for example, with this expansion at the $r\to-\infty$ end we now have 
\begin{align}\label{bpsrelscetextmin}
\tilde\alpha_{i,(s)}&=+\kappa \frac{L}{\ell} \tilde\phi_{i,(s)}\,,\qquad i=1,\dots,3\,,\nn
\tilde\beta_{1,(s)}&=\frac{1}{3}\left(\tilde\phi_{1,(s)}^2+\tilde\phi_{2,(s)}^2-2\tilde\phi_{3,(s)}^2\right)\,,\nn
\tilde\beta_{2,(s)}&=\tilde\phi_{1,(s)}^2-\tilde\phi_{2,(s)}^2\,,
\end{align}
with $\tilde\phi_{4,(s)}=\tilde\varphi_{(s)}=0$.

Furthermore, to carry out the coordinate transformation back to flat space we can use 
\eqref{asbfgghtext} with $r\to-r$ and $y_3\to -y_3$.
This will then give the relevant quantities on the $y_3<0$ part of the conformal boundary, with flat boundary metric. 
Thus, to obtain the flat boundary results for $y_3<0$ from those
for $y_3>0$, we need to make the replacements $y_3\to -y_3$ and $\kappa\to -\kappa$.

As an illustration, we can consider the sub-class of explicit $\mathcal{N}=4$ SYM Janus solutions that are symmetric under the $\mathbb{Z}_2$ symmetry given below in \eqref{z2twosymg}. For this class we have, for example, that $\phi_i(r)=-\phi_i(-r)$, while $\alpha_i(r)=+\alpha_i(-r)$. We then find for $AdS_4$ boundary metric if we have a source $\ell\phi_{i(s)}$ at the $r\to\infty$ end, we will have a source $\ell\tilde\phi_{i(s)}=-\ell\phi_{i(s)}$ at the $r\to-\infty$ end. Transforming to flat boundary coordinates we then find that for both $y_3>0$ and $y_3<0$ we have a source of the form $\ell\phi_{i(s)}/y_3$, i.e. antisymmetric in $y_3$. 
Similarly we would have source $\ell^2\alpha_{i(s)}$ at both $r\to\pm\infty$ ends and transforming to flat space source $\ell^2\alpha_{i(s)}/(y_3)^2$, i.e. symmetric in $y_3$. Similar comments apply to expectation values.

We can similarly consider the sub-class of explicit Janus solutions that are 
symmetric under the $\mathbb{Z}_2$ symmetry given below in \eqref{z2twosymgminus}. 
For this class we have, for example, that $\phi_i(r)=+\phi_i(-r)$, while $\alpha_i(r)=-\alpha_i(-r)$. We then find in $AdS_4$ slicing
we will have source $\ell\phi_{i(s)}$ at both $r\to\pm\infty$ ends and transforming to flat boundary coordinates we 
would have a source of the form $\ell\phi_{i(s)}/|y_3|$, i.e. symmetric in $y_3$.
Similarly, we would have a source $\ell^2\alpha_{i(s)}$ at the $r\to+\infty$ end and $\ell^2\tilde\alpha_{i(s)}=-\ell^2\alpha_{i(s)}$ at the $r\to -\infty$ end, and transforming to
 flat space we get a source $\ell^2\alpha_{i(s)}/(y_3)^2$ for $y_3>0$ and a source $-\ell^2\alpha_{i(s)}/(y_3)^2$ for $y_3<0$ 
 i.e. antisymmetric in $y_3$. Similar results apply to the expectation values.

\subsubsection{Constructing solutions}
Having made some general comments on how we determine the sources and expectation values for the Janus solutions, in the next subsections
we turn to summarising the solutions that we have found for the three different truncations. At this point it is worth recalling
the various general constraints on the space of solutions that are itemised at the end of the last section, below \eqref{eq:BPSStartingConditiontext}.

It is also helpful to recall that the ten scalar model, and the three further truncations, are all invariant under the $\mathbb{Z}_2$ symmetry that 
takes
\begin{align}\label{zasymg}
z^A\to -z^A\,.
\end{align}
Furthermore, the BPS equations for the Janus solutions in \eqref{Apbps},\eqref{eq:ten_scalar_BPS} are also invariant under the
$\mathbb{Z}_2$ symmetry that acts as
\begin{align}\label{z2twosymg}
r\to-r,\qquad z^A\to \bar z^A,\qquad\xi\to -\xi+\pi\,.
\end{align}
Combining these two, we conclude that the BPS equations are also invariant under
\begin{align}\label{z2twosymgminus}
r\to-r,\qquad z^A\to -\bar z^A,\qquad\xi\to -\xi+\pi\,.
\end{align}

We have utilised various approaches to solving the BPS equations numerically. One approach is to start at, say, $r\to\infty$,
and then use the expansion \eqref{asexp10scjan} to set initial conditions to integrate in to smaller values of $r$ and see where one ends up. As we will see, while some solutions
end up at a similar asymptotic region at $r\to-\infty$, and hence are Janus solutions of $\mathcal{N}=4$ SYM, there are also solutions that run off to singular behaviour.
Furthermore, there are also solutions which do not have an asymptotic region of the form \eqref{asexp10scjan}
or \eqref{asexp10scjanotherend}.
Another approach, and a more general one, is to start at a point in the bulk,
for example a turning point of the function $A_J(r)$ at say $r=0$ and then integrate out to smaller and larger values of $r$, and again see where one ends up.
In the following we will summarise the main results of these constructions. 

To simplify the discussion it will be helpful to first discuss the $\mathcal{N}=2^*$ model, which is the simplest, before discussing the two $\mathcal{N}=1^*$ models.

\subsection{$\mathcal{N}=2^*$ model}

This model was summarised in section \ref{n2modeltrunc}. There is one complex scalar field  $z^1$, which we write as
$z^1 = \tanh[ \alpha_1       -i \phi_1 ]$ and one real scalar field $\beta_1$.

Consider solutions that approach $\mathcal{N}=4$ SYM with mass sources at, say $r\to\infty$.
Following the discussion in the last subsection and using the results of appendices \ref{appb}, \ref{appc} we can summarise the source and expectation values
for the relevant operators that are active. All of the source terms are specified by $\phi_{1,(s)}$ with
\begin{align}
\alpha_{1,(s)}=-\kappa \frac{L}{\ell} \phi_{1,(s)}\,,\qquad
\beta_{1,(s)}=\frac{2}{3}\phi_{1,(s)}^2\,.
\end{align}
The field theory sources on $AdS_4$ are given by $\phi_{1,(s)}$, $\alpha_{1,(s)}$, $\beta_{1,(s)}$, with
$\ell\phi_{1,(s)}$, $\ell^2\alpha_{1,(s)}$, $\ell^2\beta_{1,(s)}$, invariant under Weyl scalings of $\ell$, while 
those on flat spacetime are given by \eqref{fspace}:
\begin{align}\label{fspace2}
\frac{\ell\phi_{1,(s)}}{y_3},\qquad \frac{\ell^2 \alpha_{1,(s)}}{y_3^2}, \qquad \frac{\ell^2\beta_{1,(s)}}{y_3^2}\,,
\end{align}
and have scaling dimensions $1,2,2$, respectively.

For the associated expectation values of the operators in flat spacetime, we have
\begin{align}
\langle\mathcal{O}_{\alpha_1}\rangle=\langle\mathcal{O}_{\alpha_2}\rangle
&=\frac{1}{4\pi GL}\frac{\ell^2}{y_3^2}\Big({\alpha}_{1,(v)}+\alpha_{1,(s)}
\log(\frac{y_3}{\ell e^{2\delta_\alpha}})\Big)\,,
\end{align}
which then, along with $\phi_{1,(s)}$, determines the remaining expectation values 
\begin{align}
\langle\mathcal{O}_{\beta_1}\rangle
&=-\frac{4\kappa\ell}{L}\langle\mathcal{O}_{\alpha_1}\rangle\,\phi_{1,(s)}+
\frac{(1+4\delta_{\alpha}-4\delta_{\beta})}{2\pi G L}\frac{\ell^2}{y_3^2}\phi_{1,(s)}^2\,,\nn
\langle\mathcal{O}_{\phi_{1}}\rangle=\langle\mathcal{O}_{\phi_{2}}\rangle&=
-\frac{2}{3}\frac{\ell}{y_3}\langle\mathcal{O}_{\beta_1}\rangle\,\phi_{1,(s)}-{2\kappa L}
\frac{1}{y_3} \langle\mathcal{O}_{\alpha_1}\rangle-\frac{L}{4\pi G}\frac{\ell}{y_3^3}\phi_{1,(s)}\,.
\end{align}
where $\delta_\alpha$, $\delta_\beta$ are unspecified finite counterterms.

An important aspect of the above summary, is that for a specific choice of finite counterterms, 
 all 
of the scalar sources and expectation values of the dual field theory can be obtained
by giving $\ell\phi_{1,(s)}$ 
as well as $\ell^2{\alpha}_{1,(v)}$. We now set $\ell=1$ (as well as $L=1$) and also fix the sign arising
in the BPS equations: $\kappa=+1$.

\begin{figure}[h!]
\centering
%\raisebox{-0.5\height}
{\includegraphics[scale=0.27]{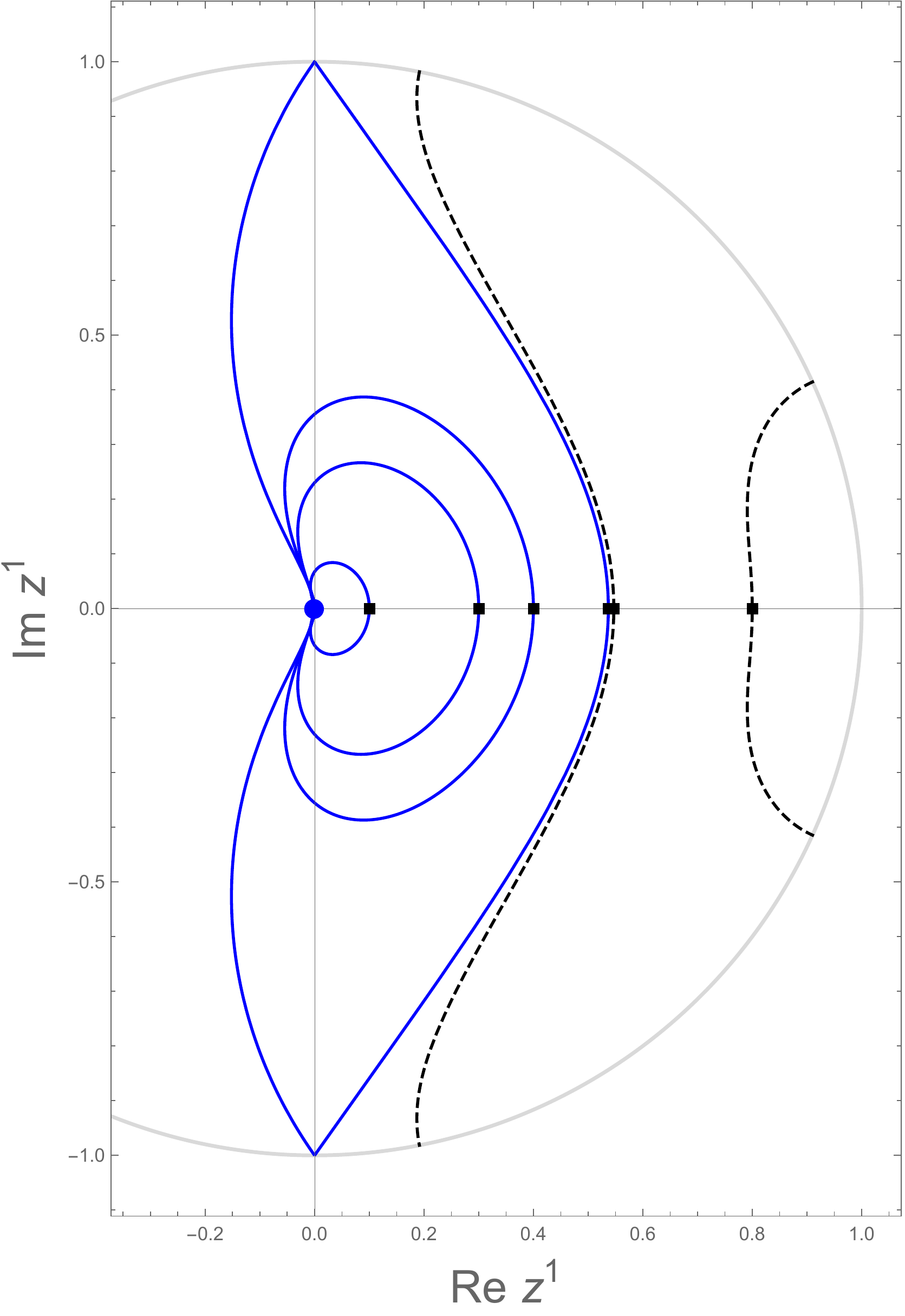}\quad
\includegraphics[scale=0.27]{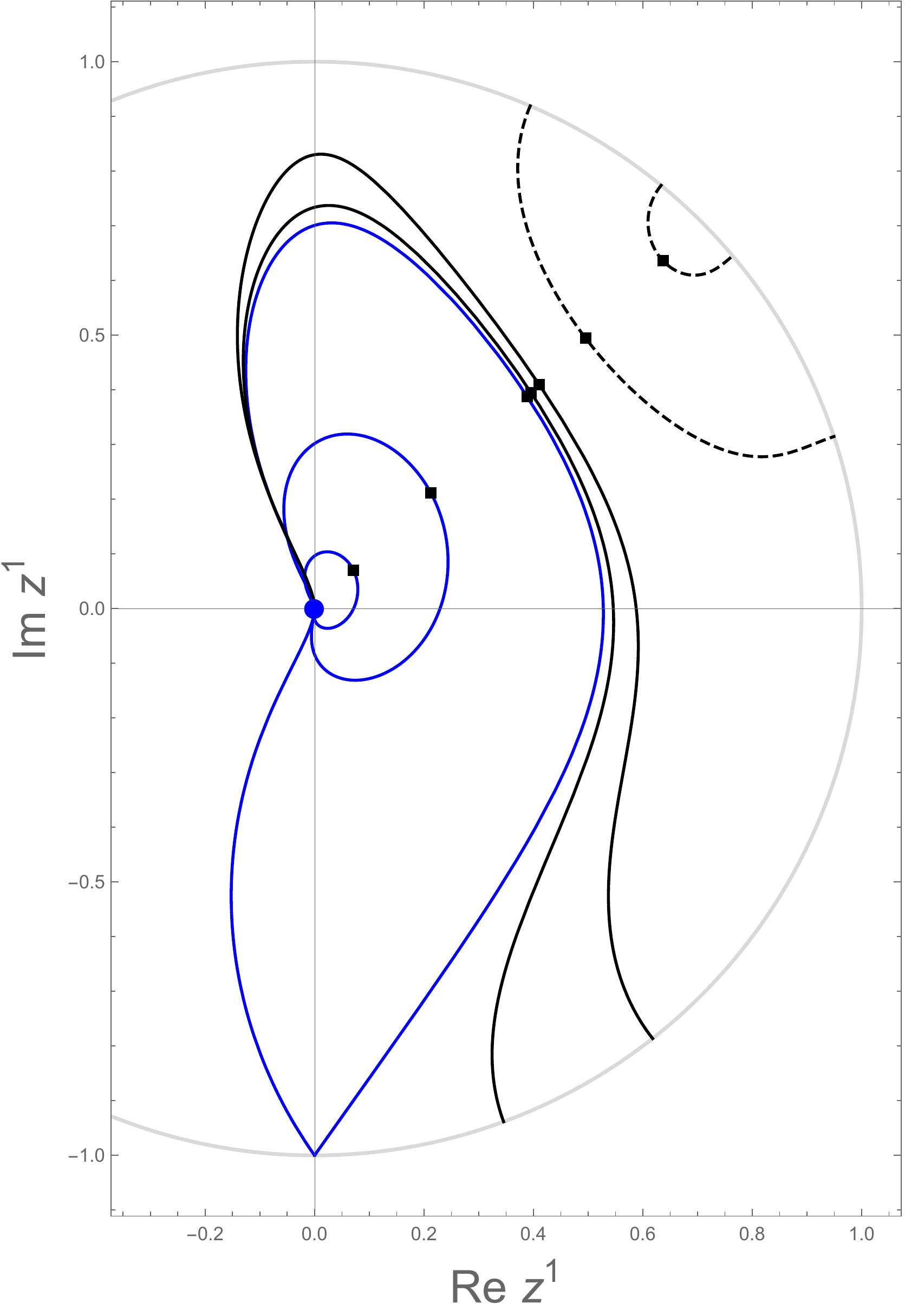}\quad
\includegraphics[scale=0.27]{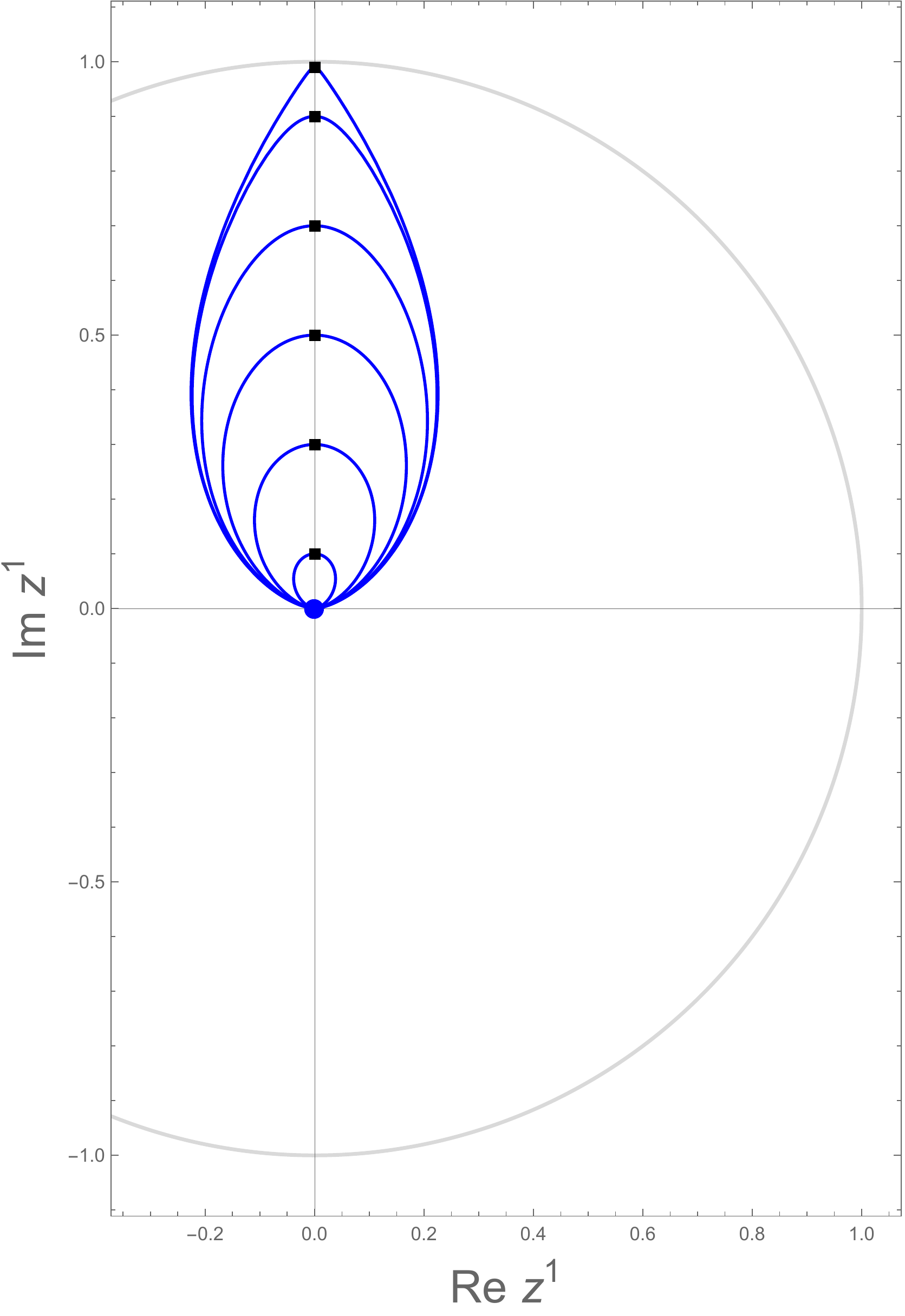}}
\caption{The family of BPS solutions for the $\mathcal{N}=2^*$ model is summarised by parametrically plotting the real and imaginary parts of 
the scalar field $z^1$. The black squares correspond to turning points of the function $A_J(r)$ and the three plots, from left to right, correspond to solutions
where the phase of the complex scalar field at the turning point is $0,\pi/4$ and $\pi/2$, respectively. 
The blue dot at the origin is the $\mathcal{N}=4$ SYM $AdS_5$ vacuum solution and the blue lines are Janus solutions. The boundary of field space is $|z^1|=1$, marked with the grey circle. As one moves from $r=-\infty$ to $r=+\infty$ one moves clockwise on the curves.
 }\label{fig:n2}
\end{figure}

Following the discussion given at the end of section \ref{bpsjanus}, we know that there is a two parameter family of solutions for this model. 
A useful way to parametrise them is to take one parameter to be the phase of the complex scalar $z^1$ at the turning point of the 
function $A_J(r)$. Due to the symmetries given in \eqref{zasymg}, \eqref{z2twosymg} we can restrict to solutions for which this phase lies in
the domain $[0,\pi/2]$. Then, fixing this phase we can construct a one parameter family of solutions that we can represent
by parametric plots of the real and imaginary parts of the complex scalar field $z^1$, as displayed in figure \ref{fig:n2}. In these plots the
black squares correspond to turning points of the function $A_J(r)$ and, from left to right, the phase is equal to $0,\pi/4,\pi/2$, respectively.
The blue dot at the origin in each of the plots corresponds to the $\mathcal{N}=4$ SYM $AdS_5$ vacuum solution.

For each fixed value of the phase, there is a one parameter family of
$\mathcal{N}=4$ SYM Janus solutions (blue curves) that approach the $\mathcal{N}=4$ SYM $AdS_5$ vacuum solution at $r\to\pm\infty$ with, generically, spatially modulated mass terms
that are parametrised by $\phi_{1,(s)}$. Furthermore, focussing on the $r\to+\infty$ end we find $0<\phi_{1,(s)}<\phi_{1,(s)}|_{crit}$ and 
$\phi_{1,(s)}|_{crit}\ne \infty$. The exception to this occurs
for the class of solutions in which the phase at the turning point is exactly $\pi/2$ (right plot in figure \ref{fig:n2}): for this class, remarkably,
we find that it is source free, $\phi_{1,(s)}= 0$, on both sides of the interface, a point we return to below. We also note that, somewhat
surprisingly, for the generic solutions as the phase approaches $\pi/2$, the critical value of the source, $\phi_{1,(s)}|_{crit}$ does not approach zero.

Another interesting feature of this model is that for each Janus solution, with phase not equal to $\pi/2$,
on either side of the interface at $r\to\pm \infty$, we always find\footnote{This 
suggests that there is some kind of conserved quantity for the BPS equations which we have yet to identify. In seeking such quantity it
is important to note, as we state below, that the expectation values are not simply related on either side of the interface.} that $\tilde\phi_{1,(s)}=-\phi_{1,(s)}$.
If we convert to sources in flat space, recalling that we have set $\ell=1$,
this means we have a source of the form $\phi_{1,(s)}/y_3$, for all $y_3$ and where here 
$\phi_{1,(s)}$ is the expansion coefficient at $r=+\infty$ (which we noted above is in the range $0<\phi_{1,(s)}<\phi_{1,(s)}|_{crit}$).

We can also determine the expectation values of various operators for the Janus solutions on each side of the interface at $r=\pm\infty$. 
With $\ell=1$, we just explain the behaviour of $\alpha_{1,(v)}$ which can be used to get all expectation values of scalar operators.
For the special case when the phase is equal to zero (left plot in figure \ref{fig:n2}), the solutions
are invariant under the symmetry \eqref{z2twosymg} and as explained in section \ref{otherendtext},
we find $\alpha_{1,(v)}$ is the same on each side of the interface. 
For this case we also find for the $r=+\infty$ end with $0<\phi_{1,(s)}<\phi_{1,(s)}|_{crit}$, that as
$\phi_{1,(s)}$ goes from $0$ to $\phi_{1,(s)}|_{crit}$, then $\alpha_{1,(v)}$ increases from $0$, hits a maximum and then decreases to
a finite negative value at $\phi_{1,(s)}|_{crit}$.

By contrast, for the class of Janus solutions when the phase is in the domain $(0,\pi/2)$ we find that $\alpha_{1,(v)}$ 
and $\tilde\alpha_{1,(v)}$ do not have the
same value at $r=\pm\infty$, respectively.

When the phase is exactly equal to $\pi/2$, there is a different picture. As we noted above there are no sources on either side of the interface.
We also find for the two sides of the interface $\alpha_{1,(v)}=-\tilde\alpha_{1,(v)}$ 
and the energy density \eqref{relsces} is zero off of the interface. The absence of sources on either side of the interface is noteworthy.
It seems most likely that there is a distributional source
that is located on the interface itself, otherwise we would have a configuration that spontaneously breaks translations,
and it would be interesting to verify this in detail.

The plots given in figure \ref{fig:n2} also reveal that there are other non-Janus solutions for this model.
When the phase is in the open domain $(0,\pi/2)$, there is also a one parameter family of solutions
that approach $\mathcal{N}=4$ SYM as $r\to -\infty$, with  $-\infty<\tilde \phi_{1,(s)}<-\phi_{1,(s)}|_{crit}$. As one moves
along the radial direction, at some finite value of the radial coordinate, past the turning point, one hits a singularity, with $|z^1|\to 1$. 
Such solutions, corresponding to the black curves in figure \ref{fig:n2} are one-sided interfaces (of a type
for which it has been suggested they describe BCFTs \cite{Gutperle:2012hy}).
Finally, there are also solutions which approach singular behaviour at both ends of the radial domain, denoted by black dashed lines
in figure \ref{fig:n2}.
When the phase is equal to $\pi/2$, all solutions are regular Janus solutions
except for the one solution in the right plot of figure \ref{fig:n2} which would have a turning point with $\text{Im}(z^1)=1$, a singular point 
in field space.

\subsection{$\mathcal{N}=1^*$ one-mass model}
This model was summarised in section \ref{sec:onemass}. There is again one complex scalar field  $z^1$, which we write as
$z^1 = \tanh[ \frac{1}{2} \big( \alpha_3       -i \phi_3  \big)]$ and one real scalar field $\beta_1$. A particularly interesting 
feature of this model, that plays an important role in the solutions, is the presence of the two LS$^\pm$ $AdS_5$ fixed point solutions
given in \eqref{lsfpts}.

Consider solutions that approach $\mathcal{N}=4$ SYM with mass sources at, say, $r\to\infty$.
Following the discussion in section \ref{prelim} and using the results of appendices \ref{appb}, \ref{appc} we can summarise the source and expectation values
for the relevant operators that are active. All of the source terms are specified by $\phi_{3,(s)}$ with 
\begin{align}
\alpha_{3,(s)}=-\kappa \frac{L}{\ell} \phi_{3,(s)}\,,\qquad
\beta_{1,(s)}=-\frac{2}{3}\phi_{3,(s)}^2\,.
\end{align}
The field theory sources on $AdS_4$ are given by $\phi_{3,(s)}$, $\alpha_{3,(s)}$, $\beta_{1,(s)}$, 
with
$\ell\phi_{3,(s)}$, $\ell^2\alpha_{3,(s)}$, $\ell^2\beta_{1,(s)}$, invariant under Weyl scalings of $\ell$, while 
for those on flat spacetime the dimensionful quantities are given by \eqref{fspace}:
\begin{align}\label{fspace2onmassm}
\frac{\ell\phi_{3,(s)}}{y_3},\qquad \frac{\ell^2 \alpha_{3,(s)}}{y_3^2}, \qquad \frac{\ell^2\beta_{1,(s)}}{y_3^2}\,,
\end{align}
and have scaling dimensions $1,2,2$, respectively.

For the associated expectation values of the operators in flat spacetime, we have
\begin{align}
\langle\mathcal{O}_{\alpha_3}\rangle
&=\frac{1}{4\pi GL}\frac{\ell^2}{y_3^2}\Big({\alpha}_{3,(v)}+\alpha_{3,(s)}
\log(\frac{y_3}{\ell e^{2\delta_\alpha}})\Big)\,,
\end{align}
which then, with along with $\phi_{3,(s)}$ determines the remaining expectation values 
\begin{align}
\langle\mathcal{O}_{\phi_{3}}\rangle&=
\frac{4}{3}\frac{\ell}{y_3}\langle\mathcal{O}_{\beta_1}\rangle\,\phi_{3,(s)}-{2\kappa L}\frac{1}{y_3} \langle\mathcal{O}_{\alpha_3}\rangle
-\frac{L}{4\pi G}\frac{\ell}{y_3^3}\phi_{3,(s)}\,,\nn
\langle\mathcal{O}_{\beta_1}\rangle
&=\frac{4\kappa\ell}{L}\langle\mathcal{O}_{\alpha_3}\rangle\,\phi_{3,(s)}-
\frac{(1+4\delta_{\alpha}-4\delta_{\beta})}{2\pi G L}\frac{\ell^2}{y_3^2}\phi_{3,(s)}^2\,.
\end{align}

An important aspect of the above summary, is that for a specific choice of finite counterterms, all 
of the scalar sources and expectation values of the dual field theory can be obtained by giving $\ell\phi_{3,(s)}$ 
as well as $\ell^2{\alpha}_{3,(v)}$. We now set $\ell=\kappa=1$.

\begin{figure}[h!]
\centering
%\raisebox{-0.5\height}
{\includegraphics[scale=0.26]{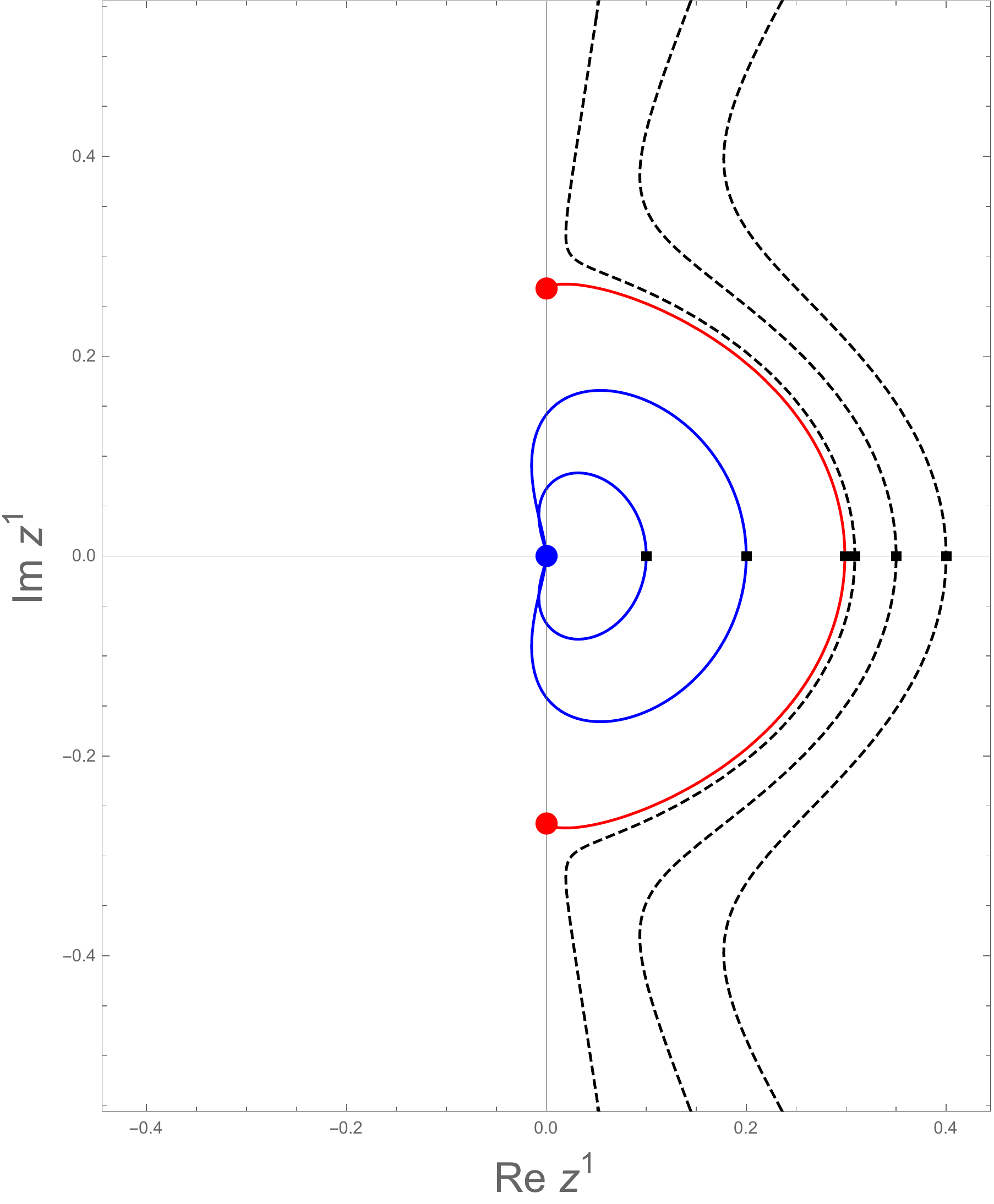}
\includegraphics[scale=0.26]{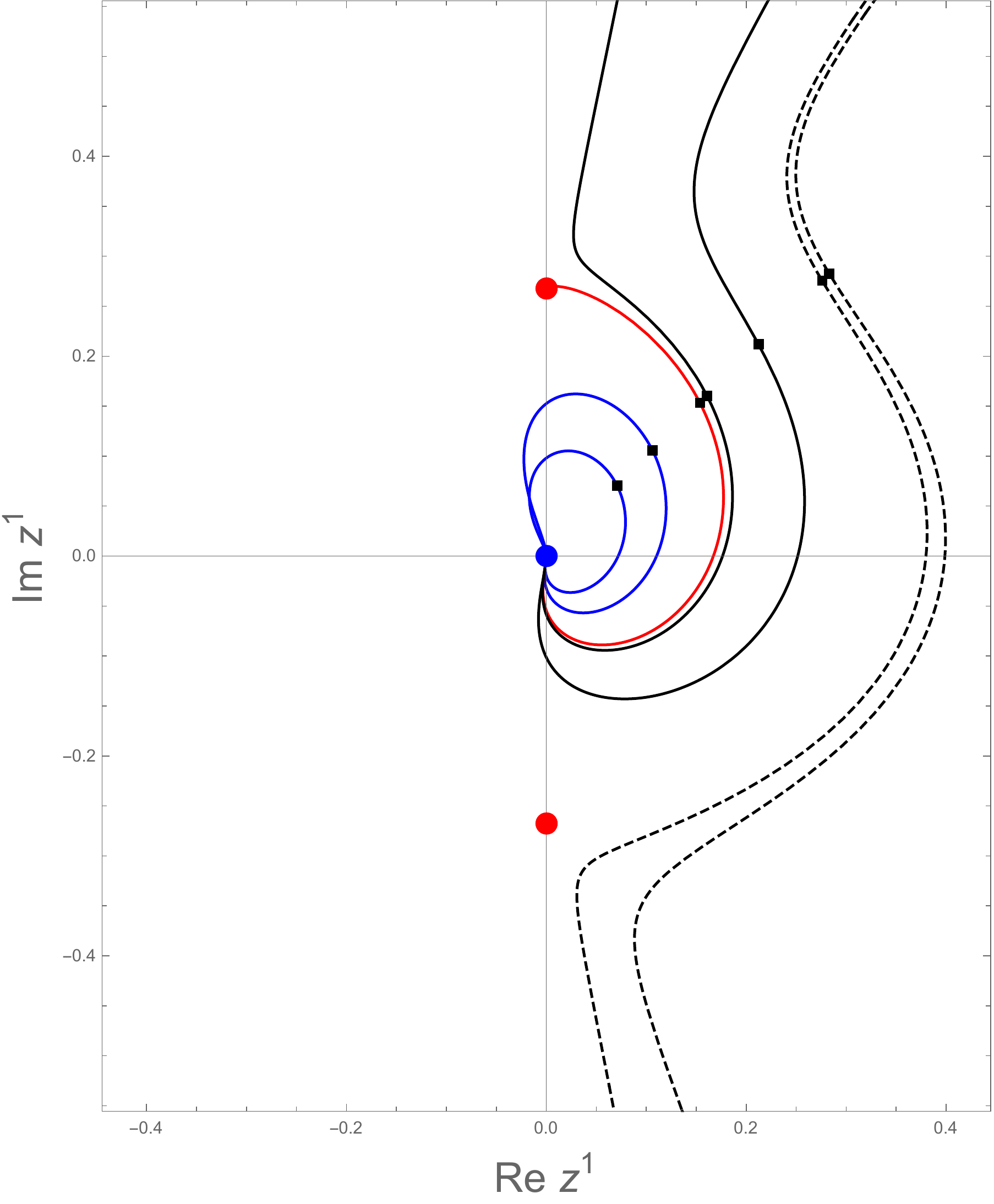}
\includegraphics[scale=0.26]{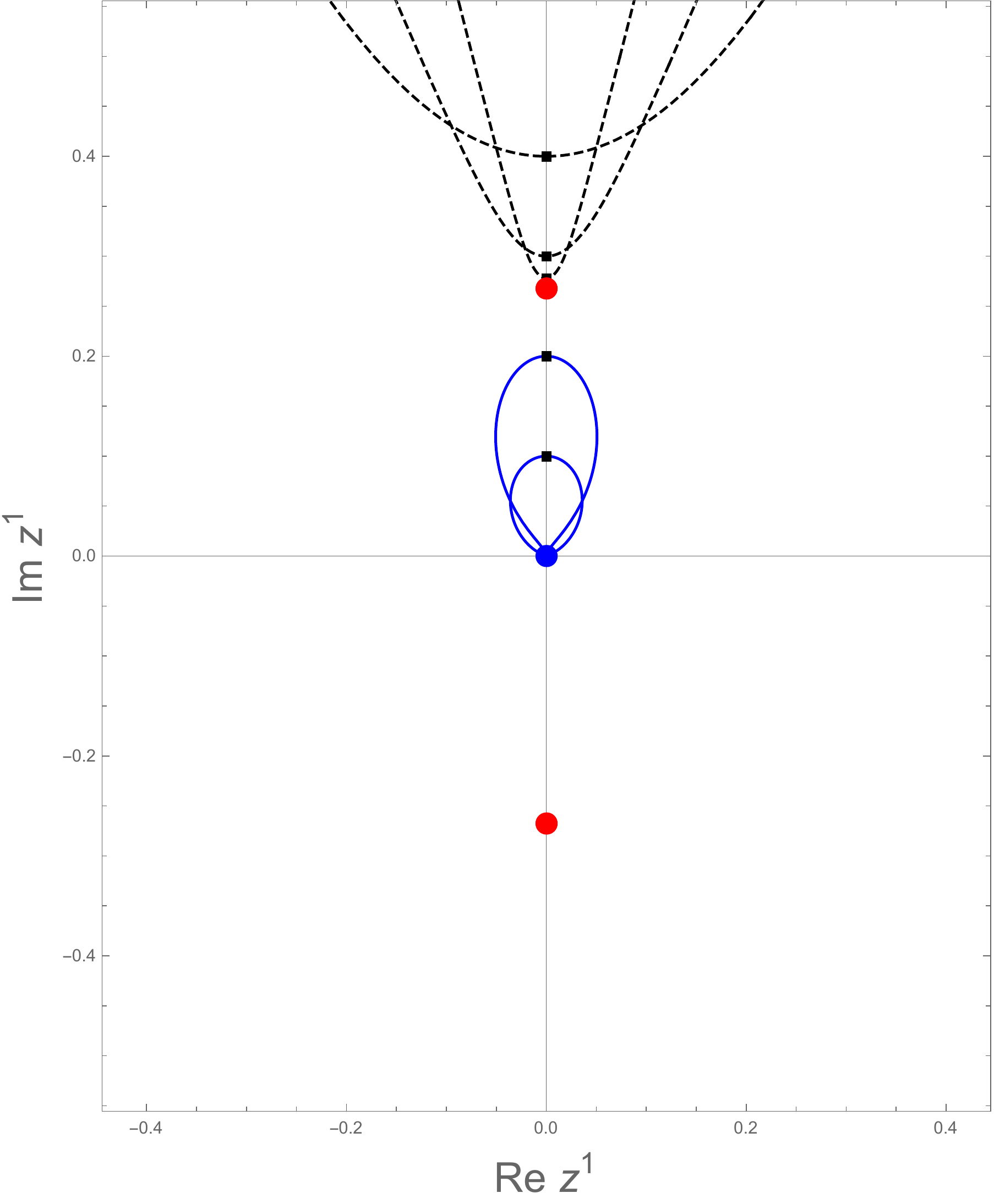}}
\caption{The family of BPS solutions for the $\mathcal{N}=1^*$ one mass model is summarised by parametrically plotting the real and imaginary parts of 
the scalar field $z^1$. 
The black squares correspond to turning points of the function $A_J(r)$ and the three plots, from left to right, correspond to solutions
where the phase of the complex scalar field at the turning point is $0,\pi/4$ and $\pi/2$, respectively. 
The blue dot at the origin is the $\mathcal{N}=4$ SYM $AdS_5$ vacuum and the blue lines are Janus solutions. 
The two red dots
are the two LS $^\pm$ $AdS_5$ solutions, each dual to the Leigh-Strassler SCFT. In the middle plot the red curve is a conformal RG interface with 
$\mathcal{N}=4$ SYM on one side of the interface and the LS SCFT on the other. In the left plot the red curve is a conformal interface with 
LS on either side of the the interface.
The boundary of field space is $|z^1|=1$ and the black curves are singular on one or both ends.
 As one moves from $r=-\infty$ to $r=+\infty$ one moves clockwise on the curves.}\label{fig:one}
\end{figure}

We next turn to the solutions which we have summarised in figure \ref{fig:one}. As before each plot corresponds to a fixed 
phase of the scalar field $z^1$ at the turning point of $A_J(r)$. From left to right we again have the phase is $0$, $\pi/4$ and $\pi/2$, respectively.
The blue dot at the origin is the $\mathcal{N}=4$ SYM $AdS_5$ vacuum solution, while the two red dots correspond to the two LS$^{\pm}$ $AdS_5$ fixed points 
given in \eqref{lsfpts}, each dual to the $\mathcal{N}=1$ LS SCFT.

First consider the left panel in figure \ref{fig:one}. There is a one parameter family
of $\mathcal{N}=4$ SYM Janus solutions (blue curves) that approach the $\mathcal{N}=4$ SYM $AdS_5$ vacuum solution with spatially modulated mass terms. 
Since the phase is zero, these solutions are invariant under the symmetry
\eqref{z2twosymg} and, as discussed in section \ref{otherendtext},
 we find that we have source $\phi_{3,(s)}$ on the 
$r\to +\infty$ side of the interface and source $\tilde\phi_{3,(s)}=-\phi_{3,(s)}$ on the $r\to -\infty$ side.
From the flat space perspective we therefore have (with $\ell=1$)
a source of the form $\phi_{3,(s)}/y_3$, for all $y_3$.
Similarly, we find that $\tilde{\alpha}_{3,(v)}={\alpha}_{3,(v)}$ on either side of the interface. These Janus solutions exist
for $0<\phi_{3,(s)}<\infty$. 

As $\phi_{3,(s)}\to \infty$, we have ${\alpha}_{3,(v)}\to \infty$ and the Janus solutions approach a new type of solution (red curve):
namely, a novel Janus solution with the LS$^+$ $AdS_5$ vacuum on one side of the interface and  
the LS$^-$ $AdS_5$ vacuum on the other. These solutions are discussed in more detail 
in \cite{1807495}. Note that there are no source terms that are active on either side of this $LS^+/LS^-$ interface;
this actually follows from the fact that once we demand that there are no sources for the irrelevant scalar operators with 
$\Delta=2+\sqrt{7}$ and $\Delta=3+\sqrt{7}$, it is not possible to source the relevant scalar operator of the LS SCFT with
dimension $\Delta=1+\sqrt{7}$ whilst preserving supersymmetry \cite{1807495}.
We also note that the irrational scaling dimensions for these operators seem to exclude
the possibility of having distributional sources for these scalar operators on this interface while still preserving conformal symmetry.
As explained in \cite{1807495} the two sides of the $LS^+/LS^-$ interface are related by a discrete automorphism.
Beyond this novel LS Janus solution there is also a one parameter family of solutions that approach singular behaviour, with $|z^1|\to 1$
at finite values of $r$. 

The middle panel of figure \ref{fig:one}
shows the set of solutions when the phase is $\pi/4$ and this provides the generic picture for phases in the open domain
$(0,\pi/2)$. There is again a one parameter family
of  $\mathcal{N}=4$ SYM Janus solutions (blue curves) with, for the $r=\infty$ end, 
$0<\phi_{3,(s)}<\phi_{3,(s)}|_{crit}$, with finite $\phi_{3,(s)}|_{crit}$. As $\phi_{3,(s)}\to \phi_{3,(s)}|_{crit}$, we have ${\alpha}_{3,(v)}$ approaching a finite
value and the Janus solutions approach another new type of solution (red curve). Before discussing that, we note that
$\phi_{3,(s)}$ at $r=\infty$ and $\tilde \phi_{3,(s)}$ at $r=-\infty$ are not simply related in general and hence we have flat space sources as in \eqref{mixedsce}.
Returning to the new solution (red curve), we see that it approaches the $\mathcal{N}=4$ SYM $AdS_5$ vacuum at $r\to\infty$ 
and the LS$^+$ $AdS_5$ solution at $r\to-\infty$. This describes 
a conformal RG interface, with $\mathcal{N}=4$ SYM on one side of the interface, with 
spatially dependent sources with $\phi_{3,(s)}=\phi_{3,(s)}|_{crit}$, and the LS SCFT on the other.  
Once again there are no sources on the $LS^+$ side of the interface.
This solution is also discussed in more detail in \cite{1807495}.
Beyond this solution, for $\phi_{3,(s)}|_{crit}<\phi_{3,(s)}<\infty$ we obtain solutions which start off at the mass deformed $\mathcal{N}=4$ SYM $AdS_5$ vacuum at $r\to \infty$ and then become singular at some finite value of $r$, as marked with the black lines in the middle panel of figure \ref{fig:one}. 
There are also solutions that become singular at both $r\to\pm\infty$ which are marked by black dashed lines in figure \ref{fig:one}.

What we described in the previous paragraph was for the phase at the turning point equal to 
$\pi/4$ and applies for the
phase in the range $(0,\pi/2)$, with one small difference. Beyond some value of the phase, we find that $\phi_{3,(s)}$ is no longer always positive
for the Janus solutions. In particular, this leads to a limiting red curve solution,
with a specific value of the phase of the turning point, 
describing an RG interface where the source $\phi_{3,(s)}|_{crit}$ vanishes on the $\mathcal{N}=4$ SYM side. 
This solution is also discussed in \cite{1807495}.

Finally, when the phase is $\pi/2$ (third plot in figure \ref{fig:one}), there is a one parameter family of $\mathcal{N}=4$ SYM Janus solutions
that exist for $-\infty<\phi_{3,(s)}<0$.
These solutions are invariant under the symmetry \eqref{z2twosymgminus} and, as discussed in section \ref{otherendtext},
we find that the source on either side of the interface at $r=\pm\infty$ takes the same value $\tilde\phi_{3,(s)}=\phi_{3,(s)}$.
From the flat space perspective we therefore have (with $\ell=1$)
a source of the form $\phi_{3,(s)}/|y_3|$, for all $y_3$. 
There is also a one parameter family of solutions that are singular at finite values of the radial coordinate in each direction and are marked by 
the dashed black lines in the right plot in figure \ref{fig:one}.

\subsection{$\mathcal{N}=1^*$ equal-mass model}

This model was summarised in section \ref{eqmassctrunc}. There are two independent complex fields
$z^1$, $z^2$ which we write
\begin{align}\label{4scmod2}
     z^1 &= \tanh \big[ \frac{1}{2} \big( 3\alpha_1 + \varphi 
       -i 3\phi_1+ i \phi_4 \big) \big] \,,\nn
            z^2 &= \tanh \big[ \frac{1}{2} \big( \alpha_1 - \varphi 
       -i \phi_1 -i\phi_4 \big) \big] \,.
     \end{align}

Consider solutions that approach $\mathcal{N}=4$ SYM with mass sources at, say $r\to\infty$. As we have mentioned several times, in this paper we focus on solutions for which the source terms for the coupling constant and the gaugino mass vanish:
\begin{align}
\varphi_{(s)}=\phi_{4,(s)}=0\,.
\end{align}
All of the source terms for BPS configurations are then specified by $\phi_{1,(s)}$ with
\begin{align}
\alpha_{1,(s)}=-\kappa \frac{L}{\ell} \phi_{1,(s)}\,.
\end{align}
The field theory sources on $AdS_4$ are given by $\phi_{1,(s)}$, $\alpha_{1,(s)}$, with
$\ell\phi_{1,(s)}$, $\ell^2\alpha_{1,(s)}$ invariant under Weyl scalings of $\ell$, while 
those on flat spacetime are given by \eqref{fspace}:
\begin{align}\label{fspace2n1seqm}
\frac{\ell\phi_{1,(s)}}{y_3},\qquad \frac{\ell^2 \alpha_{1,(s)}}{y_3^2}\,,
\end{align}
and have scaling dimensions $1$ and $2$, respectively.

For the associated expectation values of the operators in flat spacetime, we have
\begin{align}
\langle\mathcal{O}_{\alpha_1}\rangle=\langle\mathcal{O}_{\alpha_2}\rangle=\langle\mathcal{O}_{\alpha_3}\rangle
&=\frac{1}{4\pi GL}\frac{\ell^2}{y_3^2}\Big({\alpha}_{1,(v)}+\alpha_{1,(s)}
\log(\frac{y_3}{\ell e^{2\delta_\alpha}})\Big)\,,\nn
\langle\mathcal{O}_{\phi_4}\rangle&=\frac{1}{2\pi GL}\frac{\ell^3}{y_3^3}\Big(\phi_{4,(v)}-\frac{9-2\delta_{4(5)}}{3}\phi_{1,(s)}^3\Big)\,.
\end{align}
For BPS configurations the remaining expectation values are determined by these expressions,
along with $\phi_{1,(s)}$, via
\begin{align}
\langle\mathcal{O}_{\phi_1}\rangle=\langle\mathcal{O}_{\phi_2}\rangle=\langle\mathcal{O}_{\phi_3}\rangle&=-{2\kappa L}\frac{1}{y_3}\langle\mathcal{O}_{\alpha_1}\rangle
-\frac{L}{4\pi G}\frac{\ell}{y_3^3}\phi_{1,(s)}\,,\nn
y_3\langle\mathcal{O}_{\varphi}\rangle&=-\frac{3\kappa L}{2}\langle\mathcal{O}_{\phi_4}\rangle -\frac{\kappa(3-2\delta_{4(5)})}{4\pi G}\frac{\ell^3}{y_3^3}\phi_{1,(s)}^3\,.
\end{align}
Note that $\delta_\alpha$, $\delta_{4(5)}$ parametrise finite counterterms which we have not fixed.
We now set $\ell=\kappa=1$.

Following the discussion given at the end of section \ref{bpsjanus}, we know that there is a four parameter family of solutions for this model. 
Here we will just study a one parameter family of solutions, leaving a more complete exploration for future work. We also note the following technical point in solving the numerical equations. If we construct a solution with, say, the $\mathcal{N}$=4 SYM dilaton source non-vanishing at the
$r\to\infty$ end,  $\varphi_{(s)}\ne 0$, then we can obtain a solution with $\varphi_{(s)}=0$ by using the shift symmetry of the dilaton \eqref{dilshift}.  

\begin{figure}[h!]
  \centering
  \begin{subfigure}[b]{0.23\linewidth}
    \includegraphics[scale=0.27]{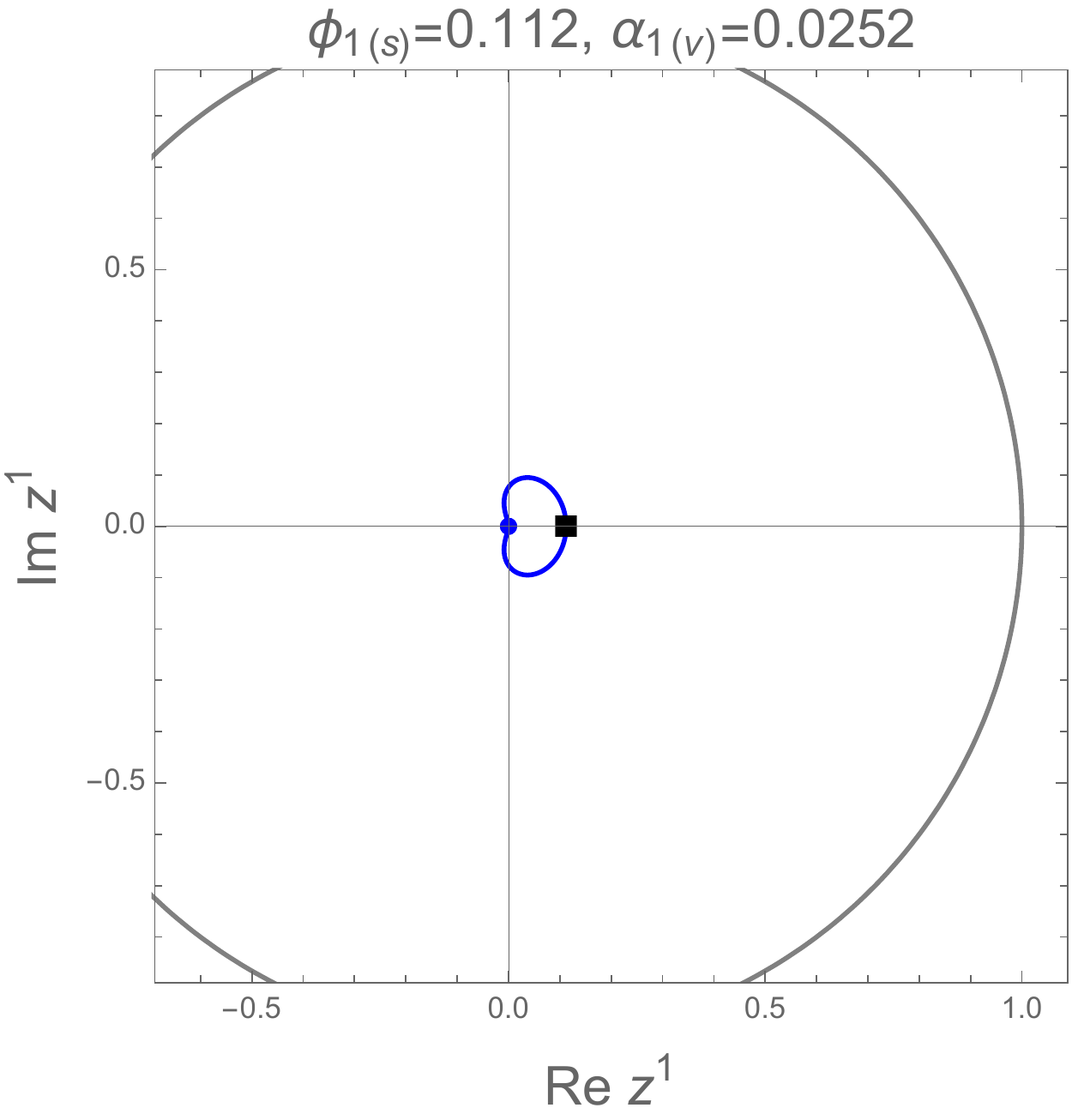}
    \caption{}
  \end{subfigure}
  \begin{subfigure}[b]{0.23\linewidth}
    \includegraphics[scale=0.27]{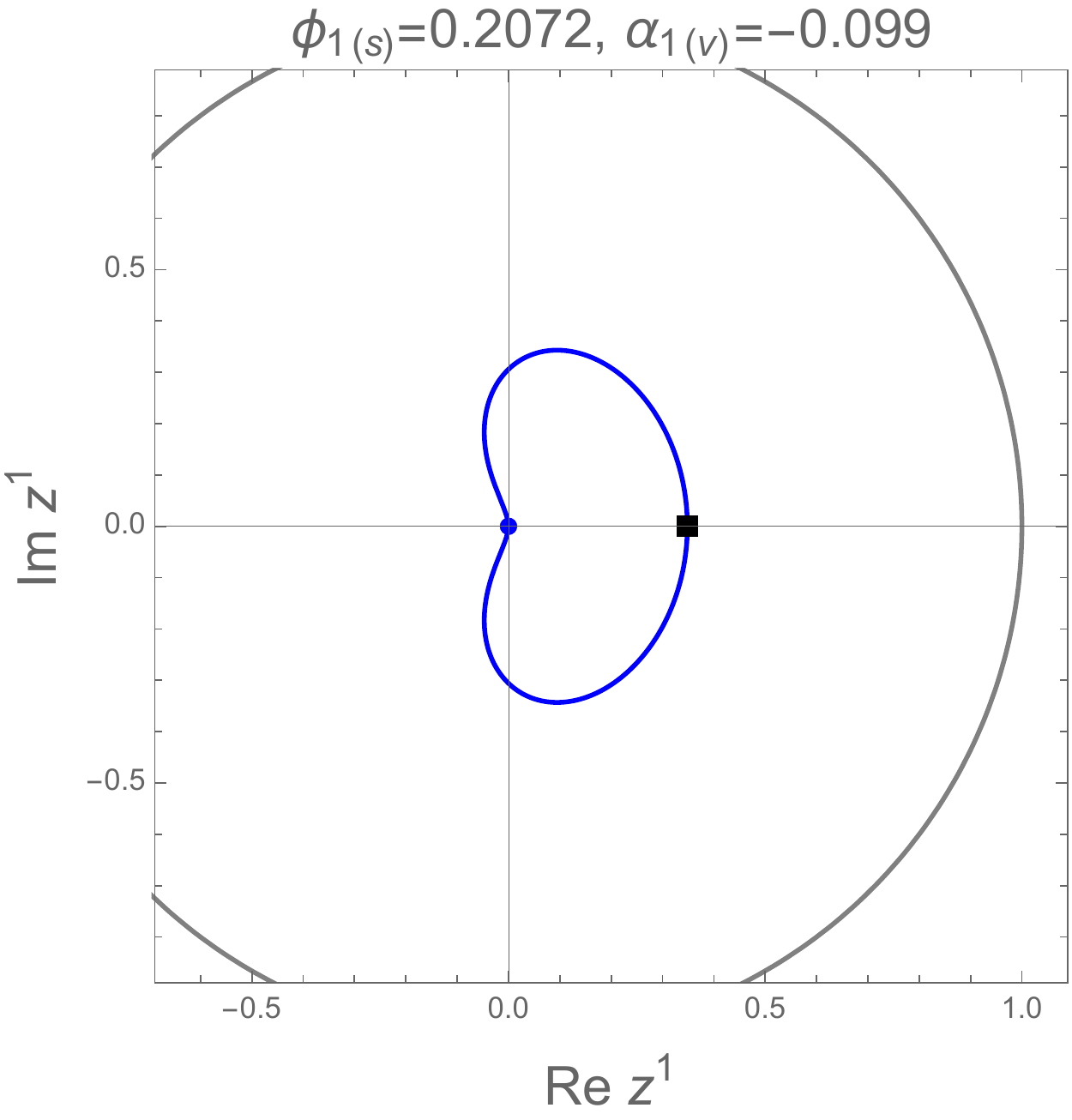}
    \caption{}
  \end{subfigure}
    \begin{subfigure}[b]{0.23\linewidth}
    \includegraphics[scale=0.27]{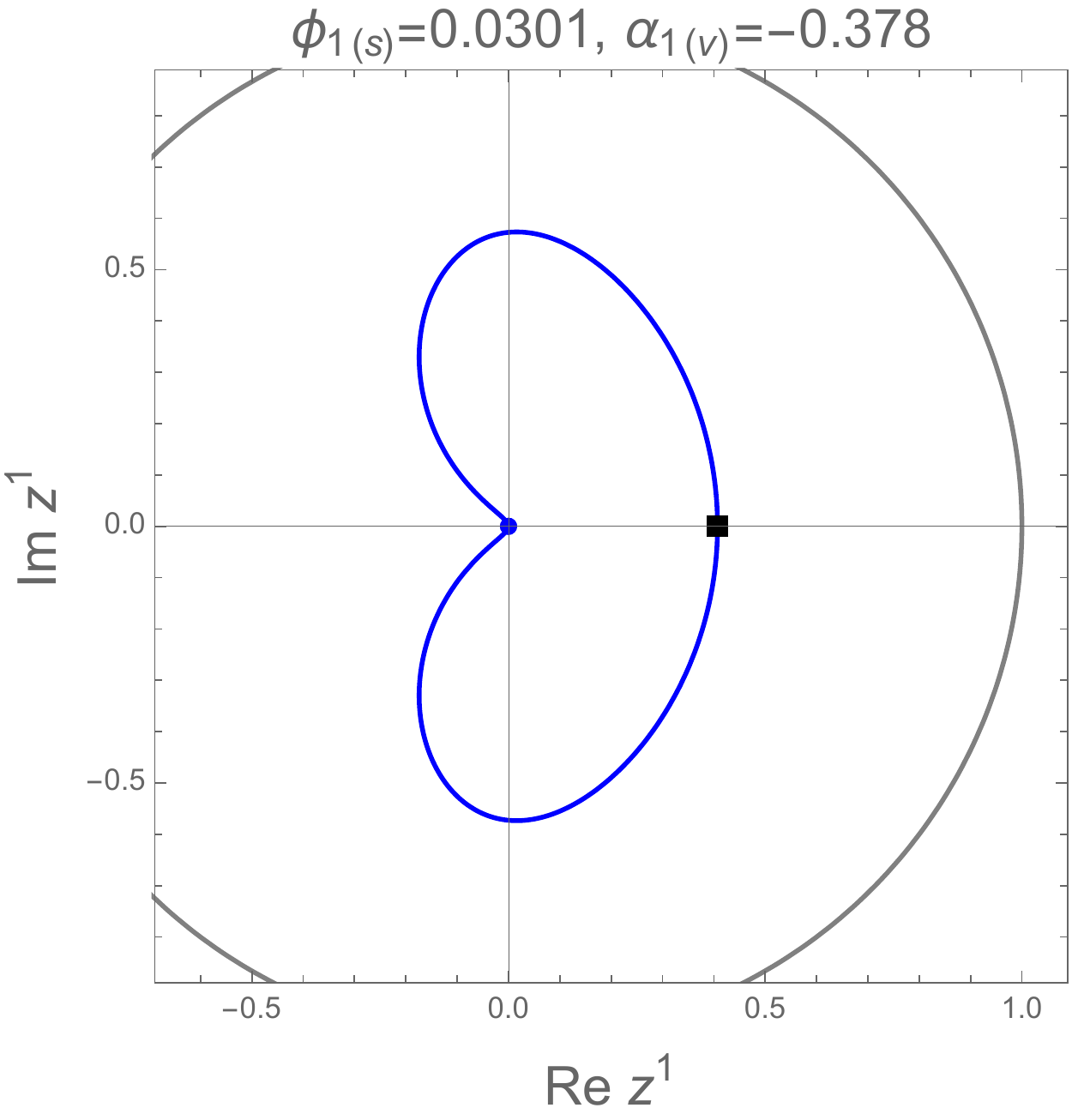}
    \caption{}
  \end{subfigure}
    \begin{subfigure}[b]{0.23\linewidth}
    \includegraphics[scale=0.27]{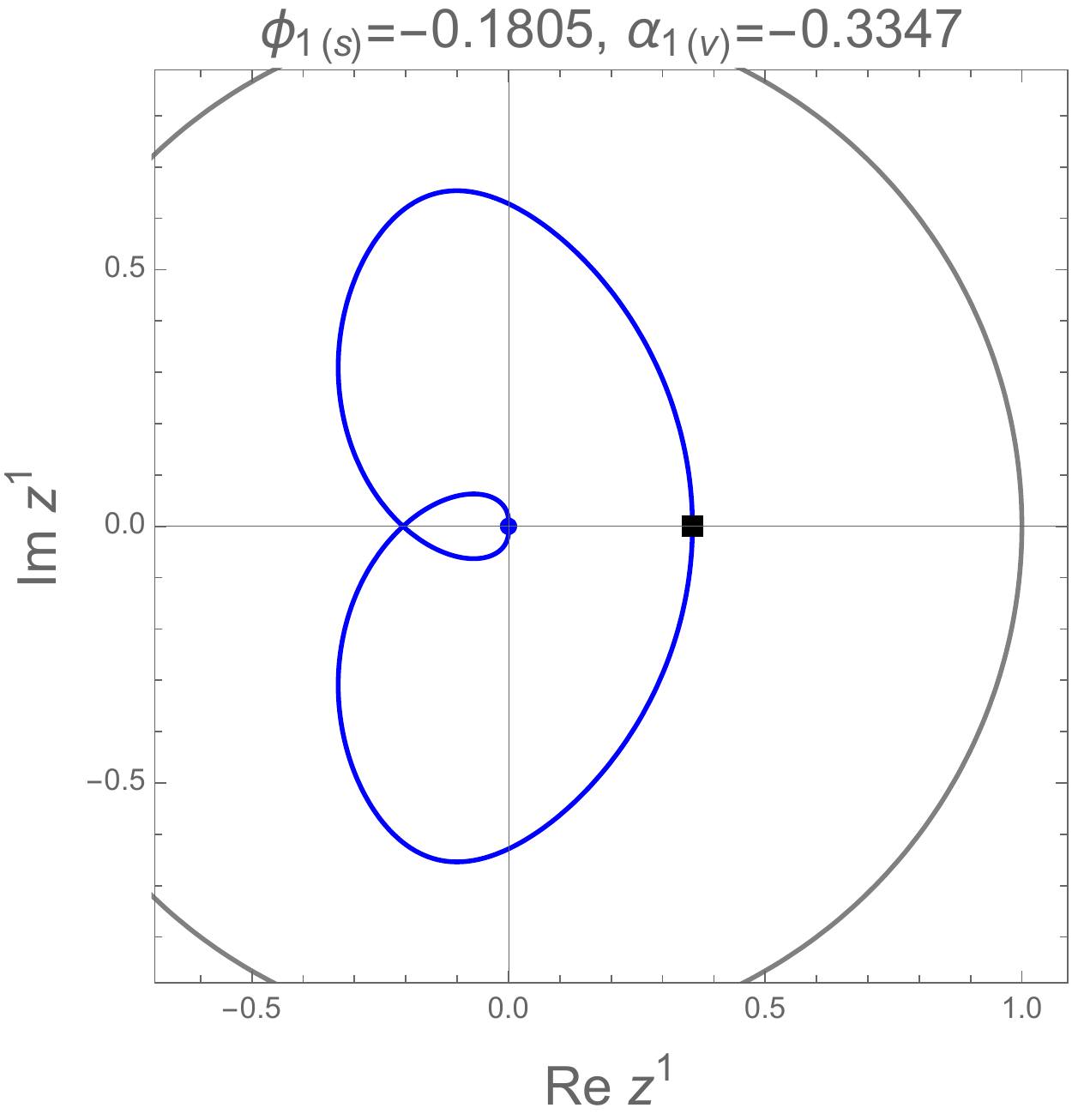}
    \caption{}
  \end{subfigure}
    \begin{subfigure}[b]{0.23\linewidth}
    \includegraphics[scale=0.27]{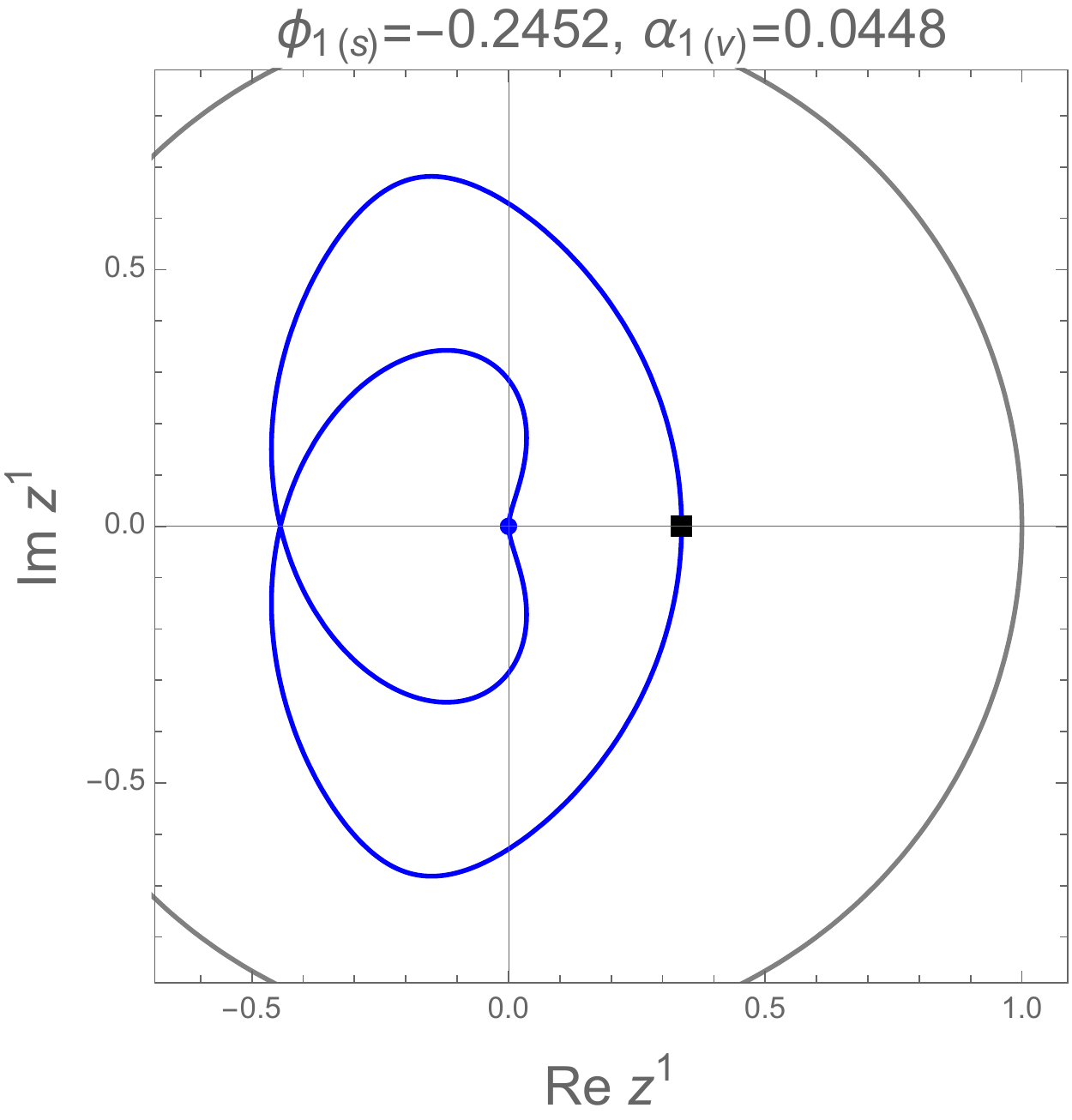}
    \caption{}
  \end{subfigure}
  \begin{subfigure}[b]{0.23\linewidth}
    \includegraphics[scale=0.27]{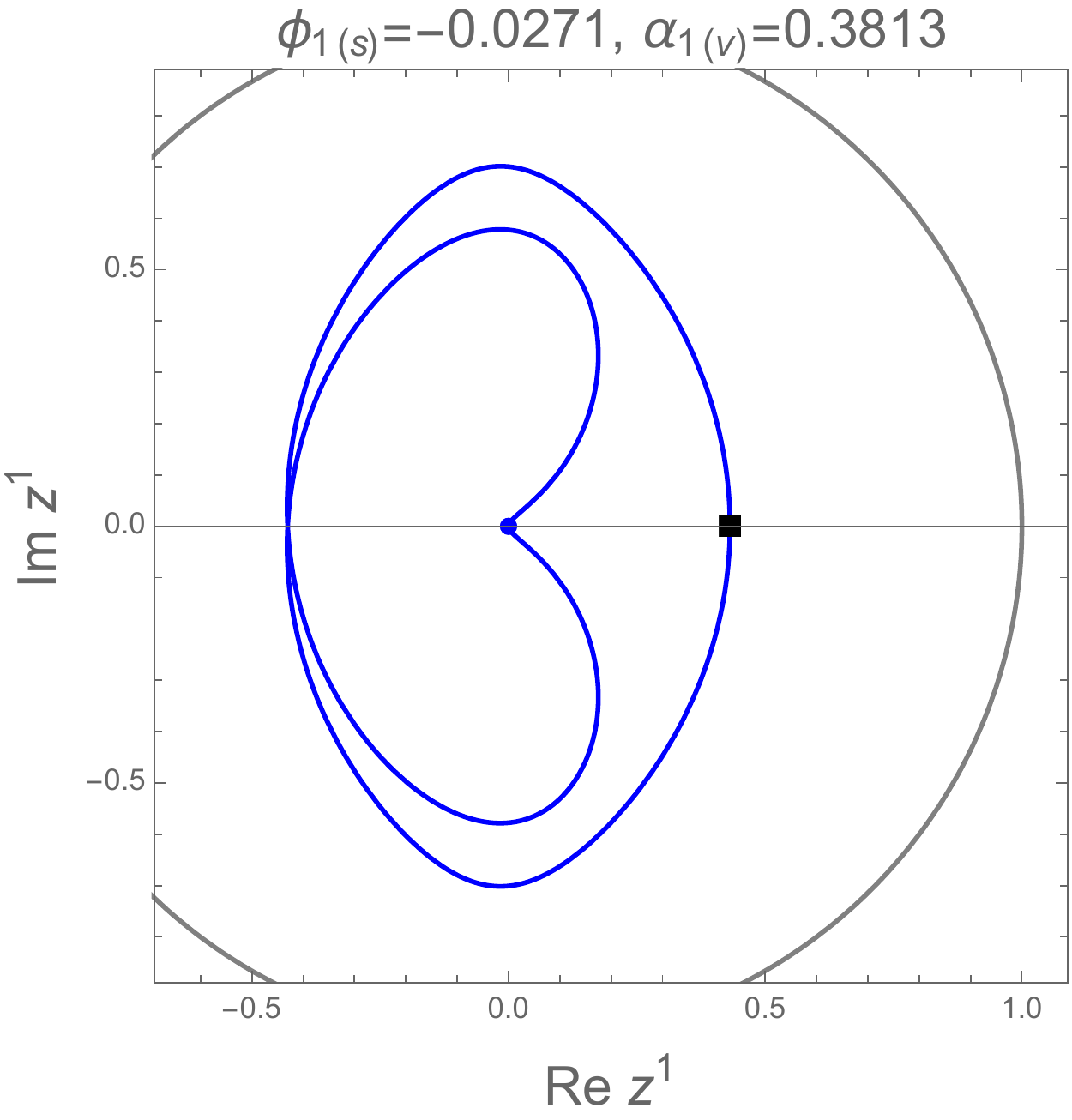}
    \caption{}
  \end{subfigure}
    \begin{subfigure}[b]{0.23\linewidth}
    \includegraphics[scale=0.27]{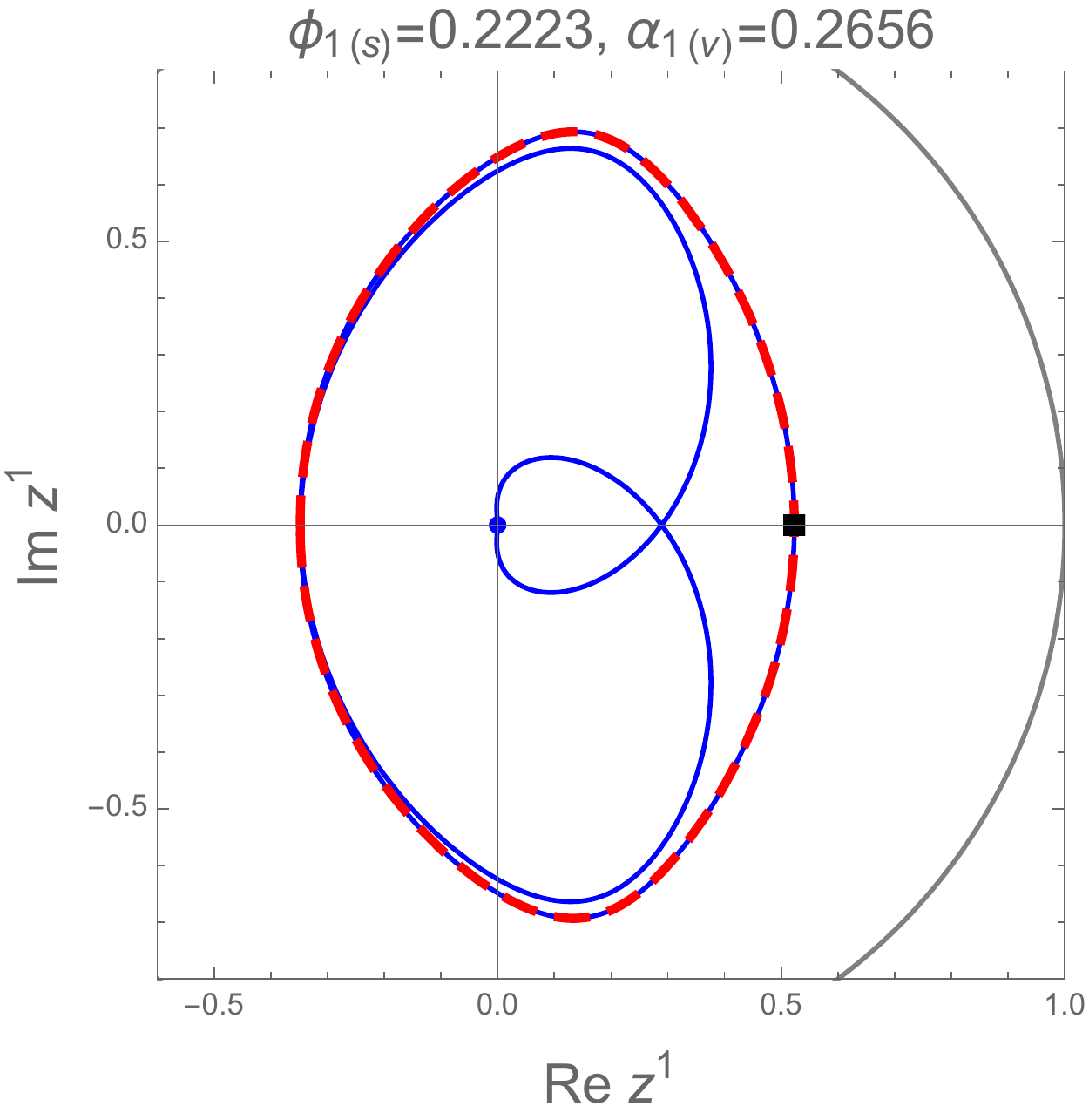}
    \caption{}
  \end{subfigure}
    \begin{subfigure}[b]{0.23\linewidth}
    \includegraphics[scale=0.27]{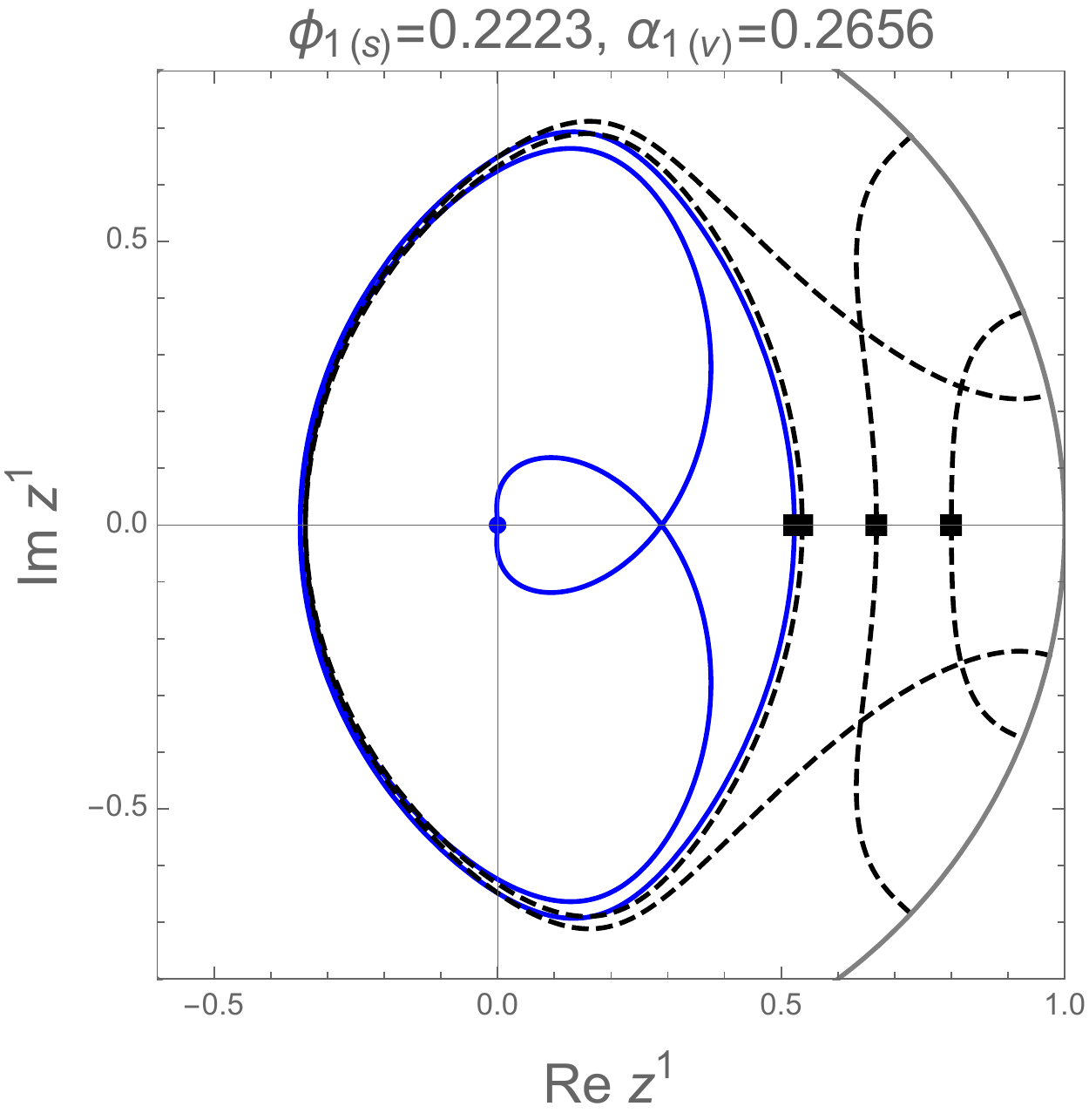}
    \caption{}
  \end{subfigure}
  \caption{A family of symmetric BPS solutions for the $\mathcal{N}=1^*$ equal mass model is summarised by 
parametrically plotting the real and imaginary parts of the scalar field $z^1$; the behaviour of the other scalar field $z^2$
is broadly similar. The black squares correspond to turning points of the function $A_J(r)$, where the phase of both scalars is zero.
The family of solutions can be labelled by $(\alpha_1)_{tp}$, a function of $z^1$ and $z^2$ at the turning point invariant under shifts of the dilaton.
The blue dot at the origin is the $\mathcal{N}=4$ SYM $AdS_5$ vacuum and the blue lines are Janus solutions. 
As $(\alpha_1)_{tp}$ increases monotonically from (a)-(g), we see the appearance of more and more loops, asymptoting to the red curve,
in figure (g), which describes a solution periodic in the radial direction. In figure (h) we have exactly the same solutions as figure (g)
but with the addition of some illustrative solutions (black dashed lines) that are singular at both ends (and without
the red curve for clarity) .
As one moves from $r=-\infty$ to $r=+\infty$ one moves clockwise on the curves.
 }\label{fig:so3janus}
\end{figure}

In figure \ref{fig:so3janus} we have summarised a one-parameter family of $\mathcal{N}=4$ SYM Janus solutions for this model (with $\varphi_{(s)}=0$ on both sides), for which the phase of both scalars is zero at the turning point and so the solutions are invariant under the symmetry \eqref{z2twosymg}. 
In contrast to previous models it is convenient to label this family of solutions not by the values of $z^i$ at the turning point but instead in terms of the value of
$\alpha_1$ at the turning point which we label as $(\alpha_1)_{tp}$. In particular, we note that this is invariant under the dilaton shift.  
For a fixed value of $(\alpha_1)_{tp}$ there is a one-parameter family of solutions for which $z^i_{tp}$ are real, all related by shifts of the dilaton and so for regular solutions we can use this symmetry to fix $\varphi_{(s)}=0$ for each value of $(\alpha_1)_{tp}$ (and using \eqref{z2twosymg} we find it is set to zero on both sides). We find that regular solutions exist for $-\alpha_{crit} <(\alpha_1)_{tp}< \alpha_{crit}$ with $\alpha_{crit} \approx 0.447$. In figure \ref{fig:so3janus} we have displayed a series of Janus solutions as blue curves, 
for various values in the range $(\alpha_1)_{tp}\in [0,\alpha_{crit}).$ Interestingly, as $(\alpha_1)_{tp}$ increases the solutions start to develop a sequence of more and more loops in the scalar field parameter space and, surprisingly,  as $(\alpha_1)_{tp} \rightarrow \alpha_{crit}$ we obtain a new solution which is exactly periodic in the radial coordinate $r$ (the red curve), which we return to below.

Note that in figure \ref{fig:so3janus} we have just plotted $z^1$; the behaviour of $z^2$ is broadly similar.
We also note in addition to the Janus solutions, there are also a host of solutions that are singular
at both ends. The last panel in figure \ref{fig:so3janus} illustrates a few such solutions. In particular
there are solutions that can wind several times around, before hitting the singularity.

\begin{figure}[h!]
\centering
%\raisebox{-0.5\height}
{\includegraphics[scale=0.45]{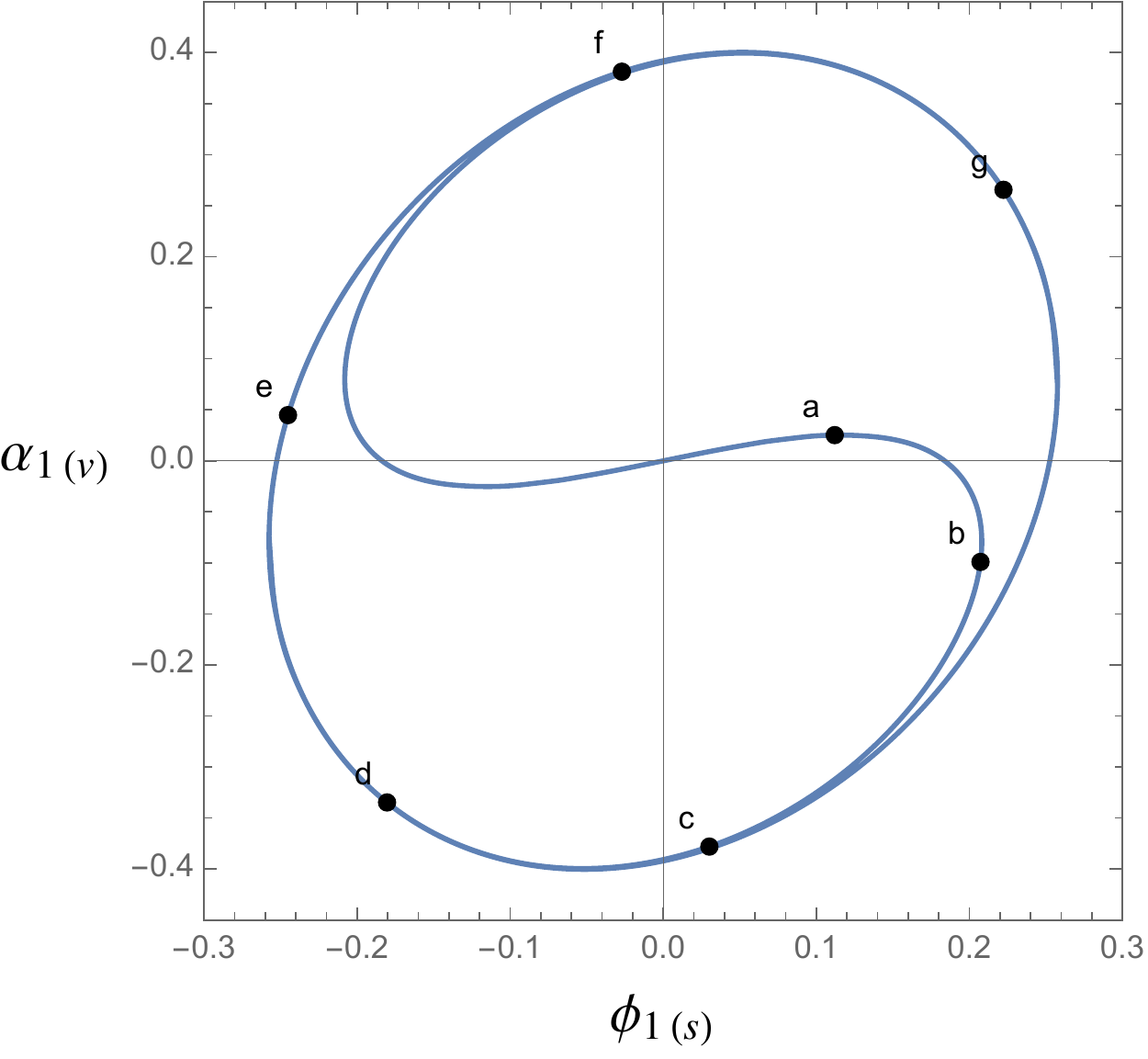}
%\qquad\includegraphics[scale=0.35]{SO3_source_vev_2}
}
\caption{The behaviour of $\alpha_{1,(v)}$ determining certain expectation values as a function of $\phi_{1,(s)}$, for the symmetric BPS solutions, for the $\mathcal{N}=1^*$ equal mass model. The labelled black dots (a)-(g) correspond to the solutions
given in figure \ref{fig:so3janus}.
 }\label{fig:so3alphvev}
\end{figure}

We next discuss the expectation values of operators for the $\mathcal{N}=4$ SYM Janus solutions. Again, note that because our solutions are invariant under \eqref{z2twosymg}, $\phi_{1,(s)}$ and $\phi_{4,(v)}$ are antisymmetric for these solutions while $\alpha_{1,(v)}$ is symmetric.
In figure \ref{fig:so3alphvev} we have plotted $\alpha_{1,(v)}$ against $\phi_{1,(s)}$ for the blue curve solutions depicted in figure \ref{fig:so3janus}  (as well as their reflected versions with $(\alpha_1)_{tp}<0$) . The figure
depicts a curve that spirals out, winding an infinite number
of times while asymptoting
to the oval shape. Thus,
we see that for each value of $\phi_{1,(s)}$ there are multiple values of $\alpha_{1,(v)}$ all of
which are associated with a different BPS Janus interface of $\mathcal{N}=4$ SYM for the same $\phi_{1,(s)}$. It is more demanding to 
extract from our numerics the value of $\phi_{4,(v)}$, which fixes the remaining expectation values, but the results we have make it clear that $\phi_{4,(v)}$ behaves in a similar manner to $\alpha_{1,(v)}$, but with just one limiting value of $\phi_{4,(v)}$ for a given $\phi_{1,(s)}$ instead of two.

\begin{figure}[h!]
\centering
%\raisebox{-0.5\height}
{\includegraphics[scale=0.35]{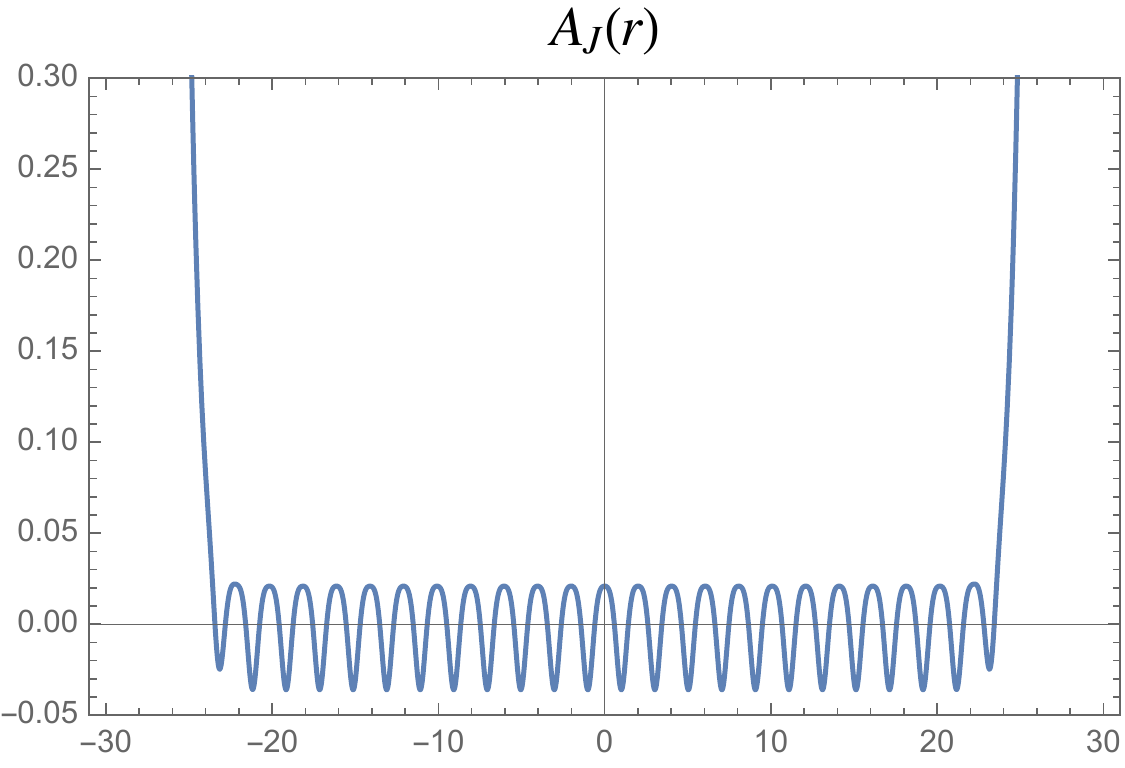}\qquad
\includegraphics[scale=0.35]{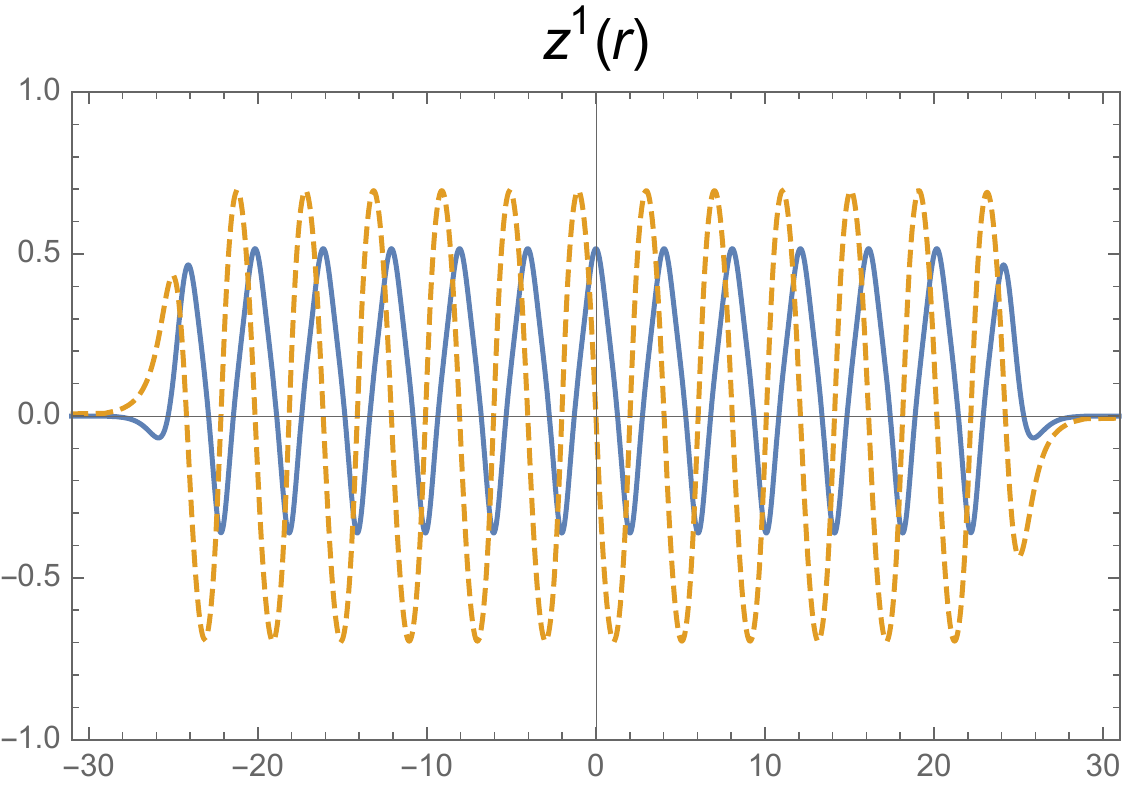}\qquad
\includegraphics[scale=0.35]{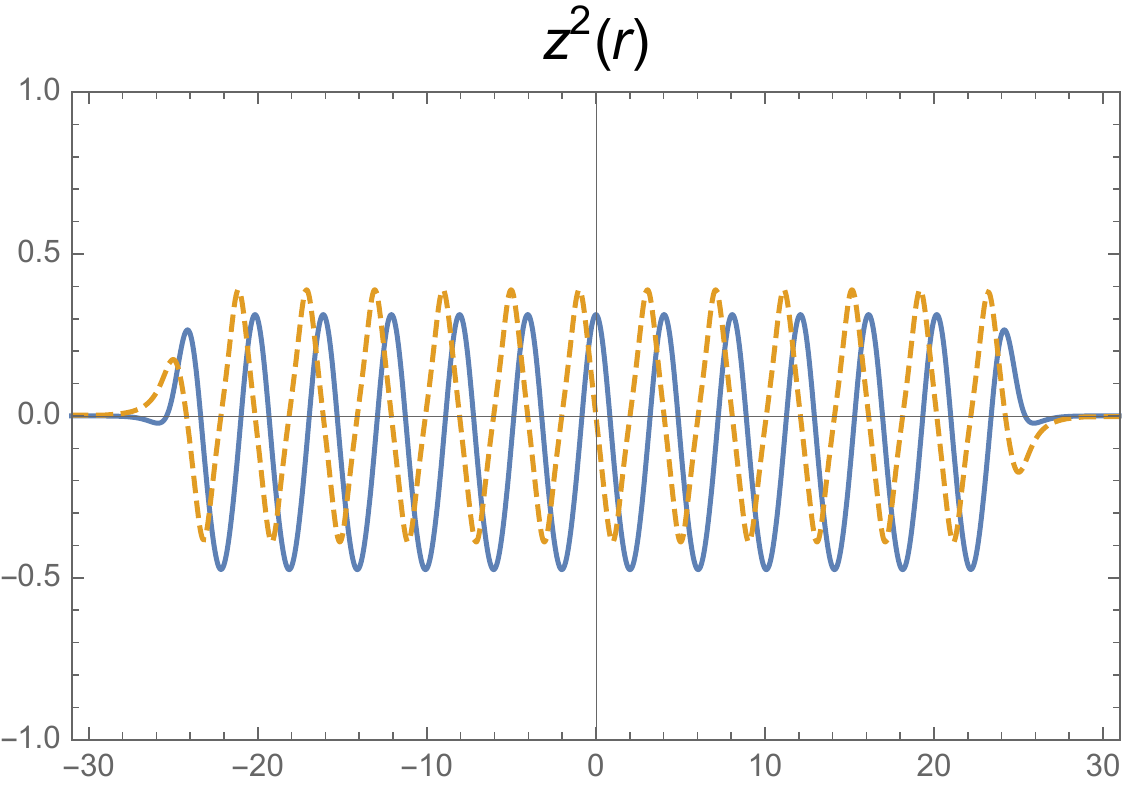}
}
\caption{For the $\mathcal{N}=1^*$ equal mass model as $(\alpha_1)_{tp}\to \alpha_{crit}$, approaching the red curve in figure 
\ref{fig:so3janus}, the $\mathcal{N}=4$ SYM Janus solutions
have a radial region approaching a solution that is periodic in the radial coordinate. For any $-\alpha_{crit}<(\alpha_1)_{tp} < \alpha_{crit}$, the solution is a Janus solution
and so $A_J\to \pm r/L$ and $z^A\to 0$ as $r\to \pm\infty$. Both the period and shape of the middle region is essentially
unchanged as we approach the critical solution, with just more periods appearing, and clearly reveals the functional form of the periodic solution. 
The blue and orange curves are the real and imaginary components of $z^1,z^2$, respectively.
 }\label{periodic}
\end{figure}

We now return to the limiting periodic solution corresponding to the red curve in figure \ref{fig:so3janus}.
As $(\alpha_1)_{tp}\to \alpha_{crit}$ all of the functions develop more and more periods in the
radial direction, with the period and shape changing very little as the limit is taken. In figure \ref{periodic} we have plotted
the metric function $A_J$ as well as the scalar functions $z^1,z^2$ as a function of $r$ for a solution close to $\alpha_{crit}$.
For any $(\alpha_1)_{tp}< \alpha_{crit}$ we have a Janus solution, so $A_J\to \pm r/L$  and
all the scalars go to zero as $r\to \pm\infty$. The region in between, however, approaches a solution
that has periodic behaviour in the entire region $r\in(-\infty,\infty)$. By compactifying the radial direction for this limiting
periodic solution, we obtain a new $AdS_4\times S^1$ solution that will be further explored in \cite{toappear3}. Note that we can also approach this critical solution from above, $(\alpha_1)_{tp} > \alpha_{crit}$, where solutions develop more and more periods before becoming singular (see figure \ref{fig:so3janus}(h)).

We conclude this section with one further comment regarding a possibly confusing feature of figure \ref{fig:so3janus}. As argued above, there exists a one parameter family of periodic solutions related by the shift symmetry of the dilaton. When constructing Janus solutions, we fix this shift symmetry by requiring our solutions have vanishing dilaton on the boundaries, but for different values of $(\alpha_1)_{tp}$ this corresponds to a different average value of the dilaton in the periodic intermediate region. Thus the enveloping ellipses of the blue curves differ for the different curves in figure \ref{fig:so3janus} (as is clear in plots (e)-(g)), though they are all related by a dilaton shift \eqref{dilshift}. Said another way, when $\varphi_{(s)}=0$ on the boundaries for each value of $(\alpha_1)_{tp}$, only the fields $\phi_1$, $\alpha_1$ and $\phi_4$ have a well-defined limit when $(\alpha_1)_{tp} \to \alpha_{crit}$, whereas $\varphi$ does not.

\section{Discussion}\label{dics}

In this paper we have analysed mass deformations of $\mathcal{N}=4$ SYM theory that depend on one
of the three spatial dimensions and preserve some amount of supersymmetry. We have focussed on configurations with constant
coupling constant. We have also explored these
deformations within the context of holography, studying both configurations that preserve $ISO(1,2)$ symmetry as well
those that in addition preserve conformal symmetry. For the latter class of deformations we have also constructed a number
of interesting new classes of explicit Janus solutions.

In section 2 we analysed the supersymmetric mass deformations of $\mathcal{N}=4$ SYM from a field theory perspective. We achieved this 
by coupling $\mathcal{N}=4$ SYM to off shell conformal supergravity and then taking the Planck mass to infinity as in \cite{Maxfield:2016lok}.
For configurations that have constant $\tau$ (i.e. constant coupling constant and theta angle)
as well as no deformations in the ${\bf 15}$ and ${\bf 6}$, parametrised by
$V_\mu^i\,_j$ and $T_{\mu\nu}^{ij}$, respectively,  we reduced the problem 
to solving the equations given in \eqref{masteqs}. We then focussed on deformations that generalised the homogeneous $\mathcal{N}=1^*$ mass deformations, studying in some detail three particular examples: the one mass model, the equal mass model and the $\mathcal{N}=2^*$ model. 
It would be interesting to further investigate other possible solutions to \eqref{masteqs}. 
In the static case, we anticipate that the examples we have studied cover the most general case of conformal interfaces after employing suitable $SU(4)$ rotations. However, there are additional classes of solutions that allow time dependence which involve a null projection condition on the Killing spinors which can be explored.

It would be interesting to analyse more general deformations that also allow $\tau$ to depend on the spatial coordinates. For the Janus class this will include the classification of \cite{DHoker:2006qeo}, which considered deformations with varying coupling constant combined with other
deformations all proportional to spatial derivatives of the coupling 
constant\footnote{The supersymmetric Janus supergravity solutions 
corresponding to \cite{DHoker:2006qeo} have recently been discussed in \cite{Bobev:2020fon}. From \cite{Bobev:2020fon} one can check that the there are no source terms for the dimension $\Delta=2,3$ operators away from the interface, consistent with \cite{DHoker:2006qeo}.}.
By relaxing this latter condition, one can anticipate that additional cases are possible, as a sort of superposition of the those of 
\cite{DHoker:2006qeo} with the ones of this paper. However, the non-linearity of the equations \eqref{masteqs} with respect to $E_{ij}$ indicates that
a detailed analysis is warranted. More generally, one can also explore supersymmetry preserving deformations that also involve $g_{\mu\nu}$, $V_\mu^i\,_j$, which have been utilised in other situations, such as D3-branes wrapping supersymmetric cycles \cite{Maldacena:2000yy}, as well as $T_{\mu\nu}^{ij}$ and, additionally, allowing for time dependence.

In the remainder of the paper we analysed the supersymmetric mass deformations, with constant $\tau$, from a holographic perspective. We utilised
a consistent truncation of $D=5$ gauged supergravity that involves 10 real scalar fields which allowed us to obtain 
BPS equations preserving $ISO(1,2)$ symmetry for real mass deformations. The natural arena to 
analyse complex mass deformations would be to utilise an $\mathcal{N}=2$ $D=5$ gauged supergravity theory coupled to two vector multiplets and four hypermultiplets, with scalar manifold as in \eqref{etnscs}. However, this supergravity theory has not yet been explicitly constructed, but has been explored recently in \cite{Bobev:2020ttg}.

For the $ISO(1,2)$ preserving configurations associated with real mass deformations we carried out in some detail
the holographic renormalisation procedure. We saw that the model admits a large number of finite counterterms. We managed to
reduce this number a little by demanding that supersymmetric configurations have vanishing energy density. It would be desirable to
identify a fully supersymmetric scheme along the lines of \cite{Papadimitriou:2017kzw}, but this could be a challenging task. Our results indicate that there will not be a unique supersymmetric scheme due to the possibility of adding finite supersymmetric invariants; a useful 
starting point to determine these invariants would be to use the results of \cite{Butter:2016mtk}. 
A complementary approach would be to generalise the field theory analysis in section 3 of \cite{Bobev:2016nua}.
For the Janus configurations, our holographic renormalisation allowed us to clearly identify sources and expectation values of operators
viewing the interface as describing $\mathcal{N}=4$ SYM on flat spacetime with spatially modulated mass sources or 
$\mathcal{N}=4$ SYM on $AdS_4$ spacetime with constant mass sources.

We showed that the deformed $\mathcal{N}=4$ SYM theory has a conformal anomaly that includes terms that are quadratic and quartic in the scalar source terms similar to
\cite{Petkou:1999fv,Bianchi:2001kw}. For Janus solutions we showed that while the sources for the scalar operators on either side of the interface transform covariantly with respect to Weyl transformations, the expectation values for the corresponding operators do not. In particular, the expectation values of the operators for interfaces of 
$\mathcal{N}=4$ SYM on flat spacetime contained novel terms logarithmic in the coordinate transverse to the interface as well as the
usual terms expected from conformal invariance. 
 
In this paper we have focussed on constructing Janus solutions of supergravity, with $d=3$ conformal invariance. However, it would be interesting to further study the more general class of solutions
that just preserve $ISO(1,2)$ symmetry. What would be most desirable is if the BPS equations can be suitably manipulated 
to give a simpler system set of equations, as was seen for the analogous constructions of \cite{Arav:2018njv} in $D=4$.

In section 6, we discussed various explicit Janus solutions of $\mathcal{N}=4$ SYM for the $\mathcal{N}=2^*$ theory as well
as one-mass and equal mass models. For all cases, our constructions also revealed solutions that approach the
$\mathcal{N}=4$ SYM $AdS_5$ as $r\to\infty$ (or $r\to -\infty$ in some cases) and then become singular at some finite value of $r$. As such, these solutions have a conformal
boundary dual to $\mathcal{N}=4$ SYM with mass deformations on a half space that ends at a singularity.
It would be interesting to examine these solutions in more detail, including elucidating the precise nature of the
singularities in type IIB supergravity, and see if they can be interpreted as BCFTs, as suggested in \cite{Gutperle:2012hy}. Perhaps they can also be interpreted as a kind of RG flow for 
$\mathcal{N}=4$ SYM on $AdS_4$. It seems even more challenging to find any physical interpretation for the singular solutions that do not have any
conformal boundary.

For the one mass model we also found some interesting special solutions which involve the two $LS^\pm$ $AdS_5$ fixed points that this model admits, each dual to the LS SCFT. We found examples of both RG interface solutions, with $\mathcal{N}=4$ SYM on one side of the interface, and the LS SCFT on the other, as well as a novel $LS^+/LS^-$ Janus solution dual to a novel conformal interface of the LS SCFT. Both of these are further discussed 
in \cite{1807495}. 

The equal mass model also revealed solutions with interesting new features. In this model we constructed a class of $\mathcal{N}=4$ SYM Janus solutions that develop a periodic structure in the bulk radial coordinate, and in the critical limit we find solutions which are exactly periodic. After compactifying the radial direction, we obtained a new supersymmetric $AdS_4\times S^1$ solution that uplifts to a new $AdS_4\times S^1\times S^5$ solution of type IIB supergravity, which will be further explored in \cite{toappear3}. 
This solution is somewhat reminiscent of the interesting $AdS_4\times S^1$ solutions in \cite{Bobev:2020fon}. An important difference, however, is that while our new solutions are simply periodic in the $S^1$ direction, the solutions of \cite{Bobev:2020fon} have non-trivial $SL(2,\mathbb{Z})$ monodromy. 
One might anticipate that there are many more Janus solutions that can be constructed in gauged supergravity that have the axion and dilaton activated as well as the mass sources that we have focussed on. It seems likely that this will also lead to a host of new $AdS_4\times S^1$ solutions
for which there is non-trivial $SL(2,\mathbb{Z})$ monodromy along the $S^1$ direction, as in the solutions of \cite{Bobev:2020fon}. 
Indeed we have already constructed some specific examples that will be reported in \cite{toappear3}.

\subsection*{Acknowledgments}
We thank Nikolay Bobev for discussions.
This work was supported by STFC grant ST/P000762/1 and with support from the European Research Council under the European Union's Seventh Framework Programme (FP7/2007-2013), ERC Grant agreement ADG 339140. 
KCMC is supported by an Imperial College President's PhD Scholarship. 
JPG is supported as a KIAS Scholar and as a Visiting Fellow at the Perimeter Institute. 
The work of CR is funded by a Beatriu de Pin\'os
Fellowship.

\appendix

\section{Derivation of the BPS equations with $ISO(1,2)$ symmetry}\label{deriveio21bps}

To discuss the supersymmetry, we will use the conventions of \cite{Gunaydin:1985cu,Bobev:2016nua}.
The $D=5$ gamma matrices obey $\{\gamma_m, \gamma_n\} = 2 \eta_{mn} = 2 \text{diag}\, \{1, -1, -1, -1, -1\}$, and we take $\gamma_0$,
$\gamma_1$, $\gamma_2$, $\gamma_3$ to be imaginary and $\gamma_4$ to be real. We also take $\gamma^{01234}=-1$.

Consider the ansatz
\begin{align}
ds^2 = e^{2A}(dt^2 - dy_1^2 - dy_2^2) - e^{2V} dx^2 - N^2 dr^2\,,
\end{align}
with $A,V,N$ and the scalars $z^A$, $\beta_1$, $\beta_2$, functions of $(x,r)$ only.
We use the orthonormal frame
$(e^{0},e^{1},e^{2},e^{3},e^{4})= 
(e^A dt, e^A dy_1, e^A dy_2, e^V dx, Ndr)$.
We assume that the Killing spinor is independent of $t,y_1,y_2$ and begin by imposing the 
projection
\begin{align}\label{basicproj}
\gamma^{34}\epsilon_1=-i\kappa\epsilon_1\,,
\end{align}
and hence $\gamma^{012}\epsilon_1=-i\kappa\epsilon_1$.
Using the Majorana condition $\epsilon_2=-i\gamma^4\epsilon_1^*$,  we also have
$\gamma^{34}\epsilon_2=i\kappa\epsilon_2$ and $\gamma^{012}\epsilon_2=i\kappa\epsilon_2$. 

From the $t,y_1,y_2$ components of the gravitino equations we get
\begin{align}\label{gravtxy}
\left(-e^{-V}\partial_x A-i\kappa N^{-1}\partial_r A\right)\epsilon_1=\frac{1}{3}e^{\mathcal{K}/2}\bar{\mathcal{W}}\gamma^3\epsilon_2\,,
\end{align}
while from the $x,r$ components we get, respectively,
\begin{align}\label{gravtxr}
e^{-V}\partial_x\epsilon_1+\frac{i\kappa}{2}N^{-1}\partial_r V \epsilon_1+e^{-V}\mathcal{A}_x\epsilon_1+\frac{1}{6}e^{\mathcal{K}/2}\bar{\mathcal{W}}\gamma^3\epsilon_2&=0\,,\nn
N^{-1}\partial_r\epsilon_1-\frac{i\kappa}{2}N^{-1}e^{-V}\partial_x N \epsilon_1+N^{-1}\mathcal{A}_r\epsilon_1+\frac{1}{6}e^{\mathcal{K}/2}\bar{\mathcal{W}}\gamma^4\epsilon_2&=0\,.
\end{align}
Taking the complex conjugate of \eqref{gravtxy} and using the Majorana condition we deduce that
\begin{align}
|(\frac{1}{3}e^{\mathcal{K}/2}\bar{\mathcal{W}})^{-1}(e^{-V}\partial_xA+i\kappa N^{-1}\partial_r A)|=1\,.
\end{align}
We therefore introduce a phase $\xi(x,r)$ via
\begin{align}\label{intphase}
e^{-V}\partial_xA=-\frac{\kappa}{3}e^{\mathcal{K}/2}\text{Im}(e^{-i\xi}\bar{\mathcal{W}})\,,\nn
N^{-1}\partial_rA=\frac{1}{3}e^{\mathcal{K}/2}\text{Re}(ie^{-i\xi}\bar{\mathcal{W}})\,,
\end{align}
and solve \eqref{gravtxy} by imposing the projection
\begin{align}
\gamma^3\epsilon_2=-i\kappa e^{-i\xi}\epsilon_1\,.
\end{align}
We also note that \eqref{intphase} implies the integrability condition
\begin{align}
-\partial_r\left[\kappa e^Ve^{\mathcal{K}/2}\text{Im}(e^{-i\xi}\bar{\mathcal{W}})\right]
=\partial_x\left[  Ne^{\mathcal{K}/2}\text{Re}(e^{-i\xi}\bar{\mathcal{W}})      \right]\,.
\end{align}
We can now rewrite \eqref{gravtxr} in the form
\begin{align}
e^{i\xi/2}e^{A/2}\partial_x\left(e^{-A/2}e^{-i\xi/2}\epsilon_1\right)&=
\frac{ie^{V}}{2}\left[
-e^{-V}\partial_x\xi-\kappa N^{-1}\partial_r V+2ie^{-V}\mathcal{A}_x
+\frac{\kappa}{3}e^{\mathcal{K}/2}\text{Re}(e^{-i\xi}\bar{\mathcal{W}})  \right]\epsilon_1\,,\nn
e^{i\xi/2}e^{A/2}\partial_r\left(e^{-A/2}e^{-i\xi/2}\epsilon_1\right)&=
\frac{iN}{2}\left[
-N^{-1}\partial_r\xi+\kappa N^{-1}e^{-V}\partial_x N+2iN^{-1}\mathcal{A}_r+\frac{1}{3}
e^{\mathcal{K}/2}\text{Im}(e^{-i\xi}\bar{\mathcal{W}})  \right]\epsilon_1\,,\nn
\end{align}
By taking the complex conjugate of these two equations and using
$\epsilon_1^*=i e^{-i\xi}\epsilon_1$, we deduce that in each expression, the
left and right hand sides each separately vanish.
We thus conclude that the Killing spinor takes the form
\begin{align}\label{fundepks}
\epsilon_1=e^{A/2}e^{i\xi/2}\eta_0,\qquad
\epsilon_2=i\kappa e^{-i\xi}\gamma^3\epsilon_1\,,
\end{align}
where $\eta_0$ is a constant spinor satisfying $\gamma^{012}\eta_0=-i\kappa\eta_0$.

The combined system of BPS equations are thus given by
\begin{align}\label{bpsderxi1appa}
e^{-V}\partial_x A+i\kappa N^{-1}\partial_r A-\frac{i\kappa}{3}e^{\mathcal{K}/2}e^{-i\xi}\bar{\mathcal{W}}&=0\,,\nn
-e^{-V}\partial_x\xi-\kappa N^{-1}\partial_r V
+2ie^{-V}\mathcal{A}_{x}
+\frac{\kappa}{3}e^{\mathcal{K}/2}\text{Re}(e^{-i\xi}\bar{\mathcal{W}})&=0\,,\nn
-N^{-1}\partial_r\xi+\kappa N^{-1}e^{-V}\partial_x N
+2iN^{-1}\mathcal{A}_{r}
+\frac{1}{3}e^{\mathcal{K}/2}\text{Im}(e^{-i\xi}\bar{\mathcal{W}})&=0\,,
\end{align}
as well as the following equations from the remaining fermion variations
\begin{align}\label{fermvsremain}
i\kappa e^{i\xi}\left(e^{-V}\partial_x +i\kappa N^{-1}\partial_r \right)z^A&=\frac{1}{2}e^{\mathcal{K}/2}\mathcal{K}^{\bar B A}\nabla_{\bar B}\bar{\mathcal{W}}\,,\nn
i\kappa e^{i\xi}\left(e^{-V}\partial_x +i\kappa N^{-1}\partial_r \right)\beta_1&=\frac{1}{12}e^{\mathcal{K}/2}
\partial_{\beta_1}\bar{\mathcal{W}}\,,\nn
i\kappa e^{i\xi}\left(e^{-V}\partial_x +i\kappa N^{-1}\partial_r \right)\beta_2&=\frac{1}{4}e^{\mathcal{K}/2}
\partial_{\beta_2}\bar{\mathcal{W}}\,.
\end{align}
We note that these equations are not all independent.
We also observe that these equations are invariant under 
\begin{align}
r\to-r\,,\qquad  x\to -x\,,\qquad \xi\to\xi+\pi\,.
\end{align}

Next we can rewrite them in a simplified manner if we choose the gauge $N=e^V$. We can define the complex coordinate
$w=r-i\kappa x$ so that $\bar\partial=d\bar w\frac{1}{2}(\partial_r-i\kappa \partial_x)$ and also the $(1,0)$ form $B$ as
\begin{align}\label{ay13}
B=\frac{1}{6}e^{i\xi+V+\mathcal{K}/2}{\mathcal{W}}d w\,.
\end{align}
The equations
\eqref{bpsderxi1appa} can be cast in the form
\begin{align}\label{parelbee}
\partial A&=B\,,\nn
\bar \partial B&= -\mathcal{F}B\wedge \bar B\,,
\end{align}
where
\begin{equation}
\mathcal{F} \equiv 1-\frac{1}{|\mathcal{W}|^2}[\frac{3}{2}\nabla_A\mathcal{W}\mathcal{K}^{A\bar B}\nabla_{\bar B}\bar{\mathcal{W}}
+\frac{1}{4}|\partial_{\beta_1}\mathcal{W}|^2
+\frac{3}{4}|\partial_{\beta_2}\mathcal{W}|^2]\,,
\end{equation}
while \eqref{fermvsremain} become
\begin{align}\label{fermvsremaingaugef}
\bar\partial z^A&=-\frac{3}{2}(\bar{\mathcal{W}})^{-1}\mathcal{K}^{\bar B A}\nabla_{\bar B}\bar{\mathcal{W}}\bar B\,,\nn
\bar\partial \beta_1&=-\frac{1}{4}(\bar{\mathcal{W}})^{-1}\partial_{\beta_1}\bar{\mathcal{W}}\bar B\,,\nn
\bar\partial \beta_2&=-\frac{3}{4}(\bar{\mathcal{W}})^{-1}\partial_{\beta_2}\bar{\mathcal{W}}\bar B\,.
\end{align}

After fixing the gauge-freedom, by fixing $N$,
the BPS equations \eqref{bpsderxi1appa} and \eqref{fermvsremain} are a set of 16 real equations for 13 real functions, 
$A,V,\xi,z^A,\beta_1,\beta_2$ with $A=1,\dots,4$, in the ten scalar truncation,
and therefore naively seem to be over constrained. To analyse
the consistency of these equations it is convenient to work in the gauge $N=e^V$, 
and analyse \eqref{parelbee}, \eqref{fermvsremaingaugef}. 
We first observe that $\mathcal{F}$ is a manifestly real quantity that 
depends only on the scalar fields.
The BPS equations are constrained due to the fact that $A$, $\beta_1$ and $\beta_2$ are all real. If one takes the holomorphic exterior
derivative $\bar \partial$ of the equations for these functions in \eqref{parelbee},\eqref{fermvsremaingaugef} one obtains
necessary conditions for the equations to be satisfied. For $A$ this condition is given by: 
\begin{equation}
\operatorname{Re}(\bar\partial B) = 0 . 
\end{equation}
which is automatically satisfied from \eqref{parelbee}. We are therefore left with two constraints to check.

To do so, it is useful to first prove the following result.
Consider any function
$\overline{\mathcal{G}}(\bar{z}^A,\beta_1,\beta_2)$ which depends only on the scalar fields and is anti-holomorphic in the four complex scalars $z^A$. Using the BPS equations \eqref{parelbee},\eqref{fermvsremaingaugef}
we deduce
\begin{equation}\label{eq:BPSAntiHolIdent}
\partial (\overline{\mathcal{G}} \bar{B}) = (\hat{\mathcal{O}}\overline{\mathcal{G}}) \, B \wedge \bar{B},
\end{equation}
where $\hat{\mathcal{O}}$ is a differential operator on the scalar manifold defined as
\begin{equation}
\hat{\mathcal{O}}\overline{\mathcal{G}} \equiv
\mathcal{F}\overline{\mathcal{G}}
-\frac{3}{2} \mathcal{K}^{\bar{A}B} \frac{\nabla_B \mathcal{W}}{\mathcal{W}} \partial_{\bar{A}} \overline{\mathcal{G}} 
-\frac{1}{4} \partial_{\beta_1}\log\mathcal{W}\partial_{\beta_1} \overline{\mathcal{G}} - \frac{3}{4} \partial_{\beta_2}\log\mathcal{W}\partial_{\beta_2} \overline{\mathcal{G}} .
\end{equation}
Then, taking the $\partial$ derivative of the last two equations in \eqref{fermvsremaingaugef}, we
obtain the following necessary conditions for these set of equations to be consistent with $\beta_i$ being real:
\begin{equation}
\label{eq:BPSConsistencyConditions}
\operatorname{Im}\left( \hat{\mathcal{O}} \partial_{\beta_i} \log \overline{\mathcal{W}} \right) = 0\,, \qquad (i=1,2).
\end{equation}
Notice that these conditions do not involve $B$, just the scalar fields, and hence they are conditions on $\mathcal{K}$ and $\mathcal{W}$.
One can explicitly check that these conditions are satisfied for \eqref{kpot} and \eqref{superpotlike} in the ten scalar model.
We expect that these conditions are sufficient conditions for consistency; while we have not proven this in general, we did for the sub-class of Janus solutions as discussed below \eqref{eq:BPSConsistencyConditionstext}.
It would be interesting if there is a way to understand these consistency conditions more directly from the underlying
$\mathcal{N}=2$ supergravity theory.

\section{Holographic Renormalisation}\label{appb}
In this appendix we provide some details on the holographic renormalisation that we use and, in particular,  
give expressions for various one point functions. We will focus on
configurations that preserve $ISO(1,2)$ symmetry, first considering general configurations before restricting to BPS configurations.
In appendix \ref{appc} we will specialise to those that, in addition, preserve conformal symmetry.

Holographic renormalisation relevant for mass deformed Euclidean $\mathcal{N}=4$ SYM theory 
was discussed in \cite{Bobev:2016nua}, and there is some overlap with our analysis below.  
In particular, finite counterterms that are consistent with the global symmetries of the four-sphere were analysed in some detail. While the analysis
in \cite{Bobev:2016nua} was sufficient in order to be able to calculate the universal part of the free energy, the observable of principle objective in that paper, it is not sufficient to calculate other observables. Our analysis will include other finite counterterms which appear in observables that we consider for our solutions. We also note in advance that while we have extended the results of \cite{Bobev:2016nua} in a manner that is sufficient for our purposes, additional work is still required in order to have a complete holographic renormalisation scheme that is consistent with $\mathcal{N}=4$ supersymmetry.

\subsection{General case}
We consider the class of solutions that are general enough to describe
sources which depend on one of the spatial directions and preserve $ISO(1,2)$ symmetry. 
Specifically, we consider metrics of the form
\begin{align}\label{iso21metansapp}
ds^2 &= e^{2A(r,x)}(dt^2 - dy_1^2 - dy_2^2) - e^{2V(r,x)} dx^2 - dr^2\,,\nn
&\equiv 
\gamma_{ab}(r,x)dx^adx^b-dr^2\,,
\end{align}
with all scalar fields functions of $(r,x)$ only.
The conformal boundary is located at
$r\to\infty$ and there we have the expansion
\begin{align}\label{defh}
\gamma_{ab}&=e^{2r/L}h_{ab}(x)+\dots\,,
\end{align}
where $h_{ab}(x)$ is the metric for the spacetime where the dual field theory lives, which we write as
\begin{align}\label{defhexpl}
h_{ab}(x)dx^a dx^b= e^{2\Omega(x)}\left(dt^2 - dy_1^2 - dy_2^2 - e^{2f(x)} dx^2\right)\,,
\end{align}
where the function $f(x)$ is included for convenience (it can be useful in utilising different gauge choices in numerically solving the equations). 
Two cases of particular interest are firstly,
when $\Omega(x)$ is constant, associated with a flat boundary metric. Secondly, when $e^\Omega=\ell/x$ and $f(x)=$ constant, associated
with an $AdS_4$ boundary metric, with radius $\ell$ (more precisely, this gives a component of the boundary for the Janus solutions
as we elaborate further in appendix \ref{appc}).

The full action can be written as the sum of four terms:
\begin{align}
S=S_{Bulk}+S_{GH}+S_{ct}+S_{finite}\,.
\end{align}
The first two terms are the bulk action and the boundary Gibbons-Hawking term, given by
\begin{align}
S_{Bulk}+S_{GH}=\frac{1}{4\pi G}\int d^5 x\sqrt{|g|}\mathcal{L}-\frac{1}{8\pi G}\int d^4x\, \sqrt{|\gamma|}K\,,
\end{align}
where the the bulk Lagrangian $\mathcal{L}$ for the ten scalar model is given in \eqref{bulklag}
and the trace of extrinsic curvature for the outward pointing normal one-form $n=dr$ is given by
$K=-\frac{1}{2}\gamma^{ab}\partial_r \gamma_{ab}$. 
As in \cite{Bobev:2016nua} we also have 
\begin{align}
16 \pi G=\frac{8 \pi^2 L^3}{N^2}\,,
\end{align}
with the $AdS_5$ vacuum solution, with vanishing scalar fields, dual to $SU(N)$ $\mathcal{N}=4$ SYM theory.
The boundary action $S_{ct}$ that is required to remove divergences takes the form
\begin{align}\label{ctact}
S_{{ct}}=\frac{1}{16\pi G}&\int d^4x\, \sqrt{|\gamma|}\Bigg\{
-\frac{6}{L}+\frac{L}{2}R-L(\nabla\varphi)^2-\frac{2}{L}\sum_{i=1}^4(\phi_i)^2
\nn
&-\frac{4}{L}\Big(1-\frac{L}{2r}\Big)\Big(6(\beta_1)^2+2(\beta_2)^2+\sum_{k=1}^3(\alpha_k)^2\Big)\nn
&-\frac{r}{L}\Big[\frac{L^3}{4}(R_{ab}R^{ab}-\frac{1}{3}R^2)+\frac{L}{3}R\sum_{i=1}^4(\phi_i)^2\nn
&\qquad-\frac{16}{3L}\sum_{i=1}^4(\phi_i)^4+\frac{16}{3L}\sum_{1\leq i<j\leq 4}^4(\phi_i)^2(\phi_j)^2
+2L\sum_{i=1}^4(\nabla \phi_i)^2\Big]\nn
&-\frac{r}{L}\Big[
\frac{L^3}{3}R(\nabla\varphi)^2+\frac{2L^3}{3}(\nabla\varphi)^4-L^3R^{ab}\partial_a\varphi\partial_b\varphi\nn
&\qquad\qquad\qquad\qquad\qquad+\frac{L^3}{2}(\Box\varphi)^2+\frac{4L}{3}\sum_{i=1}^4(\phi_i)^2(\nabla\varphi)^2
\Big]\Bigg\}\,,
\end{align}
where all quantities are evaluated with respect to $\gamma_{ab}$ evaluated in the limit $r\to \infty$.
Finally, the finite counterterms that we shall consider are given by 
\begin{align}\label{finitectact}
S_{{finite}}=&\frac{1}{16\pi G}\int d^4x\, \sqrt{|\gamma|}\Big\{
-\delta_{R^2}\frac{L^3}{4}\Big(R_{ab}R^{ab}-\frac{1}{3}R^2\Big)-\delta_{\Delta R^2}\frac{L^3}{4}\Big(R_{ab}R^{ab}+\frac{1}{3}R^2\Big)\nn
&-\delta_{R\phi^2(1)}\frac{L}{3}R\sum_{i=1}^3(\phi_i)^2-\delta_{R\phi^2(2)}\frac{L}{3}R(\phi_4)^2+\delta_{4 (1)}\frac{16}{3L}\sum_{i=1}^3(\phi_i)^4+\delta_{4 (2)}\frac{16}{3L}(\phi_4)^4\nn
&-\delta_{4 (3)}\frac{16}{3L}\sum_{1\leq i<j\leq 3}^3(\phi_i)^2(\phi_j)^2-\delta_{4 (4)}\frac{16}{3L}\sum_{i=1}^3(\phi_i)^2(\phi_4)^2+\delta_{4(5)}\frac{16}{3L}\phi_1\phi_2\phi_3\phi_4\nn
&-\delta_\alpha\frac{4}{L}\Big(\frac{L}{r}\Big)^2\sum_{k=1}^3(\alpha_k)^2-\delta_{\beta}\frac{4}{L}\Big(\frac{L}{r}\Big)^2\Big(6(\beta_1)^2+2(\beta_2)^2\Big)\nn
&-\delta_{\partial\phi^2(1)}2L\sum_{i=1}^3(\nabla\phi_{i})^2-\delta_{\partial\phi^2(2)}2L(\nabla\phi_{4})^2
\nn
&+\delta_{\tilde{\beta}}\Big[24L\Big(\frac{\beta_1}{r}-\frac{1}{3L}\left[(\phi_1)^2+(\phi_2)^2-2(\phi_3)^2\right]\Big)^2+8L\Big(\frac{\beta_2}{r}-\frac{1}{L}\left[(\phi_1)^2-(\phi_2)^2\right]\Big)^2\Big]\Big\}
\end{align}
which depends on 14 coefficients $\{\delta_{R^2},\delta_{\Delta R^2},\dots,\}$, which we take to be constants, and again we have utilised
the boundary metric $\gamma_{ab}$ evaluated in the limit $r\to \infty$. 

There are a number of comments concerning this choice of finite counterterms, which defines a renormalisation scheme.
We first note that
we have not included a Riemann squared term, $R_{abcd}R^{abcd}$, since we are only considering conformally flat backgrounds
as in \eqref{defhexpl} and hence they can be expressed in terms of $R_{ab}R^{ab}$ and $R^2$.
We next note 
$S_{{finite}}$ respects the discrete symmetries \eqref{genz2}-\eqref{gens33} of the $D=5$ theory. 
We now recall that the scalar field $\varphi$ is dual to a marginal operator in $\mathcal{N}=4$ SYM theory, as in \eqref{opfieldmapz}, and its boundary value can be identified with changing the coupling constant of the theory. Being a source for a marginal operator there are many
additional finite counterterms that one might consider, including allowing the $\delta$ coefficients appearing in $S_{{finite}}$ to be functions
of $\varphi$ as well as including terms with derivatives of $\varphi$. These additional counterterms could be significantly simplified if we impose that they respect the 
shift symmetry $\varphi\to \varphi+const.$ of the bulk $D=5$ gravitational theory, but this is not a natural scheme to consider.

However, for the purpose of this paper, where will assume that there are no sources active for $\varphi$, as we make precise below, and the fact that we will only be calculating one point functions, $S_{{finite}}$ is in fact general enough, with one extra assumption.
In particular, if any of the $\delta$'s did depend on $\varphi$ it would only be the
terms linear in $\varphi$ that could affect the one-point functions, and we exclude such terms using the symmetry  \eqref{genz2}.
In fact this is not quite true: we could still consider terms of the schematic form $\varphi\alpha\phi^2$. Taking into 
account \eqref{gens31}-\eqref{gens33}, we could have terms $\varphi(\alpha_1\phi_2\phi_3+\alpha_2\phi_3\phi_1+\alpha_3\phi_1\phi_2)$
and $\varphi(\alpha_1\phi_1+\alpha_2\phi_2+\alpha_3\phi_3)\phi_4$ and so we
exclude\footnote{If they were included they would only affect the expectation value of $\mathcal{O}_\varphi$ for the equal mass model.} these
by making the extra assumption that the finite counterterms respect $\varphi\to-\varphi$.
Demanding that this renormalisation scheme is supersymmetric also places certain restrictions on $\{\delta_{R^2},...,\}$ some of which we will 
obtain below.

Before continuing, we highlight that $\varphi\to-\varphi$ and 
\eqref{genz2} are not symmetries of the perturbative field theory since they involve a $\mathbb{Z}_2\subset SL(2,\mathbb{Z})$ duality transformation\footnote{Note that the field theory $SL(2,\mathbb{Z})$ acts in type IIB on the ten-dimensional axion and dilaton.
The transformation on the $D=5$ fields is discussed in appendix C of \cite{Bobev:2016nua} and is rather involved; all we need here is the
fact that near the boundary the $\mathbb{Z}_2$ symmetry acts on the sources as $\varphi\to-\varphi$.}.
In other words we are invoking this non-perturbative symmetry as part of our scheme. It is clear that, more generally, one might invoke invariance under
the full $SL(2,\mathbb{Z})$ as a starting principle and this will impose restrictions on the various $\delta$'s. In fact this point of view was considered
in section 3 of \cite{Bobev:2016nua}, from a field theory perspective, in the specific case of $\mathcal{N}=4$ SYM on the $S^4$; it would be interesting to extend that analysis to the present setup of spatially modulated masses. We also note that we have included finite counter
terms in \eqref{finitectact} that were not needed for the holographic analysis of \cite{Bobev:2016nua}. These include some
with spatial dependence (e.g. $\delta_{\partial\phi^2(1)}$) as well as $\delta_\alpha$, $\delta_\beta$ which were not needed
in the calculation of the universal part of the free energy on the four-sphere (see also footnote \ref{alphsbets} below).

Using the bulk equations of motion, we can develop the following, schematic,
asymptotic expansion as $r\to \infty$: 
\begin{align}\label{asexp10sc}
A&=\frac{r}{L}+\Omega+\cdots+{A}_{(v)}e^{-4r/L}+\cdots\,,\nn
V&=\frac{r}{L}+\Omega+f+\cdots+{V}_{(v)}e^{-4r/L}+\cdots\,,\nn
\phi_i&=\phi_{i,(s)}e^{-r/L}+\dots+{\phi}_{i,(v)}e^{-3r/L}+\cdots\,,\qquad i=1,\dots,4\,,\nn
\alpha_i&=\alpha_{i,(s)}\frac{r}{L}e^{-2r/L}+{\alpha}_{i,(v)}e^{-2r/L}+\cdots\,,,\qquad i=1,\dots,3\,,\nn
\beta_i&=\beta_{i,(s)}\frac{r}{L}e^{-2r/L}+{\beta}_{i,(v)}e^{-2r/L}+\cdots\,,\qquad i=1,\dots,2\,,\nn
\varphi&=\varphi_{(s)}+\cdots+{\varphi}_{(v)}e^{-4r/L}+\cdots\,,
\end{align}
where all coefficients, except for $\varphi_{(s)}$, can depend on the coordinate $x$.
In this expansion 
$\phi_{i,(s)}$, $\alpha_{i,(s)}$, $\beta_{i,(s)}$ and $\varphi_{(s)}$ provide sources for the corresponding dual
operators in $\mathcal{N}=4$ SYM given in \eqref{opfieldmapz}. Our interest in this paper is spatially dependent mass deformations
and hence we allow $\phi_{i,(s)}$, $\alpha_{i,(s)}$, $\beta_{i,(s)}$ to depend on $x$ but we take
\begin{align}
\varphi_{(s)}=0\,.
\end{align}
Note however, that in general ${\varphi}_{(v)}$ does depend on $x$ 
and is related to the operator dual to $\varphi$ acquiring a spatially dependent expectation value. We also note that in developing the asymptotic expansion, there is
one algebraic and one differential constraint relating ${A}_{(v)}$ and ${V}_{(v)}$ which ensure the
Ward identities in the boundary theory, given below, are satisfied.

The expectation value for the stress tensor is given by
\begin{align}\label{eq:stress_tensor_definition}
\langle\mathcal{T}^{ab}\rangle=\lim_{r\rightarrow\infty}\Big\{e^{6r/L}\frac{-2}{\sqrt{|\gamma|}}\frac{\delta S}{\delta \gamma_{ab}}\Big\}\,.
\end{align}
The expectation value of the operators dual to the scalar fields are given by
\begin{align}
\langle\mathcal{O}_{\Psi}\rangle=\lim_{r\rightarrow\infty}\Big\{e^{\Delta_{\Psi}r/L}\frac{1}{\sqrt{|\gamma|}}\frac{\delta S}{\delta \Psi}\Big\}\quad \text{or}\quad \lim_{r\rightarrow\infty}\Big\{\Big(\frac{r}{L}\Big)e^{\Delta_{\Psi}r/L}\frac{1}{\sqrt{|\gamma|}}\frac{\delta S}{\delta \Psi}\Big\}\,,
\end{align}
where the former expression is for $\varphi,\phi_i,\phi_4$, dual to operators with $\Delta=4,3,3$
and the latter for $\alpha_i,\beta_i$, dual to operators with $\Delta=2$. We will not present these expressions
for general finite counterterms as the expressions are lengthy. Instead we just note that we have checked
that the following Ward identity is satisfied
\begin{align}
\nabla_{a}\langle\mathcal{T}^{a}{}_{b}\rangle+\sum_{i=1}^4\langle\mathcal{O}_{\phi_i}\rangle\,\partial_b\phi_{i,(s)}+\sum_{i=1}^3\langle\mathcal{O}_{\alpha_i}\rangle\,\partial_b\alpha_{i,(s)}+\sum_{i=1}^2\langle\mathcal{O}_{\beta_i}\rangle\,\partial_b\beta_{i,(s)}=0\,,
\end{align}
where here the covariant derivative is defined with respect to the field theory metric $h_{ab}$ in \eqref{defh},\eqref{defhexpl} and this
metric has been used to raise the index on $\langle\mathcal{T}^{a}{}_{b}\rangle$. We also
recall here that we have assumed that the source $\varphi_{(s)}$ vanishes.

Furthermore, the trace of the stress tensor can be expressed as
 \begin{align}\label{tracetee}
\langle\mathcal{T}^{a}{}_{a}\rangle+\sum_{i=1}^4\langle\mathcal{O}_{\phi_i}\rangle\,\phi_{i,(s)}+2\sum_{i=1}^3\langle\mathcal{O}_{\alpha_i}\rangle\,\alpha_{i,(s)}+2\sum_{i=1}^2\langle\mathcal{O}_{\beta_i}\rangle\,\beta_{i,(s)}=\mathcal{A}\end{align}
where $\mathcal{A}$ is given by 
 \begin{align}\label{traceteeconfanom}
8\pi GL\mathcal{A}
&=-\frac{L^4}{8}\big(R_{ab}R_{}^{ab}-\frac{1}{3}R^2\big)-\delta_{\Delta R^2}{L^4}\Box R\nn
&-\sum_{i=1}^3\alpha_{i,(s)}^2-6\beta_{1,(s)}^2-2\beta_{2,(s)}^2+\frac{8}{3 }\sum_{i=1}^4\phi_{i,(s)}^4-\frac{8}{3}\sum_{1\leq i<j\leq 4}^4\phi_{i,(s)}^2\phi_{j,(s)}^2\nn
&-{L^2}\sum_{i=1}^4\Big[(\nabla\phi_{i,(s)})^2   +\frac{1}{6}R  \phi_{i,(s)}^2   \Big]\nn
&\phantom{=}
+{2L^2(\delta_{\partial\phi^2(1)}-\delta_{R\phi^2(1)})}\sum_{i=1}^3
\nabla(\phi_{i,(s)}\nabla\phi_{i,(s)})  \nn
&\phantom{=}
+{2L^2(\delta_{\partial\phi^2(2)}-\delta_{R\phi^2(2)})}
\left[\nabla(\phi_{4,(s)}\nabla\phi_{4,(s)}) \right]\,,
\end{align}
where, again, the geometric quantities are written here with respect to the field theory metric $h_{ab}$ in 
\eqref{defh}, \eqref{defhexpl}.
Here $\mathcal{A}$ is the conformal anomaly for $\mathcal{N}=4$ SYM on a curved $ISO(1,2)$ invariant boundary
in the presence of spatially dependent sources.
The first line of the conformal anomaly is the standard term involving the Ricci tensor along with a familiar contribution coming from the finite counterterm parametrised
by $\delta_{\Delta R^2}$. The remaining contributions are terms involving the sources for the scalar operators \cite{Petkou:1999fv,Bianchi:2001kw}.
The integrated anomaly should be invariant under Weyl transformations and we can see that this is true after recalling that $\alpha_{i(s)}$, $\beta_{i(s)}$ have scaling dimension two, $\phi_{i(s)}$, $\phi_{4(s)}$ have dimension one as well
as the expression for a conformally coupled scalar in four spacetime dimensions appearing in the third line. 
The presence of these source terms crucially relies
on the fact that they are sourcing operators with integer conformal dimensions and hence are not be present for generic CFTs.

It is illuminating to see how the source and expectations values change under a class of Weyl transformations of the boundary metric \eqref{defhexpl}.
Specifically, we consider the transformation
\begin{align}\label{weylmet}
h_{ab}\to \Lambda^{2}h_{ab}\,,
\end{align}
with $\Lambda=e^{-\Omega}$, which takes the boundary metric \eqref{defhexpl} (with $f(x)=0$) to a flat space metric.
As $r\rightarrow \infty$, we can achieve this by implementing the following schematic coordinate transformation:
\begin{align}
x&\to x-\frac{L^2}{2}(\partial_{x}\Omega) e^{-2r/L}+\cdots\,,\nn
e^{r/L}&\to e^{-\Omega}e^{r/L}+\frac{L^2}{4}e^{-\Omega}(\partial_{x}\Omega)^2e^{-r/L}+\cdots\,.
\end{align}
We can quickly conclude that the source terms all transform covariantly, as one expects:
\begin{align}
\alpha_{i,(s)}\to   \Lambda^{-2}\alpha_{i,(s)},\quad\beta_{i,(s)}\to   \Lambda^{-2}\beta_{i,(s)},\quad\phi_{i,(s)}\to   \Lambda^{-1}\phi_{i,(s)},\quad \varphi_{(s)}\to \varphi_{(s)}\,,
\end{align}
(though we will set $\varphi_{(s)}=0$). The transformation on the $``(v)"$ expansion coefficients in
\eqref{asexp10sc}
is more elaborate and this leads to the following non-covariant transformation properties for the associated
expectation values. For the $\Delta=2$ operators we find 
\begin{align}\label{dim2weylt}
\langle\mathcal{O}_{\alpha_{i}}\rangle
\to &\,\Lambda^{-2}\langle\mathcal{O}_{\alpha_{i}}\rangle+\frac{1}{4\pi GL}\alpha_{1,(s)}\Lambda^{-2}\log\Lambda\,,\nn
\langle\mathcal{O}_{\beta_1}\rangle
\to &\,\Lambda^{-2}\langle\mathcal{O}_{\beta_{1}}\rangle+\frac{3}{2\pi GL}\beta_{1,(s)}\Lambda^{-2}\log\Lambda\,,\nn
\langle\mathcal{O}_{\beta_2}\rangle
\to &\,\Lambda^{-2}\langle\mathcal{O}_{\beta_{2}}\rangle+\frac{1}{2\pi GL}\beta_{2,(s)}\Lambda^{-2}\log\Lambda\,,
\end{align}
for the $\Delta=3$ operators we have
\begin{align}\label{phiiweylt}
\langle\mathcal{O}&_{\phi_{i=1,2,3}}\rangle\to \,\Lambda^{-3}\langle\mathcal{O}_{\phi_{i=1,2,3}}\rangle
+\frac{L}{4\pi G}\Lambda^{-2}\partial_{x}\phi_{i,(s)}\partial_{x}\Lambda
-\frac{L}{8\pi G}\phi_{i,(s)}\Lambda^{-3}(\partial_{x}\Lambda)^2\nn
&+\frac{1}{2\pi GL}\Lambda^{-3}\log\Lambda\Big(-4\phi_{i,(s)}^3+\frac{4}{3}\phi_{i,(s)}\sum_{j=1}^4\phi_{j,(s)}^2
-L^2\Lambda\,\partial_{x}\phi_{i,(s)}\partial_{x}\Lambda\nn
&\phantom{\,-\frac{1}{2\pi GL}e^{3\Omega}\Omega\Bigg(}
+{L^2}\phi_{i,(s)}(\partial_{x}\Lambda)^2+\frac{L^2}{2}\Lambda^{2}\partial_{x}\partial_{x}\phi_{i,(s)}
-\frac{L^2}{2}\Lambda\,\phi_{i,(s)}\partial_{x}\partial_{x}\Lambda\Big)\nn
&\phantom{\,}+(\delta_{R\phi^2(1)}-\delta_{\partial\phi^2(1)})\frac{L}{4\pi G}\phi_{i,(s)}\Big(
2\Lambda^{-3}(\partial_{x}\Lambda)^2-\Lambda^{-2}\partial_{x}\partial_{x}\Lambda\Big)\,,
\end{align}
and $\langle\mathcal{O}_{\phi_{i=4}}\rangle$ transforms as in \eqref{phiiweylt}, but with the coefficient $(\delta_{R\phi^2(1)}-\delta_{\partial\phi^2(1)})$
in the last line replaced with $(\delta_{R\phi^2(2)}-\delta_{\partial\phi^2(2)})$. 
In particular, if $\phi_{4,(s)}=0$, as it is for BPS configurations when $\varphi_{(s)}=0$, then $\langle\mathcal{O}_{\phi_{i=4}}\rangle$
transforms covariantly.
Finally, the $\Delta=4$ operator transforms covariantly when $\varphi_{(s)}=0$:
\begin{align}\label{dim4weylt}
\langle\mathcal{O}_{\varphi}\rangle\to &\,\Lambda^{-4}\langle\mathcal{O}_{\varphi}\rangle\,.
\end{align}
The presence of the $\log\Lambda$ terms appearing in the expressions for $\Delta=2,3$ are a consequence of the conformal anomaly
\eqref{traceteeconfanom}.

\subsection{BPS configurations}
We now restrict to $ISO(1,2)$ configurations which satisfy the BPS equations in 
\eqref{bpsderxi1appatext},\eqref{fermvsremaintext}.
Continuing to assume
that $\varphi_{(s)}=0$ we find the following constraints on the sources
\begin{align}\label{consces}
\varphi_{(s)}&=0\,,\nn
\phi_{4,(s)}&=0\,,\nn
\alpha_{i,(s)}&=\kappa L e^{-\Omega-f}\Big(\partial_x\phi_{i,(s)}+\phi_{i,(s)}\partial_x\Omega\Big)\,,\qquad i=1,\dots,3\,,\nn
\beta_{1,(s)}&=\frac{1}{3}\Big(\phi_{1,(s)}^2+\phi_{2,(s)}^2-2\phi_{3,(s)}^2\Big)\,,\nn
\beta_{2,(s)}&=\phi_{1,(s)}^2-\phi_{2,(s)}^2\,.
\end{align}
In particular, we see that the sources $\alpha_{i,(s)}$, $\beta_{i,(s)}$ are determined by $\phi_{i,(s)}$ with $ i=1,\dots,3$.
We also find an additional set of relations amongst the expansion functions $A_{(v)},
{V}_{(v)}$ and ${\phi}_{i,(v)},{\phi}_{4,(v)}, {\alpha}_{i,(v)},
{\beta}_{i,(v)},{\varphi}_{(v)}$ which provide relations between the expectation values of the various dual 
scalar operators as well as the stress tensor. As they are rather long, we will not record them here, but
we will below for each of the sub-truncations that 
we study.

It is now illuminating to use these results to calculate the energy density for flat field theory metric, $h_{ab}=\eta_{ab}$ (i.e. $\Omega=f=0$ in 
\eqref{defhexpl}). Firstly, we find that stress energy tensor itself takes the form 
\begin{align}
&\pi G\langle\mathcal{T}^{ab}\rangle=\eta^{ab}\Bigg[-\frac{1+4\delta_{4(1)}-8\delta_{\beta}}{12L}\sum_{i=1}^3\phi_{i,(s)}^4-\frac{3-4\delta_{4(3)}+8\delta_{\beta}}{12L}\sum_{1\leq i<j\le 3}^3\phi_{i,(s)}^2\phi_{j,(s)}^2\nn
&\phantom{=g^{ab}_{(0)}\Bigg[}+\sum_{i=1}^3\Big\{-\frac{\kappa}{16}\phi_{i,(s)}\partial_x{\alpha}_{i,(v)}-\frac{\kappa}{8}\partial_x\phi_{i,(s)}{\alpha}_{i,(v)}+\frac{L(\delta_{R\phi^2(1)}-\delta_{\partial\phi^2(1)}+4\delta_{\alpha})}{16}(\partial_x\phi_{i,(s)})^2
\nn
&\phantom{=g^{ab}_{(0)}\Bigg[+\sum_{i=1}^3\Bigg\{}+\delta_{R\phi^2(1)}\frac{L}{16}\phi_{i,(s)}\partial^2_x\phi_{i,(s)}\Big\}\Bigg]\nn
&\phantom{=}+\sigma^{ab}\Bigg[\sum_{i=1}^3\Big\{-\frac{\kappa}{48}\phi_{i,(s)}\partial_x{\alpha}_{i,(v)}+\frac{\kappa}{24}\partial_x\phi_{i,(s)}{\alpha}_{i,(v)}+\frac{L(\delta_{R\phi^2(1)}-3\delta_{\partial\phi^2(1)})}{48}(\partial_x\phi_{i,(s)})^2\nn
&\phantom{=g^{ab}_{(0)}\Bigg[+\sum_{i=1}^3\Bigg\{}+\delta_{R\phi^2(1)}\frac{L}{48}\phi_{i,(s)}\partial^2_x\phi_{i,(s)}\Big\}\Bigg]\,,
\end{align}
where the matrix $\sigma^{ab}=\text{diag}(1,-1,-1,3)$ satisfies $\eta_{ab}\sigma^{ab}=0$ and hence does not contribute to
the conformal anomaly.
Using this, we obtain the following expression for the local energy density for BPS configurations in flat space:
\begin{align}\label{relsces}
&8\pi GL\langle\mathcal{T}^{t}{}_{t}\rangle  
=\sum_{i=1}^3\Big[\frac{2}{3}\partial_{x}\Big(\delta_{R\phi^2(1)}{L^2\phi_{i,(s)}\partial_x\phi_{i,(s)}}-{\kappa L \phi_{i,(s)}{\alpha}_{i,(v)}}\Big)-\frac{2}{3}(1+4\delta_{4(1)}-8\delta_{\beta})\phi_{i,(s)}^4\nn
&\qquad-{L^2(\delta_{\partial\phi^2(1)}-2\delta_{\alpha})}(\partial_x\phi_{i,(s)})^2\Big]-\frac{2}{3}(3-4\delta_{4(3)}+8\delta_{\beta})\sum_{1\leq i<j\le 3}^3\phi_{i,(s)}^2\phi_{j,(s)}^2\,.
\end{align}
For a supersymmetric renormalisation scheme, we demand that the finite counterterms
are such that the right hand side is a total spatial derivative, in order that 
the total energy for spatially modulated supersymmetric sources (with compact support) is exactly zero.
This implies the following conditions must be satisfied
\begin{align}\label{susyscheme}
\delta_{4(1)}&=-\frac{1}{4}+2\delta_{\beta},\nn
\delta_{4(3)}&=\frac{3}{4}+2\delta_{\beta},\nn 
\delta_{\partial\phi^2(1)}&=2\delta_{\alpha}\,.
\end{align}
These conditions are similar those one would get if one used the ``Bogomol'nyi trick" used in \cite{Freedman:2013ryh,Bobev:2013cja,Bobev:2016nua}, but we note that
the analysis of \cite{Bobev:2016nua} did not include the possibility of $\delta_{\alpha}$ and $\delta_{\beta}$.

Since we would like to work with a scheme that preserves supersymmetry we will impose \eqref{susyscheme}.
Notice in the above energy analysis we have set $\varphi_{(s)}=0$ which, for supersymmetric configurations implies 
$\phi_{4,(s)}=0$. It seems likely that if we consider zero energy BPS configurations when these sources are active that we
would be able to constrain\footnote{From the results of \cite{Bobev:2016nua} we anticipate that we would get
$\delta_{4(2)}= -3/4+\dots$,  $\delta_{4(4)}= 3/2+\dots$, $\delta_{4(5)}= 9/2+\dots$, where the dots refer to terms involving
 $\delta_\alpha$ and $\delta_\beta$.} $\delta_{4(2)}$, $\delta_{4(4)}$ and $\delta_{4(5)}$. In order to 
fully determine the
14 coefficients appearing in the finite counterterm action, one would
like to implement a fully supersymmetric holographic renormalisation scheme, along the lines of \cite{Papadimitriou:2017kzw}, including imposing $SL(2,\mathbb{Z})$ invariance, but we leave that for
future work (see also e.g. \cite{Butter:2016mtk}). We will explicitly see that the terms $\delta_\alpha$, $\delta_\beta$, in particular, 
appear\footnote{\label{alphsbets}
Observe that if we substitute the supersymmetry condition \eqref{susyscheme} as well as the BPS conditions on the sources 
\eqref{consces} into the finite counter term action \eqref{finitectact}, then $\delta_\beta$ drops out; this is relevant for evaluating
the free energy of a given configuration, but we reiterate that $\delta_\beta$ does appear in our one point functions. Finally, it would be interesting to make a connection with the $\mathcal{N}=1$ supersymmetric field theory analysis in section 3 of \cite{Bobev:2016nua}. Here we simply note that this would appear to involve the invariant $I_2$ in equation (3.12) of \cite{Bobev:2016nua} as well as an additional counterterm involving background gauge supermultiplets that was not considered (nor needed) in \cite{Bobev:2016nua}.}
in novel contributions to the expectation values
of operators for Janus solutions (e.g. see \eqref{delalphalog}).

\subsubsection{$\mathcal{N}=1^*$ one-mass model}\label{appb11}
This model is obtained from the 10-scalar model by setting $\phi_1=\phi_2=0$, $\alpha_1=\alpha_2=0$ as well as 
$\varphi=\phi_4=0$ and $\beta_2=0$. Thus, we have
\begin{align}
z^1=z^2=-z^3=-z^4\,, \qquad \text{and}\qquad \qquad \beta_2=0\,,\end{align}
and we write (as in \cite{Bobev:2016nua})
\begin{align}
z^1=\tanh\big[\frac{1}{2}\big(\alpha_3-i\phi_3\big)\big]\,.
\end{align}
For the general $ISO(1,2)$ configurations, with boundary field theory metric \eqref{defhexpl}, we use the expansion
\begin{align}\label{onsbpsiso}
A&=\frac{r}{L}+\Omega+\cdots+{A}_{(v)}e^{-4r/L}+\cdots\,,\nn
V&=\frac{r}{L}+\Omega+f+\cdots+{V}_{(v)}e^{-4r/L}+\cdots\,,\nn
\phi_3&=\phi_{3,(s)}e^{-r/L}+\dots+{\phi}_{3,(v)}e^{-3r/L}+\cdots\,,\nn
\alpha_3&=\alpha_{3,(s)}\frac{r}{L}e^{-2r/L}+{\alpha}_{3,(v)}e^{-2r/L}+\cdots\,,\nn
\beta_1&=\beta_{1,(s)}\frac{r}{L}e^{-2r/L}+{\beta}_{1,(v)}e^{-2r/L}+\cdots\,,
\end{align}
where $\phi_{3,(s)}$, $\alpha_{3,(s)}$, $\beta_{1,(s)}$ are the source terms for the scalar operators in \eqref{opfieldmapz}.
Using the renormalisation scheme \eqref{susyscheme} we find that 
the one-point functions of the scalar operators are given by
\begin{align}\label{onemassalphvev}
\langle\mathcal{O}_{\alpha_3}\rangle
&=\frac{1}{4\pi GL}\Big({\alpha}_{3,(v)}-2\delta_{\alpha}\alpha_{3,(s)}\Big)\,,\nn
\langle\mathcal{O}_{\beta_1}\rangle
&=\frac{3}{2\pi GL}\Big({\beta}_{1,(v)}-2\delta_{\beta}\beta_{1,(s)}
+2\delta_{\tilde\beta}( \beta_{1,(s)}+\frac{2}{3}\phi_{3,(s)}^2   )\Big)\,,\nn
\langle\mathcal{O}_{\phi_{3}}\rangle
&=\frac{1}{2\pi GL}\Big({\phi}_{3,(v)}+\frac{1}{6}(7+32\delta_\beta)\phi_{3,(s)}^3
+8\delta_{\tilde{\beta}}\phi_{3,(s)}( \beta_{1,(s)}+\frac{2}{3}\phi_{3,(s)}^2   )
\nn&\qquad\qquad+\frac{L^2}{4}(1+4\delta_\alpha)\Box\phi_{3,(s)}-\frac{L^2}{24}(1+2\delta_{R\phi^2(1)})R\phi_{3,(s)}\Big)\,,
\end{align}
where $\Box$ and $R$ refer to 
the field theory metric $h_{ab}$ in \eqref{defhexpl},
along with the expected results
\begin{align}
\langle\mathcal{O}_{\alpha_1}\rangle&=\langle\mathcal{O}_{\alpha_2}\rangle=\langle\mathcal{O}_{\beta_2}\rangle=\langle\mathcal{O}_{\phi_{1}}\rangle=\langle\mathcal{O}_{\phi_{2}}\rangle=\langle\mathcal{O}_{\phi_{4}}\rangle=\langle\mathcal{O}_{\varphi}\rangle=0\,.
\end{align}

Focussing now on the $ISO(1,2)$ configurations that also solve the BPS equations 
\eqref{bpsderxi1appatext}, \eqref{fermvsremaintext}, the relation between the sources in \eqref{consces}
is given by
\begin{align}\label{onemassscerels}
\alpha_{3,(s)}&=\kappa L e^{-\Omega-f}\Big(\partial_x\phi_{3,(s)}+\phi_{3,(s)}\partial_x\Omega\Big)\,,\nn
\beta_{1,(s)}&=-\frac{2}{3}\phi_{3,(s)}^2\,.
\end{align}
The BPS equations also impose relations between the coefficients with $``(v)"$ subscript in \eqref{onsbpsiso}, which are explicitly given by
\begin{align}
{\phi}_{3,(v)}&=4{\beta}_{1,(v)}\phi_{3,(s)}-\frac{7}{6}\phi_{3,(s)}^3-\frac{L^2}{4}\Box\phi_{3,(s)}+\frac{L^2}{24}R\phi_{3,(s)}\nn
&\phantom{=}+\kappa Le^{-\Omega-f}\Big(\frac{1}{2}\partial_x{\alpha}_{3,(v)}+{\alpha}_{3,(v)}\partial_x\Omega\Big)-\frac{L^2}{4}e^{-2\Omega-2f}\phi_{3,(s)}(\partial_x \Omega)^2\,,\nn
-2{\alpha}_{3,(v)}\phi_{3,(s)}&=\frac{\kappa L}{2}e^{-\Omega-f}\Big(3\partial_x{\beta}_{1,(v)}+[6{\beta}_{1,(v)}+2\phi_{3,(s)}^2]\partial_x\Omega\Big)\,.
\end{align}
Under the renormalisation scheme \eqref{susyscheme}  
these are equivalent to the following set of relationships between the one point functions of the scalar operators
for the BPS configurations
\begin{align}\label{onemassvrels}
\langle\mathcal{O}_{\alpha_3}\rangle\,\phi_{3,(s)}=&-\frac{\kappa L}{8}e^{-\Omega-f}\Big(\partial_x\langle\mathcal{O}_{\beta_1}\rangle+
2\langle\mathcal{O}_{\beta_1}\rangle\partial_x\Omega+\frac{1}{\pi GL}\phi_{3,(s)}^2\partial_x\Omega\Big)\nn
&+(\delta_\beta-\delta_\alpha)\frac{\kappa}{4\pi G}e^{-\Omega-f}[\partial_x(\phi_{3,(s)}^2)+2\phi_{3,(s)}^2\partial_x\Omega]\,,\nn
\langle\mathcal{O}_{\phi_{3}}\rangle=&\frac{4}{3}\langle\mathcal{O}_{\beta_1}\rangle\,\phi_{3,(s)}+\kappa L e^{-\Omega-f}\Big(\partial_x\langle\mathcal{O}_{\alpha_3}\rangle+2\langle\mathcal{O}_{\alpha_3}\rangle\partial_x\Omega\Big)\nn
&\phantom{=}-\frac{L}{8\pi G}e^{-2\Omega-2f}\phi_{3,(s)}\Big(\partial_x \Omega\Big)^2
\nn
&+\delta_\alpha\frac{L}{2\pi G}e^{-\Omega-f}\partial_x\Big[e^{-\Omega-f}(\partial_x \phi_{3,(s)}+\phi_{3,(s)}\partial_x\Omega)\Big]\nn
&+\delta_\alpha\frac{L}{2\pi G}e^{-\Omega-f}\Big[2e^{-\Omega-f}(\partial_x \phi_{3,(s)}+\phi_{3,(s)}\partial_x\Omega)\partial_x\Omega\Big]\nn
&+\delta_\alpha\frac{L}{2\pi G}\Box \phi_{3,(s)}-\delta_{R\phi^2(1)}\frac{L}{24\pi G}R\phi_{3,(s)}\,,
\end{align}
where $R$ is again the Ricci scalar for the boundary metric $h_{ab}$ given in \eqref{defhexpl}.

\subsubsection{$\mathcal{N}=1^*$ equal-mass model}\label{appb12}
This model is obtained from the 10-scalar model by setting
 ${\phi}_1={\phi}_2={\phi}_3$ as well as ${\alpha}_1 = {\alpha}_2= {\alpha}_3$.
 In addition we set $\beta_1=\beta_2 = 0$. Thus, we have
 \begin{align}
   z^4=-z^3=-z^2\,,~~~ \text{and}\qquad
   \beta_1=\beta_2 = 0\,,
\end{align}
and we parametrise
$(z^1,z^2)$ via
\begin{align}\label{4scmodap}
     z^1 &= \tanh \big[ \frac{1}{2} \big( 3\alpha_1 + \varphi 
       -i 3\phi_1+ i \phi_4 \big) \big] \,,\nn
            z^2 &= \tanh \big[ \frac{1}{2} \big( \alpha_1 - \varphi 
       -i \phi_1 -i\phi_4 \big) \big] \,.
     \end{align}

For the general $ISO(1,2)$ configurations, with boundary field theory metric \eqref{defhexpl}, we use the expansion
\begin{align}\label{asexp10sceqmass}
A&=\frac{r}{L}+\Omega+\cdots+{A}_{(v)}e^{-4r/L}+\cdots\,,\nn
V&=\frac{r}{L}+\Omega+f+\cdots+{V}_{(v)}e^{-4r/L}+\cdots\,,\nn
\phi_2=\phi_3=\phi_1&=\phi_{1,(s)}e^{-r/L}+\dots+{\phi}_{1,(v)}e^{-3r/L}+\cdots\,,\nn
\phi_4&=\phi_{4,(s)}e^{-r/L}+\dots+{\phi}_{4,(v)}e^{-3r/L}+\cdots\,,\nn
\alpha_2=\alpha_3=\alpha_1&=\alpha_{1,(s)}\frac{r}{L}e^{-2r/L}+{\alpha}_{1,(v)}e^{-2r/L}+\cdots\,,\nn
\varphi&=\varphi_{(s)}+\cdots+{\varphi}_{(v)}e^{-4r/L}+\cdots\,,
\end{align}
where $\phi_{1,(s)}$, $\phi_{4,(s)}$, $\alpha_{1,(s)}$, $\varphi_{(s)}$ determine the source terms for the scalar operators in \eqref{opfieldmapz}
and we again emphasise that we focus on $\varphi_{(s)}=0$.
Using the renormalisation scheme \eqref{susyscheme} we find that
the one-point functions of the scalar operators are given by
\begin{align}\label{equalmassalphvev}
\langle\mathcal{O}_{\alpha_1}\rangle
&=\langle\mathcal{O}_{\alpha_2}\rangle=\langle\mathcal{O}_{\alpha_3}\rangle=\frac{1}{4\pi GL}\Big({\alpha}_{1,(v)}-2\delta_{\alpha}\alpha_{1,(s)}\Big)\,,\nn
\langle\mathcal{O}_{\phi_1}\rangle
&=\langle\mathcal{O}_{\phi_2}\rangle=\langle\mathcal{O}_{\phi_3}\rangle
=\frac{1}{2\pi GL}\Big(\phi_{1,(v)}+\frac{5}{6}\phi_{1,(s)}^3-\frac{9-2\delta_{4(5)}}{3}\phi_{1,(s)}^2\phi_{4,(s)}
\nn&\qquad+\frac{5-8\delta_{4(4)}}{6}\phi_{1,(s)}\phi_{4,(s)}^2+\frac{L^2}{4}(1+4\delta_{\alpha})\Box\phi_{1,(s)}-\frac{L^2}{24}(1+2\delta_{R\phi^2(1)})R\phi_{1,(s)}\Big)\,,\nn
\langle\mathcal{O}_{\phi_4}\rangle&=\frac{1}{2\pi GL}\Big(\phi_{4,(v)}-\frac{9-2\delta_{4(5)}}{3}\phi_{1,(s)}^3-\frac{5-8\delta_{4(4)}}{2}\phi_{1,(s)}^2\phi_{4,(s)}+\frac{11+16\delta_{4(2)}}{6}\phi_{4,(s)}^3\nn
&\phantom{=\frac{1}{2\pi GL}\Big(}+\frac{L^2}{4}(1+2\delta_{\partial\phi^2(2)})\Box\phi_{4,(s)}-\frac{L^2}{24}(1+2\delta_{R\phi^2(2)})R\phi_{4,(s)}\Big)\,,\nn
\langle\mathcal{O}_{\varphi}\rangle&=\frac{1}{\pi GL}\Big({\varphi}_{(v)}-\frac{3}{4}(\alpha_{1,(s)}-4{\alpha}_{1,(v)})(\phi_{1,(s)}^2-\phi_{1,(s)}\phi_{4,(s)})\Big)\,,
\end{align}
where $\Box$ and $R$ refer to 
the field theory metric $h_{ab}$ in \eqref{defhexpl},
along with the expected results
\begin{align}
\langle\mathcal{O}_{\beta_1}\rangle
&=\langle\mathcal{O}_{\beta_2}\rangle=0\,.
\end{align}

Focussing now on the $ISO(1,2)$ configurations that also solve the BPS equations 
\eqref{bpsderxi1appatext}, \eqref{fermvsremaintext}, the relation between the sources in \eqref{consces}
is given by
\begin{align}\label{consceseqm}
\varphi_{4,(s)}&=0\,,\nn
\phi_{4,(s)}&=0\,,\nn
\alpha_{1,(s)}&=\kappa L e^{-\Omega-f}\Big(\partial_x\phi_{1,(s)}+\phi_{1,(s)}\partial_x\Omega\Big)\,.
\end{align}
The BPS equations also impose relations between the coefficients with $``(v)"$ subscript in \eqref{asexp10sceqmass}, which are explicitly
\begin{align}
{\varphi}_{(v)}&=-3{\alpha}_{1,(v)}\phi_{1,(s)}^2-\frac{\kappa L}{4}e^{-\Omega-f}\partial_x\Big(\phi_{1,(s)}^3-{\phi}_{4,(v)}\Big)-\frac{3\kappa L}{4}e^{-\Omega-f}\Big(\phi_{1,(s)}^3-{\phi}_{4,(v)}\Big)\partial_x\Omega\,,\nn
{\phi}_{1,(v)}&=-\frac{5}{6}\phi_{1,(s)}^3-\frac{L^2}{4}\Box\phi_{1,(s)}+\frac{L^2}{24}R\phi_{1,(s)}+\kappa Le^{-\Omega-f}\Big(\frac{1}{2}\partial_x{\alpha}_{1,(v)}+{\alpha}_{1,(v)}\partial_x\Omega\Big)\nn
&\phantom{=}-\frac{L^2}{4}e^{-2\Omega-2f}\phi_{1,(s)}(\partial_x \Omega)^2\,.
\end{align}
Under the renormalisation scheme \eqref{susyscheme} 
these are equivalent to the following set of relationships between the one point functions of the scalar operators
for the BPS configurations 
\begin{align}\label{equalmassvrels}
\langle\mathcal{O}_{\phi_1}\rangle&=\kappa Le^{-\Omega-f}\Big(\partial_x\langle\mathcal{O}_{\alpha_1}\rangle+2\langle\mathcal{O}_{\alpha_1}\rangle\partial_x\Omega\Big)
-\frac{L}{8\pi G}e^{-2\Omega-2f}\phi_{1,(s)}(\partial_x \Omega)^2\nn
&+\delta_\alpha\frac{L}{2\pi G}e^{-\Omega-f}\partial_x\Big[e^{-\Omega-f}(\partial_x \phi_{1,(s)}+\phi_{1,(s)}\partial_x\Omega)\Big]\nn
&+\delta_\alpha\frac{L}{2\pi G}e^{-\Omega-f}\Big[2e^{-\Omega-f}(\partial_x \phi_{1,(s)}+\phi_{1,(s)}\partial_x\Omega)\partial_x\Omega\Big]\nn
&+\delta_\alpha\frac{L}{2\pi G}\Box \phi_{1,(s)}
-\delta_{R\phi^2(1)}\frac{L}{24\pi G}R\phi_{1,(s)}\,,\nn
\langle\mathcal{O}_{\varphi}\rangle&=
\frac{\kappa L}{2}e^{-\Omega-f}\Big(\partial_x\langle\mathcal{O}_{\phi_4}\rangle+3\langle\mathcal{O}_{\phi_4}\rangle\partial_x\Omega\Big)
\nn&
+\frac{\kappa(3-2\delta_{4(5)})}{12\pi G}e^{-\Omega-f}\Big(\partial_x(\phi_{1,(s)}^3)+3\phi_{1,(s)}^3\partial_x\Omega \Big)\,,
\end{align}
where $R$ is again the Ricci scalar for the boundary metric $h_{ab}$ given in \eqref{defhexpl}.

\subsubsection{$\mathcal{N}=2^*$ model}\label{appb13}
This model is obtained from the 10-scalar model by setting
${\phi}_1={\phi}_2$, ${\alpha}_1 = {\alpha}_2$  and $\beta_1\ne 0$, while imposing ${\alpha}_3 = {\phi}_3={\phi}_4={\varphi}=\beta_2=0$. 
Thus, we set
\begin{align}
z^1=z^3,\qquad z^2=z^4=\beta_2=0\,,
\end{align}
with
\begin{align}
z^1=\tanh[\alpha_1-i\phi_1]\,.
\end{align}
The expansion for the general $ISO(1,2)$ configurations is given by
\begin{align}\label{asexp10scn2case}
A&=\frac{r}{L}+\Omega+\cdots+{A}_{(v)}e^{-4r/L}+\cdots\,,\nn
V&=\frac{r}{L}+\Omega+f+\cdots+{V}_{(v)}e^{-4r/L}+\cdots\,,\nn
\phi_2=\phi_1&=\phi_{1,(s)}e^{-r/L}+\dots+{\phi}_{1,(v)}e^{-3r/L}+\cdots\,,\nn
\alpha_2=\alpha_1&=\alpha_{1,(s)}\frac{r}{L}e^{-2r/L}+{\alpha}_{1,(v)}e^{-2r/L}+\cdots\,,\nn
\beta_1&=\beta_{1,(s)}\frac{r}{L}e^{-2r/L}+{\beta}_{1,(v)}e^{-2r/L}+\cdots\,,
\end{align}
where $\phi_{1,(s)}$, $\alpha_{1,(s)}$ and $\beta_{1,(s)}$ are the source terms for the scalar operators.
The one point functions are given by
\begin{align}\label{n2alphvev}
\langle\mathcal{O}_{\alpha_1}\rangle
=\langle\mathcal{O}_{\alpha_2}\rangle&=\frac{1}{4\pi GL}\Big({\alpha}_{1,(v)}-2\delta_{\alpha}\alpha_{1,(s)}\Big)\,,\nn
\langle\mathcal{O}_{\beta_1}\rangle
&=\frac{3}{2\pi GL}\Big({\beta}_{1,(v)}
-2\delta_{\beta}\beta_{1,(s)}+ 2\delta_{\tilde{\beta}}(\beta_{1,(s)} -\frac{2}{3}\phi_{1,(s)}^2)
\Big)\,,\nn
\langle\mathcal{O}_{\phi_{1}}\rangle
=\langle\mathcal{O}_{\phi_{2}}\rangle&=\frac{1}{2\pi GL}\Big({\phi}_{1,(v)}+\frac{3+8\delta_{\beta}}{3}\phi_{1,(s)}^3
-4\delta_{\tilde{\beta}}\phi_{1,(s)}(\beta_{1,(s)} -\frac{2}{3}\phi_{1,(s)}^2)\nn
&\qquad\qquad+\frac{L^2}{4}(1+4\delta_{\alpha})\Box\phi_{1,(s)}
-\frac{L^2}{24}(1+2\delta_{R\phi^2(1)})R\phi_{1,(s)}\Big)\,,
\end{align}
where $\Box$ and $R$ refer to 
the field theory metric $h_{ab}$ in \eqref{defhexpl},
along with the expected results
\begin{align}
\langle\mathcal{O}_{\alpha_3}\rangle&=\langle\mathcal{O}_{\beta_2}\rangle=\langle\mathcal{O}_{\phi_{3}}\rangle=\langle\mathcal{O}_{\phi_{4}}\rangle=\langle\mathcal{O}_{\varphi}\rangle=0\,.
\end{align}

Turning to the supersymmetric $ISO(1,2)$ BPS configurations satisfying 
\eqref{bpsderxi1appatext}, \eqref{fermvsremaintext}, the relation between the sources is given by
\begin{align}\label{conscesn2}
\alpha_{1,(s)}&=\kappa L e^{-\Omega-f}\Big(\partial_x\phi_{1,(s)}+\phi_{1,(s)}\partial_x\Omega\Big)\,,\nn
\beta_{1,(s)}&=\frac{2}{3}\phi_{1,(s)}^2\,.
\end{align}
The BPS equations also impose relations between the coefficients with $``(v)"$ subscript in \eqref{asexp10scn2case}
given by
\begin{align}
{\phi}_{1,(v)}&=-2{\beta}_{1,(v)}\phi_{1,(s)}-\phi_{1,(s)}^3-\frac{L^2}{4}\Box\phi_{1,(s)}+\frac{L^2}{24}R\phi_{1,(s)}\nn
&\phantom{=}+\kappa Le^{-\Omega-f}\Big(\frac{1}{2}\partial_x{\alpha}_{1,(v)}+{\alpha}_{1,(v)}\partial_x\Omega\Big)-\frac{L^2}{4}e^{-2\Omega-2f}\phi_{1,(s)}(\partial_x \Omega)^2\,,\nn
2{\alpha}_{1,(v)}\phi_{1,(s)}&=\frac{\kappa L}{2}e^{-\Omega-f}\Big(3\partial_x{\beta}_{1,(v)}+[6{\beta}_{1,(v)}-2\phi_{1,(s)}^2]\partial_x\Omega\Big)\,.
\end{align}
Under the renormalisation scheme \eqref{susyscheme} 
these are equivalent to the following set of relationships between the one point functions of the scalar operators
for the BPS configurations 
\begin{align}\label{n2vrels}
\langle\mathcal{O}_{\alpha_1}\rangle\,\phi_{1,(s)}&=\frac{\kappa L}{8}e^{-\Omega-f}\Big(\partial_x\langle\mathcal{O}_{\beta_1}\rangle+
2\langle\mathcal{O}_{\beta_1}\rangle \partial_x\Omega
-\frac{1}{\pi GL}\phi_{1,(s)}^2\partial_x\Omega\Big)\nn
&+(\delta_{\beta}-\delta_{\alpha})\frac{\kappa}{2\pi G}e^{-\Omega-f}\phi_{1,(s)}\Big(\partial_x\phi_{1,(s)}+\phi_{1,(s)}\partial_x\Omega\Big)\,,\nn
\langle\mathcal{O}_{\phi_{1}}\rangle&=-\frac{2}{3}\langle\mathcal{O}_{\beta_1}\rangle\,\phi_{1,(s)}+\kappa L e^{-\Omega-f}\Big(\partial_x\langle\mathcal{O}_{\alpha_1}\rangle+2\langle\mathcal{O}_{\alpha_1}\rangle\partial_x\Omega\Big)\nn
&-\frac{L}{8\pi G}e^{-2\Omega-2f}\phi_{1,(s)}\Big(\partial_x \Omega\Big)^2\nn
&+\delta_\alpha\frac{L}{2\pi G}e^{-\Omega-f}\partial_x\Big[e^{-\Omega-f}(\partial_x \phi_{1,(s)}+\phi_{1,(s)}\partial_x\Omega)\Big]\nn
&+\delta_\alpha\frac{L}{2\pi G}e^{-\Omega-f}\Big[2e^{-\Omega-f}(\partial_x \phi_{1,(s)}+\phi_{1,(s)}\partial_x\Omega)\partial_x\Omega\Big]\nn
&+\delta_\alpha\frac{L}{2\pi G}\Box \phi_{1,(s)}
-\delta_{R\phi^2(1)}\frac{L}{24\pi G}R\phi_{1,(s)}\,,
\end{align}
where again $R$ refer to 
the field theory metric $h_{ab}$ in \eqref{defhexpl}.

\section{One point functions for Janus solutions}\label{appc}
For orientation we first recall the metric for $AdS_{5}$ written in ``Janusian" coordinates which makes manifest the foliation by $AdS_4$
spaces. We then discuss how the results of the previous appendix can be employed to obtain holographic data for the
Janus solutions discussed in section \ref{susyjansol}.

\subsection{Janusian coordinates for $AdS_{5}$}\label{junsian}

\begin{figure}[h!]
\centering
{\includegraphics[width= .5 \textwidth]{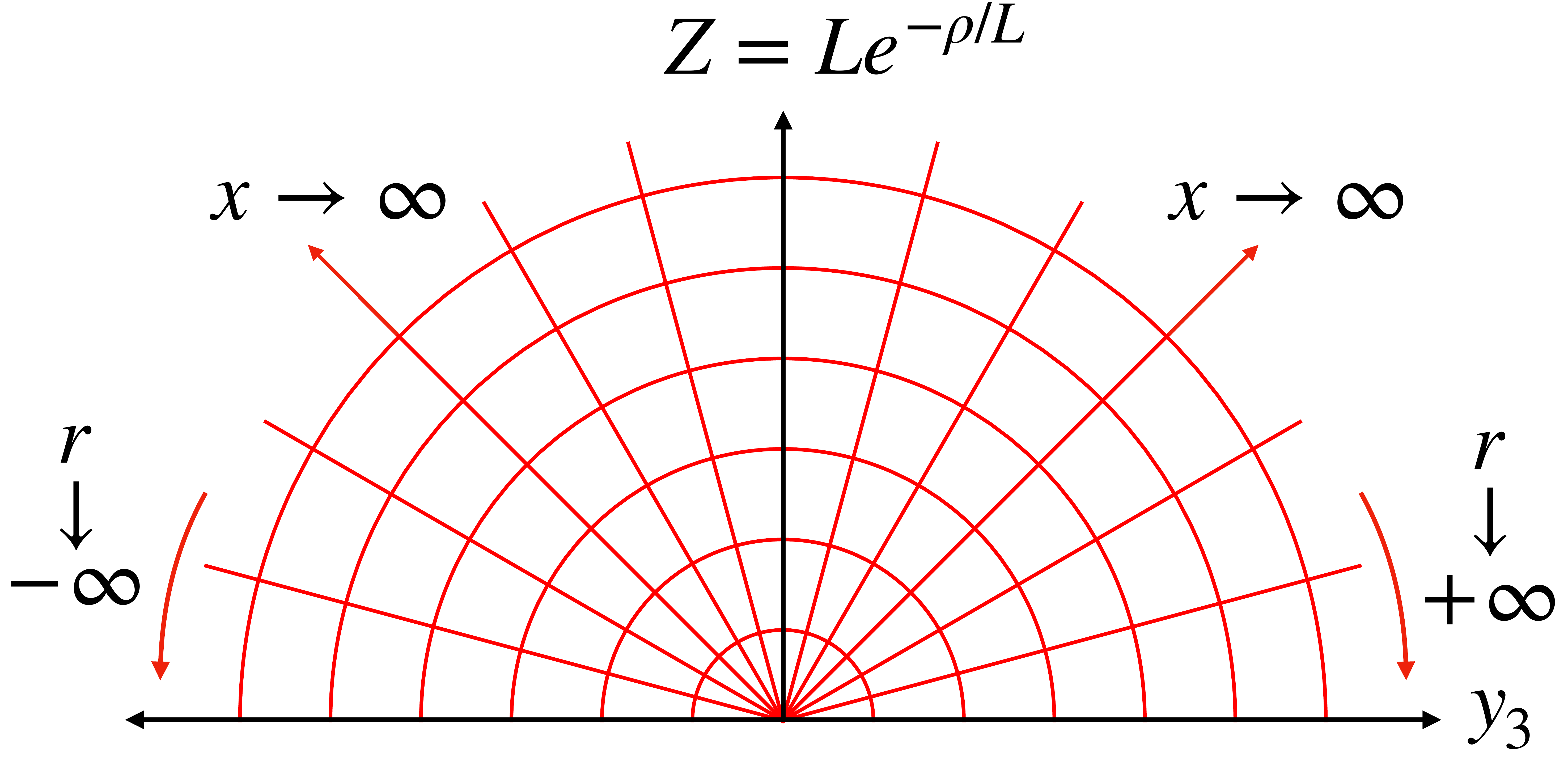}}
\caption{Coordinates for Janus configurations, with $(t,y_1,y_2)$ suppressed. We have 
$\rho\in(-\infty,+\infty)$ and $y_3\in(-\infty,\infty)$, with the conformal boundary located at $\rho\to\infty$, parametrised by $(t,y_1,y_2,y_3)$, which naturally comes with a flat space metric. We can also use coordinates with $x\in (0,\infty)$, $r\in(-\infty,\infty)$, with the straight lines of constant $r$ parametrising $AdS_4$ spacetime. In these coordinate the conformal boundary consists of three components: the two half spaces at $r=\pm\infty$, parametrised by $(t,y_1,y_2,x)$ which naturally come with an $AdS_4$ metric, and $x\to 0$ which is the interface $y_3=0$.}\label{fig:janus}
\end{figure}

Consider writing $AdS_{5}$, with radius $L$, in Poincar\'e coordinates, in mostly minus signature, singling out a preferred spatial direction $y_3$:
\begin{align}\label{poinccs}
ds^2=\frac{L^2}{Z^2}\left[-dZ^2-dy_3^2+ (dt^2-dy_1^2-dy_2^2)\right]\,,
\end{align}
with $Z\in (0,\infty)$. Notice that $(y_3,Z)$ parametrise a half plane as in figure \ref{fig:janus}. 
We can switch to polar coordinates for this half plane
via $y_3=x\sin\mu$, $Z=x\cos\mu$, with $x\in (0,\infty)$ and $\mu\in[-\pi/2,\pi/2]$ to get
\begin{align}
ds^2=\frac{L^2}{\cos^2\mu}\left[-d\mu^2+\frac{1}{x^2}\left(-dx^2+dt^2-dy_1^2-dy_2^2\right)\right]\,.
\end{align}
We can also do a further coordinate change, by setting $\cos\mu=[\cosh ({r}/{L})]^{-1}$ and keeping $x$ fixed to get
\begin{align}
ds^2=-dr^2+\cosh^2({r}/{L})\left[ \frac{L^2}{x^2}\left(-dx^2+dt^2-dy_1^2-dy_2^2\right)\right]\,,
\end{align}
with $x\in (0,\infty)$, $r\in(-\infty,\infty)$. These $(x,r)$ coordinates are related to the original Poincare coordinates via
$x=\sqrt{y_3^2+Z^2}$, $e^{{r}/{L}}=\frac{y_3+\sqrt{y_3^2+Z^2}}{Z}$ and are also illustrated in figure 
\ref{fig:janus}.
We also note that $r\to\pm\infty$ are associated with
$y_3>0$ and $y_3<0$, respectively. 
Finally, after writing $Z=Le^{-\rho/L}$ in the original metric \eqref{poinccs} we have
\begin{align}
ds^2=-d\rho^2+e^{2\rho/{L}}\left[ dt^2-dy_1^2-dy_2^2-dy_3^2\right]\,,
\end{align}
with $\rho\in(-\infty,+\infty)$ and the conformal boundary at $\rho\to\infty$. We heavily utilise the $(\rho,y_3)$ coordinates and the $(r,x)$ coordinates in this paper, with
the former associated with flat spacetime boundary metric and the latter associated with $AdS_4$ boundary metric.

\subsection{BPS Janus  solutions: field theory on $AdS_4$}\label{appc2}

The BPS Janus solutions discussed in sections \ref{bpsjanus} and \ref{susyjansol} are special sub-classes of the $ISO(1,2)$ BPS
solutions discussed in appendix \ref{appb} with
\begin{align}
e^{A(r,x)}=e^{V(r,x)}=e^{A_J(r)}\frac{\ell}{x}\,,
\end{align}
and all scalar fields taken to be a function of $r$ only. As $r\to\pm\infty$ the $\mathcal{N}=4$ SYM Janus solutions
approach the $\mathcal{N}=4$ SYM $AdS_5$ vacuum but
with additional mass sources. Like the $\mathcal{N}=4$ SYM $AdS_5$ vacuum solution itself, the conformal boundary again consists of three components, with two half spaces
(with $AdS_4$ metrics) that are joined at a planar interface. Let us first consider the $r\to\infty$ end, returning to the $r\to-\infty$ end 
in section \ref{otherend}.
After recalling \eqref{asexp10sc}, as $r\to\infty$
we have the schematic expansion of the BPS equations  \eqref{Apbps},\eqref{eq:ten_scalar_BPS} (with $N=1$) given 
by
\begin{align}\label{asexp10scjanapp}
A_J&=\frac{r}{L}+A_0+\cdots+{A}_{(v)}e^{-4r/L}+\cdots\,,\nn
\phi_i&=\phi_{i,(s)}e^{-r/L}+\dots+{\phi}_{i,(v)}e^{-3r/L}+\cdots\,,\qquad i=1,\dots,4\,,\nn
\alpha_i&=\alpha_{i,(s)}\frac{r}{L}e^{-2r/L}+{\alpha}_{i,(v)}e^{-2r/L}+\cdots\,,,\qquad i=1,\dots,3\,,\nn
\beta_i&=\beta_{i,(s)}\frac{r}{L}e^{-2r/L}+{\beta}_{i,(v)}e^{-2r/L}+\cdots\,,\qquad i=1,\dots,2\,,\nn
\varphi&=\varphi_{(s)}+\cdots+{\varphi}_{(v)}e^{-4r/L}+\cdots\,.
\end{align}
The various constant coefficients in this expansion are constrained by the BPS equations, as discussed below.
We have highlighted a constant term $A_0$ that can appear in the expansion for $A_J$. By shifting the radial coordinate via
$r\to r-A_0 L$ we can always remove this term and we shall do so in the following. In particular all the expressions for the expectation values and sources given below are obtained with
\begin{align}
A_0=0\,.
\end{align}

The terms $\phi_{i,(s)}$, $\alpha_{i,(s)}$, $\beta_{i,(s)}$, $\varphi_{(s)}$ give rise to source terms for 
$\mathcal{N}=4$ SYM on this component of the conformal boundary with $AdS_4$ metric. 
Recalling that these are sources for operators of conformal dimension $\Delta=3,2,2,4$  respectively, it
is helpful to note that the field theory sources on $AdS_4$, that are invariant under Weyl scalings of $\ell$, 
are given by $\ell\phi_{i,(s)}$, $\ell^2\alpha_{i,(s)}$, $\ell^2\beta_{i,(s)}$.
We are always assuming that
$\varphi_{(s)}=0$ and from \eqref{consces}, the BPS conditions relating the sources are given for the ten scalar model 
by
\begin{align}\label{tenscerels}
\alpha_{i,(s)}&=-\kappa\frac{L}{\ell} \phi_{i,(s)}\,,\qquad i=1,\dots,3\,,\nn
\beta_{1,(s)}&=\frac{1}{3}\Big(\phi_{1,(s)}^2+\phi_{2,(s)}^2-2\phi_{3,(s)}^2\Big)\,,\nn
\beta_{2,(s)}&=\phi_{1,(s)}^2-\phi_{2,(0)}^2\,,\nn
\phi_{4,(s)}&=0\,.
\end{align}

In a similar manner
${\phi}_{i,(v)}$, ${\alpha}_{i,(v)}$, ${\beta}_{i,(v)}$ and 
${\varphi}_{(v)}$, with suitable contributions from the sources, give rise to the expectation values of the scalar
operators. We can obtain these results for each of the three truncations considered in appendix \ref{appb}, after using
$e^\Omega=\ell/x$ and $f(x)=0$,
as we summarise below.

\subsubsection{$\mathcal{N}=1^*$ one-mass model: $AdS_4$ boundary}\label{bliponenv}
We use the renormalisation scheme \eqref{susyscheme}. From \eqref{onemassalphvev} we have
\begin{align}
\langle\mathcal{O}_{\alpha_3}\rangle
&=\frac{1}{4\pi GL}\Big({\alpha}_{3,(v)}-2\delta_{\alpha}\alpha_{3,(s)}\Big)\,.
\end{align}
For BPS Janus configurations, from \eqref{onemassvrels} we can then express the remaining non-trivial expectation values in terms of 
$\langle\mathcal{O}_{\alpha_3}\rangle$ along with $\phi_{3,(s)}$ as follows:
\begin{align}
\langle\mathcal{O}_{\beta_1}\rangle
&=\frac{4\kappa\ell}{L}\langle\mathcal{O}_{\alpha_3}\rangle\,\phi_{3,(s)}-
\frac{(1+4\delta_{\alpha}-4\delta_{\beta})}{2\pi G L}\phi_{3,(s)}^2\,,\nn
\langle\mathcal{O}_{\phi_{3}}\rangle&=\frac{4}{3}\langle\mathcal{O}_{\beta_1}\rangle\,\phi_{3,(s)}-\frac{2\kappa L}{\ell} \langle\mathcal{O}_{\alpha_3}\rangle-\frac{L}{8\pi G\ell^2}{(1-8\delta_{\alpha}+4\delta_{R\phi^2(1)})}\phi_{3,(s)}\,.
\end{align}
Notice that these expressions depend on the $\delta_\alpha$, $\delta_\beta$, $\delta_{R\phi^2(1)}$ which parametrise finite counterterms
that we haven't fixed.
We also have
\begin{align}
\langle\mathcal{O}_{\alpha_1}\rangle&=\langle\mathcal{O}_{\alpha_2}\rangle=\langle\mathcal{O}_{\beta_2}\rangle=\langle\mathcal{O}_{\phi_{1}}\rangle=\langle\mathcal{O}_{\phi_{2}}\rangle=\langle\mathcal{O}_{\phi_{4}}\rangle=\langle\mathcal{O}_{\varphi}\rangle=0\,,
\end{align}
independent of the counterterms. Notice that 
for a fixed choice of $\delta_\alpha$, $\delta_\beta$, $\delta_{R\phi^2(1)}$, we can therefore specify all of the scalar sources and expectation values
of the dual field theory by giving $\phi_{3,(s)}$ and ${\alpha}_{3,(v)}$.

\subsubsection{$\mathcal{N}=1^*$ equal-mass model: $AdS_4$ boundary}\label{eqmsads4}
We use the renormalisation scheme \eqref{susyscheme}. From \eqref{equalmassalphvev} and
\eqref{tenscerels}
for BPS configurations we have
\begin{align}
\langle\mathcal{O}_{\alpha_1}\rangle=\langle\mathcal{O}_{\alpha_2}\rangle=\langle\mathcal{O}_{\alpha_3}\rangle
&=\frac{1}{4\pi GL}\Big({\alpha}_{1,(v)}-2\delta_{\alpha}\alpha_{1,(s)}\Big)\,,\nn
\langle\mathcal{O}_{\phi_4}\rangle&=\frac{1}{2\pi GL}\Big(\phi_{4,(v)}-\frac{9-2\delta_{4(5)}}{3}\phi_{1,(s)}^3\Big)\,.
\end{align}
For BPS Janus configurations, from \eqref{equalmassvrels} we can then express the other expectation values in terms of 
$\langle\mathcal{O}_{\alpha_1}\rangle$, $\langle\mathcal{O}_{\phi_4}\rangle$ along with $\phi_{1,(s)}$ as follows
\begin{align}
\langle\mathcal{O}_{\phi_1}\rangle&=-\frac{2\kappa L}{\ell}\langle\mathcal{O}_{\alpha_1}\rangle-\frac{L(1+4\delta_{R\phi^2(1)}-8\delta_{\alpha})}{8\pi G\ell^2}\phi_{1,(s)}\,,\nn
\langle\mathcal{O}_{\varphi}\rangle&=-\frac{3\kappa L}{2\ell}\langle\mathcal{O}_{\phi_4}\rangle -\frac{\kappa(3-2\delta_{4(5)})}{4\pi G\ell}\phi_{1,(s)}^3\,.
\end{align}

Notice that these expressions depend on the $\delta_\alpha$, $\delta_{R\phi^2(1)}$, $\delta_{4(5)}$ which parametrise finite counterterms
which we haven't fixed.
We also have
\begin{align}
\langle\mathcal{O}_{\beta_1}\rangle
&=\langle\mathcal{O}_{\beta_2}\rangle=0\,,
\end{align}
independent of the counterterms. Notice that 
for a fixed choice of $\delta_\alpha$, $\delta_{R\phi^2(1)}$, $\delta_{4(5)}$ we can therefore specify all of the scalar sources and expectation values
of the dual field theory by giving $\phi_{1,(s)}$, ${\alpha}_{1,(v)}$ and $\phi_{4,(v)}$.

\subsubsection{$\mathcal{N}=2^*$ model: $AdS_4$ boundary}\label{bliptwonv}

We use the renormalisation scheme \eqref{susyscheme}. From \eqref{n2alphvev} we have
\begin{align}
\langle\mathcal{O}_{\alpha_1}\rangle=\langle\mathcal{O}_{\alpha_2}\rangle
&=\frac{1}{4\pi GL}({\alpha}_{1,(v)}-2\delta_{\alpha}\alpha_{1,(s)})\,,
\end{align}
For BPS Janus configurations, from \eqref{n2vrels} we can then express the other expectation values in terms of 
$\langle\mathcal{O}_{\alpha_1}\rangle$ along with $\phi_{1,(s)}$ as follows
\begin{align}
\langle\mathcal{O}_{\phi_{1}}\rangle=\langle\mathcal{O}_{\phi_{2}}\rangle&=-\frac{2}{3}\langle\mathcal{O}_{\beta_1}\rangle\,\phi_{1,(s)}-\frac{2\kappa L}{\ell} \langle\mathcal{O}_{\alpha_1}\rangle-\frac{L}{8\pi G\ell^2}{(1-8\delta_{\alpha}+4\delta_{R\phi^2(1)})}\phi_{1,(s)}\,,\nn
\langle\mathcal{O}_{\beta_1}\rangle
&=-\frac{4\kappa\ell}{L}\langle\mathcal{O}_{\alpha_1}\rangle\,\phi_{1,(s)}+
\frac{(1+4\delta_{\alpha}-4\delta_{\beta})}{2\pi G L}\phi_{1,(s)}^2\,.
\end{align}
Notice that these expressions depend on the $\delta_\alpha$, $\delta_\beta$, $\delta_{R\phi^2(1)}$ which parametrise finite counterterms
which we haven't fixed.
We also have
\begin{align}
\langle\mathcal{O}_{\alpha_3}\rangle&=\langle\mathcal{O}_{\beta_2}\rangle=\langle\mathcal{O}_{\phi_{3}}\rangle=\langle\mathcal{O}_{\phi_{4}}\rangle=\langle\mathcal{O}_{\varphi}\rangle=0\,,
\end{align}
independent of the counterterms. Notice that 
for a fixed choice of $\delta_\alpha$, $\delta_\beta$, $\delta_{R\phi^2(1)}$, we can therefore specify all of the scalar sources and expectation values
of the dual field theory by giving $\phi_{1,(s)}$ and ${\alpha}_{1,(v)}$.

\subsubsection{Results for the $r\to-\infty$ end, $AdS_4$ boundary}\label{otherend}
We now discuss analogous results, for the sources and expectation values, for the conformal boundary, with $AdS_4$ metric, at the $r\to-\infty$ end.
Here we can develop an asymptotic expansion to the BPS equations
\eqref{Apbps},\eqref{eq:ten_scalar_BPS} (with $N=1$) of the form
\begin{align}\label{asexp10scjanappoto}
A_J&=\frac{-r}{L}+\tilde A_0+\cdots+\tilde{A}_{(v)}e^{4r/L}+\cdots\,,\nn
\phi_i&=\tilde\phi_{i,(s)}e^{r/L}+\dots+\tilde{\phi}_{i,(v)}e^{3r/L}+\cdots\,,\qquad i=1,\dots,4\,,\nn
\alpha_i&=\tilde\alpha_{i,(s)}\frac{-r}{L}e^{2r/L}+\tilde{\alpha}_{i,(v)}e^{2r/L}+\cdots\,,,\qquad i=1,\dots,3\,,\nn
\beta_i&=\tilde\beta_{i,(s)}\frac{-r}{L}e^{2r/L}+\tilde{\beta}_{i,(v)}e^{2r/L}+\cdots\,,\qquad i=1,\dots,2\,,\nn
\varphi&=\tilde\varphi_{(s)}+\cdots+\tilde{\varphi}_{(v)}e^{4r/L}+\cdots\,,
\end{align}
and we will always set $\tilde A_0=0$, which can be achieved by a shift of the radial coordinate.
This has exactly the same form as in \eqref{asexp10scjanapp} after the interchange $r\to -r$. The BPS equations will then relate various coefficients.
We can easily deduce these relations using the following argument. We first recall that the BPS equations
\eqref{Apbps},\eqref{eq:ten_scalar_BPS} are invariant under the transformation $r\to -r$, $\xi\to \xi+\pi$ and $\kappa\to -\kappa$.
Second, we want to use the result that if a solution has $\xi=0$ at $r=+\infty$ then necessarily it will have $\xi=\pi$ at $r=-\infty$.
This can be seen from \eqref{eq:RewrittenJanusBPSAeq}: at $r\to\pm\infty$ the scalars are approaching zero so the phase of $\mathcal{W}$ is going to zero. Thus, the phase of $B_r$ at $r\to\pm\infty$ is $\xi$ and from \eqref{eq:RewrittenJanusBPSAeq} we see that $\xi$ must change by $\pi$ in going from $r=+\infty$ to $r=-\infty$.
Taking these two results together, we can then deduce that all of the results that we obtained for the $r\to+\infty$ end can
be taken over to the $r\to-\infty$ end provided that wherever $\kappa$ appears in the former, it is replaced\footnote{Recall that $\kappa=\pm1$ enters the Killing spinor projections \eqref{bpstextprojection}.
To avoid possible confusion, we emphasise that we are holding this projection fixed in developing the asymptotic expansion 
\eqref{asexp10scjanappoto} at $r\to-\infty$; the argument we have given is just a way of getting at the result.} with $-\kappa$ in the 
latter.

Thus, for example, we can conclude that the BPS equations in \eqref{Apbps},\eqref{eq:ten_scalar_BPS} (with $N=1$) imply that
in the expansion \eqref{asexp10scjanappoto} at $r\to-\infty$ we now have
\begin{align}\label{tenscerelstwo}
\tilde\alpha_{i,(s)}&=+2\kappa\frac{L}{\ell} \tilde\phi_{i,(s)}\,,\qquad i=1,\dots,3\,,\nn
\tilde\beta_{1,(s)}&=\frac{1}{3}\Big(\tilde\phi_{1,(s)}^2+\tilde\phi_{2,(s)}^2-2\tilde\phi_{3,(s)}^2\Big)\,,\nn
\tilde\beta_{2,(s)}&=\tilde\phi_{1,(s)}^2-\tilde\phi_{2,(0)}^2\,,\nn
\tilde\phi_{4,(s)}&=0\,,
\end{align}
and we note the sign flip in the first line as compared to \eqref{tenscerels}.
Similarly, for all  the results for the expectation values at the $r\to \infty$ end that we gave in the previous subsections 
\ref{bliponenv}-\ref{bliptwonv}, we can take over
to analogous results at the $r\to -\infty$ end, after replacing $\kappa$ with $-\kappa$.

\subsection{BPS Janus  solutions: field theory on flat spacetime}

For the Janus solutions, we are primarily interested in obtaining the sources and expectation values
for operators of $\mathcal{N}=4$ SYM in flat spacetime. To do this\footnote{Note that 
the results in this section can also obtained from our results
\eqref{weylmet}-\eqref{dim4weylt}.} we 
carry out a bulk coordinate transformation as we approach the $r\to\infty$ component of the conformal boundary, that we are focussing on, 
so that it has a flat metric. 
For the $r \to\infty$ component of the conformal boundary
 we can use the coordinate transformation of the form
\begin{align}\label{asbfggh}
e^{r/L} &= \frac{y_3}{\ell} e^{\rho/L}  + \frac{L^2}{4 \ell y_3}e^{-\rho/L} + \cO(e^{-3\rho/L}/y_3^3)\,,\nn
x &= y_3 + \frac{L^2}{2 y_3}e^{-2\rho/L} + \cO(e^{-4\rho/L}/y_3^3)\,,
\end{align}
with $y_3>0$. Substituting this into \eqref{asexp10scjan} then leads to an expansion as $\rho\to\infty$
with the metric asymptoting to
\begin{align}
ds^2\to -d\rho^2+e^{2\rho/{L}}\left( dt^2-dy_1^2-dy_2^2-dy_3^2\right)\,,
\end{align}
and recalling the discussion in section \ref{junsian}, this component of the conformal boundary is for $y_3>0$.
As $\rho\to\infty$ we find that the expansion for the scalars given in \eqref{asexp10scjanapp} then becomes
\begin{align}
\phi_i&=\frac{\ell}{y_3}\phi_{i,(s)}e^{-\rho/L}+\frac{\ell^3}{y_3^3}\Bigg\{{\phi}_{i,(v)}-\frac{L^2}{4\ell^2}\phi_{i,(s)}\nn
&+\Big(\frac{L^2}{\ell^2}\phi_{i,(s)}-4\phi_{i,(s)}^3+\frac{4}{3}\phi_{i,(s)}\sum_{j=1}^4\phi_{j,(s)}^2\Big)\Big[\frac{\rho}{L}+\log\Big(\frac{y_3}{\ell}\Big)\Big]\Bigg\}e^{-3\rho/L}+\cdots\,,\quad i=1,2,3\,,\nn
\alpha_i&=\frac{\ell^2}{y_3^2}\Big\{{\alpha}_{i,(v)}+\alpha_{i,(s)}\Big[\frac{\rho}{L}+\log\Big(\frac{y_3}{\ell}\Big)\Big]\Big\}e^{-2\rho/L}+\cdots\,,\qquad i=1,2,3\nn
\beta_i&=\frac{\ell^2}{y_3^2}\Big\{{\beta}_{i,(v)}+\beta_{i,(s)}\Big[\frac{\rho}{L}+\log\Big(\frac{y_3}{\ell}\Big)\Big]\Big\}e^{-2\rho/L}+\cdots\,,\qquad i=1,2\nn
\varphi&=\frac{\ell^4}{y_3^4}\Big\{{\varphi}_{(v)}-\Big(\alpha_{1,(s)}\phi_{2,(s)}\phi_{3,(s)}+\alpha_{2,(s)}\phi_{1,(s)}\phi_{3,(s)}\nn
&\qquad\qquad\qquad\qquad+\alpha_{3,(s)}\phi_{1,(s)}\phi_{2,(s)}\Big)\Big[\frac{\rho}{L}+\log\Big(\frac{y_3}{\ell}\Big)\Big]\Big\}e^{-4\rho/L}+\cdots\,,
\end{align}
and we note that we have set $\phi_{4,(s)}=0$ as implied by the BPS relations
\eqref{tenscerels}.

To proceed, we now notice that this form of the solution is a special case of the $ISO(1,2)$ invariant configurations
discussed in appendix \ref{appb}, with $\Omega(x)=f(x)=0$,
provided that we replace the coordinates $(r,x)$ in that appendix with $(\rho, y_3)$. As a consequence
we can immediately read off the sources and the expectation values for the various operators.
The non-zero scalar sources in flat spacetime are of the form
\begin{align}\label{fspaceappc}
\frac{\ell\phi_{i,(s)}}{y_3},\quad &\frac{\ell^2 \alpha_{i,(s)}}{y_3^2}, \quad i=1,\dots,3\,\nn
&\frac{\ell^2\beta_{i,(s)}}{y_3^2}\,,\quad i=1,\dots,2\,,
\end{align}
with $\phi_{4,(s)}=\varphi_{(s)}=0$.
Recalling that the numerators in these expression are scale invariant parameters, we see that these quantities have the correct field theory 
scaling dimensions of $1,2,2$, for sources of operators with conformal dimension $\Delta=3,2,2$, respectively.

We can also use the results in sections \ref{appb11}-\ref{appb13}, to deduce the expectation values of the operators for the BPS configurations
and the results are recorded in
the next subsections. A general point we can notice is the presence of the novel terms of the form $\sim \log(y_3/\ell)$.

\subsubsection{$\mathcal{N}=1^*$ one-mass model: flat boundary}
Transforming the results from section \ref{bliponenv} to flat space boundary we obtain
\begin{align}
\langle\mathcal{O}_{\alpha_3}\rangle
&=\frac{1}{4\pi GL}\frac{\ell^2}{y_3^2}\Big({\alpha}_{3,(v)}+\alpha_{3,(s)}
\log(\frac{y_3}{\ell e^{2\delta_\alpha}})\Big)\,.
\end{align}
The BPS relations between the remaining expectation values are given by
\begin{align}
\langle\mathcal{O}_{\phi_{3}}\rangle&=
\frac{4}{3}\frac{\ell}{y_3}\langle\mathcal{O}_{\beta_1}\rangle\,\phi_{3,(s)}-{2\kappa L}\frac{1}{y_3} \langle\mathcal{O}_{\alpha_3}\rangle
-\frac{L}{4\pi G}\frac{\ell}{y_3^3}\phi_{3,(s)}\,,\nn
\langle\mathcal{O}_{\beta_1}\rangle
&=\frac{4\kappa\ell}{L}\langle\mathcal{O}_{\alpha_3}\rangle\,\phi_{3,(s)}-
\frac{(1+4\delta_{\alpha}-4\delta_{\beta})}{2\pi G L}\frac{\ell^2}{y_3^2}\phi_{3,(s)}^2\,.
\end{align}

\subsubsection{$\mathcal{N}=1^*$ equal-mass model: flat boundary}
Transforming the results from section \ref{eqmsads4} to flat space boundary we obtain for the BPS configurations
\begin{align}
\langle\mathcal{O}_{\alpha_1}\rangle=\langle\mathcal{O}_{\alpha_2}\rangle=\langle\mathcal{O}_{\alpha_3}\rangle
&=\frac{1}{4\pi GL}\frac{\ell^2}{y_3^2}\Big({\alpha}_{1,(v)}+\alpha_{1,(s)}
\log(\frac{y_3}{\ell e^{2\delta_\alpha}})\Big)\,,\nn
\langle\mathcal{O}_{\phi_4}\rangle&=\frac{1}{2\pi GL}\frac{\ell^3}{y_3^3}\Big(\phi_{4,(v)}-\frac{9-2\delta_{4(5)}}{3}\phi_{1,(s)}^3\Big)\,.
\end{align}
The BPS relations between the remaining expectation values are given by
\begin{align}
\langle\mathcal{O}_{\phi_1}\rangle=\langle\mathcal{O}_{\phi_2}\rangle=\langle\mathcal{O}_{\phi_3}\rangle&=-{2\kappa L}\frac{1}{y_3}\langle\mathcal{O}_{\alpha_1}\rangle
-\frac{L}{4\pi G}\frac{\ell}{y_3^3}\phi_{1,(s)}\,,\nn
y_3\langle\mathcal{O}_{\varphi}\rangle&=-\frac{3\kappa L}{2}\langle\mathcal{O}_{\phi_4}\rangle -\frac{\kappa(3-2\delta_{4(5)})}{4\pi G}\frac{\ell^3}{y_3^3}\phi_{1,(s)}^3\,.
\end{align}

\subsubsection{$\mathcal{N}=2^*$ model: flat boundary}
Transforming the results from section \ref{bliptwonv} to flat space boundary we obtain
\begin{align}
\langle\mathcal{O}_{\alpha_1}\rangle=\langle\mathcal{O}_{\alpha_2}\rangle
&=\frac{1}{4\pi GL}\frac{\ell^2}{y_3^2}\Big({\alpha}_{1,(v)}+\alpha_{1,(s)}
\log(\frac{y_3}{\ell e^{2\delta_\alpha}})\Big)\,.
\end{align}
The BPS relations between the remaining expectation values are given by 
\begin{align}
\langle\mathcal{O}_{\beta_1}\rangle
&=-\frac{4\kappa\ell}{L}\langle\mathcal{O}_{\alpha_1}\rangle\,\phi_{1,(s)}+
\frac{(1+4\delta_{\alpha}-4\delta_{\beta})}{2\pi G L}\frac{\ell^2}{y_3^2}\phi_{1,(s)}^2\,,\nn
\langle\mathcal{O}_{\phi_{1}}\rangle=\langle\mathcal{O}_{\phi_{2}}\rangle&=
-\frac{2}{3}\frac{\ell}{y_3}\langle\mathcal{O}_{\beta_1}\rangle\,\phi_{1,(s)}-{2\kappa L}
\frac{1}{y_3} \langle\mathcal{O}_{\alpha_1}\rangle-\frac{L}{4\pi G}\frac{\ell}{y_3^3}\phi_{1,(s)}\,.
\end{align}

\subsubsection{Results for the $r\to-\infty$ end, flat boundary}\label{otherendflat}

The above analysis concerning sources and expectation values was for the conformal boundary end located at $r\to \infty$ ($AdS_4$ boundary metric) or $y_3>0$ (flat boundary metric). In section \ref{otherend} we discussed the asymptotic expansion of the solution, with $AdS_4$ boundary,
for the conformal boundary end located at $r\to -\infty$. For this end we can then employ the coordinate transformation to flat space, as given 
in \eqref{asbfggh} but switching $r\to-r$ and $y_3\to -y_3$. This will then give the relevant quantities on the $y_3<0$ part of the conformal boundary, with flat boundary metric. Recalling the discussion in section \ref{otherend}, we can therefore obtain the flat boundary results for $y_3<0$ from those
for $y_3>0$, by making the replacements $y_3\to -y_3$ and $\kappa\to -\kappa$.

%\bibliographystyle{utphys}
%\bibliography{helical}{}

\providecommand{\href}[2]{#2}\begingroup\raggedright\endgroup

\end{document}